# The BOS-TMC Dataset: DFT Properties of 159k Experimentally Characterized Transition Metal Complexes Spanning Multiple Charge and Spin States

Aaron G. Garrison[1,#], Jacob W. Toney[1,#], Tatiana Nikolaeva[1,2,3], Roland G. St. Michel[1,4], Christopher J. Stein[2,5,6], and Heather J. Kulik[1,3,7]*

[1]*Department of Chemical Engineering, Massachusetts Institute of Technology, Cambridge, MA 02139, USA*
[2]*Department of Chemistry, TUM School of Natural Sciences, Technical University of Munich, Garching, Germany*
[3]*Institute for Advanced Study, Technical University of Munich, Lichtenbergstrasse 2 a, D-85748 Garching, Germany*
[4]*Department of Materials Science and Engineering, Massachusetts Institute of Technology, Cambridge, MA 02139, USA*
[5]*Catalysis Research Center, Technical University of Munich, Garching, Germany*
[6]*Atomistic Modeling Center, Technical University of Munich, Garching, Germany*
[7]*Department of Chemistry, Massachusetts Institute of Technology, Cambridge, MA 02139, USA*
#These authors contributed equally
*corresponding author email: hjkulik@mit.edu

ABSTRACT: We present the Boston Open-Shell Transition Metal Complex (BOS-TMC) dataset, a set of density functional theory (DFT) properties for 159k experimentally characterized mononuclear transition metal complexes (TMCs) in multiple spin states with a range of formal charges derived from the Cambridge Structural Database (CSD). To curate this set, we carried out an iterative procedure to confidently assign overall TMC charge. From this information, we then obtained properties in up to three spin states, i.e., low-, intermediate-, and high-spin for 3d metals and low- and intermediate-spin for 4d and 5d metals, depending on compatibility with the metal electron configuration, for a total of 343.8k TMC/spin combinations. At odds with prior sets, we preserved experimental heavy-atom coordinates in these structures during optimization. We report all properties using PBE0/def2-TZVP single-point energies on these structures. We introduce a scheme for computing metal-spin-dependent atomization energies, which we report for each TMC. Alongside electronic energies, we report up to seven additional properties including: HOMO, LUMO, HOMO-LUMO gap, atomic partial charges, dipole moments, atomization energies, and spin-splitting energies for a total of over 2.9M TMC-associated properties. For a representative subset of over 10k complexes chosen based on size, we evaluate the sensitivity of computed properties to exchange-correlation (xc) functional choice from a set of twelve xc functionals spanning multiple rungs of "Jacob's ladder", highlighting hotspots of TMC space that have the greatest uncertainty with respect to xc choice. In comparison to prior transition-metal datasets, BOS-TMC is both larger and more diverse in terms of charge and spin configurations and, as a result, more diverse in its range of properties. This dataset is expected to provide a high-fidelity foundation for machine-learning model development, DFT benchmarking, and exploration of transition-metal chemistry.

# 1. Introduction.

The advent of machine learning (ML) has brought to the forefront the need of ever larger datasets, from applications in ML property prediction to machine learned interatomic potentials (MLIPs).[1-4] One of the earliest and most widely used sets, QM9[5], consists of 134k structures derived from the GDB set of closed-shell organic molecules consisting of up to nine heavy organic (i.e., C, N, O, or F) atoms. This and follow-on datasets, such as those for reactions of small organic molecules[6,7] or those distorted from equilibrium[8], have transformed machine learning. For larger and more diverse molecules, such as transition metal complexes (TMCs), the Cambridge Structural Database[9] (CSD) is a centralized repository of experimental (i.e., primarily from X-ray diffraction) structural data. In comparison to QM9[5], the diversity of transition metal complexes in the CSD is much higher, both in terms of size and elemental composition[9,10] as well as chemical diversity within constituent TMC ligands in comparison to molecules in QM9.

Structures in the CSD have long provided essential insight into transition metal complex chemical bonding. Pioneering work by Orpen, Fey, and coworkers[11-17] in the 1990s used CSD structures to reveal chemical bonding trends. Nevertheless, many of these TMC mapping efforts were localized to specific metals[18-20] or ligand classes (e.g., on phosphines[21-25]). Recent data-driven studies have also leveraged experimental TMC structures in the CSD to understand spin crossover[26], redox behavior in bimetallic complexes[27], oxidation state[28] and charge[18] assignment, and electronic properties of TMCs.[29] A key consideration for typically graph-based machine learning is that the TMCs in the CSD can exist in multiple spin and oxidation states, which can be distinguished based on 3D structural properties[26,30], but are indistinguishable in terms of their molecular graphs.

An alternative approach to experimental structures is combinatorial enumeration. It has been shown that even a small sample of ligands (ca. 50) corresponds to hundreds of thousands of possible TMC structures, and this number grows rapidly with ligand pool[31]. While this strategy has been successfully used to create large sets of 100k[32-34] to millions[35,36] or billions[37] of TMCs from hundreds of ligands, these structures are not necessarily synthetically realizable. Furthermore, from additivity principles[38-41] (e.g., as has been demonstrated for low denticity ligands[31,42]), many of the structures are isomers with similar properties, making the raw number of structures higher than their effective diversity.

Over the past several years, the CSD has been leveraged by computing electronic structure properties to provide training data for machine learning models. However, previously curated datasets of TMCs from the CSD analyzed either only a subset of metals (e.g., cell2mol[18]), 86k complexes that were closed-shell in nature (e.g., tmQM[29] or the more recent 61k-subset in tmQMg[43]), or specific TMCs for spin-crossover applications.[26] For the most extensive dataset, tmQM[29], numerous improvements from the initial semi-empirical, closed-shell structures have been carried out, including data cleaning for erroneous structures[44], refinement of geometry optimizations to DFT level[44,45] (e.g., in tmQMg[43]), and expansion to excited state properties[45] (e.g., in tmQMg*). In a complementary effort, a subset of the larger OMol25[35] set contains over 70k experimental complexes curated from the Crystallography Open Database (COD)[46], which has the benefit of fewer restrictions on reuse of the data than the CSD. Nevertheless, the CSD remains a greater untapped resource for chemical exploration. Earlier work estimated that from a 2020 sample of the CSD of over 204k mononuclear TMCs, roughly 118k unique ligands were present.[10]

Despite their value, existing datasets based on experimental TMCs, such as tmQM[29,43,44] or the COD-derived OMol[35] set, have limitations. While OMol contains high-spin states for 3d metals from the COD, no expansive dataset based on CSD structures has been released in multiple spin states. The tmQM set also lacks strongly charged ligands that are a substantial component of the more varied and novel chemistry in the CSD.[29] Beyond chemical diversity, structural fidelity also remains a concern. DFT optimized geometries may deviate from experiment[47-49], so an approach that restraints geometries to those deposited in the CSD may have the benefit of capturing more realistic chemical bonding. All datasets to date have used a single flavor of DFT, but properties such as spin splitting[50-63] or frontier orbital energies[61,64] are strongly sensitive to DFT functional choice. These gaps in spin state, structure, chemical diversity, and exchange correlation functional dependence motivate the creation of a new CSD-derived dataset that addresses these limitations.

In this work, we introduce the Boston Open-Shell Transition Metal Complex (BOS-TMC) dataset of DFT-computed properties of 159k TMCs (i.e., from 126k unique molecular graphs). All TMCs were extracted from the CSD with known metal oxidation states and charges, and we algorithmically assigned the complex total charge based on decomposition of the unit cell. We report 2.9 M quantities obtained from seven PBE0/def2-TZVP properties ranging from spin-dependent atomization energies to dipole moments and HOMO-LUMO gaps of these TMCs in up to three spin states, depending upon the accessible spin states afforded by the metal d-electron configuration, for a total of 343.8k spin/TMC combinations and over 228.2k vertical spin-splitting energy pairs. Finally, over a subset of greater than 10k TMCs, we report the sensitivity of computed properties to exchange-correlation functional, providing an understanding of how xc uncertainty can propagate in large datasets.

**2. Methods.**

## 2a. Data Curation and Spin State Assignment.

Structures were obtained from the March 2024 release of the Cambridge Structural Database[9] (CSD). The CSD Python API was used to extract the relevant molecular component(s) from the associated entry. On a component-wise basis, results were excluded if they contained polymeric species, zero bonds (i.e., isolated atoms), or structures containing unknown elements. We extracted only mononuclear TMCs, which we defined as those that had i) a single transition metal atom, ii) at least 1 heavy atom coordinated to the metal center and iii) at most one atom with greater than four bonds (Supporting Information Table S1). From these initial filters, we obtained 299,035 mononuclear TMCs, which includes some structures with identical molecular graphs (Supporting Information Table S1). To ensure chemical completeness, hydrogens were added using the CSD Python API, and a quality check was performed on the hydrogen additions, leading to 270,953 retained TMCs (Supporting Information Table S1). The details of the algorithm for the quality check are provided in Supporting Information Text S1. Molecular components from unit cells corresponding to TMCs were converted into the mol2 file format, which preserves bond order.

To assign individual TMC charges, we used an iterative charge calculation workflow based on the procedure introduced in Ref. 65 with some modifications outlined in Ref. 66 (Supporting Information Text S2). To assign oxidation state, we parsed formal charges for metals from the deposited compound names, omitting any invalid oxidation states (i.e., corresponding to a negative d-electron count), yielding 165,305 complexes (Supporting Information Table S1). From this set with assigned charges and oxidation states, we carried out three additional filtering steps: i) we removed TMCs that contained deuterium, lanthanides, or actinides as they were incompatible with our DFT workflow and otherwise out of scope, ii) we eliminated any TMC where the metadata formula for the structure did not match what was deposited in the structure (i.e., typically corresponding to either duplicate or missing atoms in the crystal structure), and iii) we eliminated any TMCs that contained at least 5 post-transition metal or other inorganic atoms (a full list of elements in this category is provided in Supporting Information Table S1). This leads to a final set of 162,109 complexes (Supporting Information Table S1). Of these complexes, 128,357 correspond to unique molecular graphs (Supporting Information Table S1).

The accessible spin states for a transition metal complex (TMC) in the curated dataset were initially determined based on the metal center's oxidation state assuming ligand innocence (i.e., all ligands are treated as closed-shell singlets). Based on the metal oxidation state and nominal 3d and 4s electron count, the initial likely spin states were determined and then corroborated with information from the National Institute of Standards and Technology (NIST) atomic spectra database[9,67] (Supporting Information Table S2). The only exception to this assignment was for cases where the TMC contained open-shell ligands, namely ligands with an odd number of electrons (Supporting Information Text S3). In these cases, the accessible spin states were incremented up by one unit of spin multiplicity to reflect the additional spin contribution from the unpaired electron on the ligand (e.g., singlets became doublets). Unpaired electrons on ligands were found in 2,234 TMCs in the initial set and are labeled in the dataset in our Zenodo repository.[68] Across the whole dataset, for 3d transition metals, all d-electron count compatible spin states were included (Supporting Information Table S2). For early or late 3d metals as well as 4d and 5d metals, only intermediate-spin states, up to triplet or quartet, were included (Supporting Information Table S2).

**2b. Initial Geometry Optimizations.**

Initial density functional theory (DFT) geometry optimizations were performed with a developer version of the GPU-accelerated program TeraChem[69,70] version 1.9. In all cases, constrained geometry optimizations where coordinates of all heavy atoms are fixed were performed in the low-spin state (i.e., singlet or doublet) with the PBE0 functional[71], def2-SV(P) basis set[72], and semiempirical D3 dispersion correction[73] with Becke–Johnson damping[74-77]. This choice was made to avoid spurious changes in bonding either due to the lack of crystal packing as well as computational cost, but we note that if structures were solved poorly, this choice prevents us from correcting heavy atom positions with DFT. The hydrogen atom positions were optimized using the translation rotation internal coordinate (TRIC) optimizer[78] with default convergence criteria for maximum energy gradient (4.5 ×$10^{-4}$ hartree/bohr) and energy difference between optimization steps ($10^{-6}$ hartree). Calculations were performed in implicit solvent with a dielectric of 10 to mimic a crystal environment, as in prior work[79], using the conductor-like polarizable continuum model (C-PCM)[80,81] with cavity discretization by the improved switching/Gaussian (ISWIG)[82,83] procedure with smooth analytical first-order energy derivatives and 1.2 times the van der Waals radii[84]. A restricted formalism was used for the singlet calculations and an unrestricted one for all other spin states. All calculations used level shifting with shift values of 0.25 hartree for both the virtual alpha and beta spin orbitals, where applicable, and the hybrid DIIS/A-DIIS scheme for SCF convergence[85,86].

We computed the molecular graph hash of the complexes before and after optimization using molSimplify[87,88] version 1.7.6 and found these to be changed in 1,026 complexes. In 22 of these cases, hydrogen atoms dissociated from the structure and were labeled as problematic, whereas the other 1,004 were retained because they retained bonding but fall just outside our default bonding threshold (Supporting Information Text S4). These complexes are flagged in the csv file of our dataset provided in the Zenodo repository.[68]

Optimized geometries and molecular orbital coefficients from the converged low-spin calculations were used to initialize single-point energy calculations in the intermediate-spin state (i.e., triplet and quartet) for all metals with compatible electron configurations (Supporting Information Table S2). The same singlet or doublet geometries were used to perform high-spin state (i.e., quintet and sextet) single-point energy calculations for complexes with 3d metals in compatible electron configurations but with the molecular orbital coefficients from the converged intermediate-spin calculations as an initial guess (Supporting Information Table S2).

## 2c. Higher-Fidelity Single-Point Energy Evaluations.

Higher-fidelity single-point energy evaluations were performed using Psi4 version 1.9.1[89]. To accomplish this, basis set projections to PBE0/def2-TZVP[72] were carried out, using a slight modification of the workflow developed in previous work[61]. First, the molecular orbital coefficients of the TeraChem PBE0/def2-SV(P) calculation were used to initialize a Psi4 PBE0/def2-SV(P) single-point calculation, the output of which was then projected to PBE0/def2-TZVP in Psi4. These calculations use 590 spherical points, 99 radial points, and a $3\cdot10^{-5}$ hartree threshold for both energy and density. All basis set projections were initialized using the same structure, spin state, and restricted/unrestricted formalism as the corresponding TeraChem calculation. From these resulting single-point energies, properties were extracted, including HOMO level, LUMO level, HOMO-LUMO gap, dipole moment, vertical spin-splitting energies, Mulliken and Löwdin partial charges, and atomization energies. The vertical spin-splitting energies were calculated as the energy of the higher-spin state minus the energy of the lower-spin state (e.g., the low to intermediate vertical spin-splitting energy, $E_{IS}$ - $E_{LS}$). The dipole moment was computed automatically in Psi4, which places the center of mass of the molecule at the origin and computes the dipole as the expectation value of the dipole operator. For HOMO-LUMO gaps of open-shell systems, we adopted an energetic criterion to select between majority- or minority-spin orbitals. That is, the HOMO was the highest-energy occupied orbital and the LUMO was the next orbital expected to be filled based on it being the lowest in energy. Nevertheless, we provide the frontier orbital energies of both majority- and minority-spin orbitals of all complexes in the Zenodo repository.[68]

For atomization energies, we adopted a convention to compute approximate atomization energies from the difference between isolated atom electronic energies and the total energy of each TMC. Energies of isolated ligand atoms were calculated using PBE0/def2-TZVP, where the spin state was chosen such that the PBE0/def2-TZVP energy is the lowest (Supporting Information Table S3). The metal center's charge and spin state were always assigned to be compatible with the metal oxidation and spin state chosen for the overall complex, meaning that we report spin-state dependent atomization energies depending upon the number of accessible spin states of the TMC. Atomic PBE0/def2-TZVP energies were obtained in the same manner as energies for the intact transition metal complexes wherever possible, but alternate convergence strategies were used, when needed, to assist with convergence (Supporting Information Text S5 and Figure S1). In the case of charged TMCs, electrons were added or removed from the specific ligand atom expected to be most readily reduced or ionized. That is, we assigned excess electrons to the atom with the lowest computed electron affinity and we removed electrons from species with the lowest ionization potential (Supporting Information Text S6 and Tables S4–S5). A script for assigning individual atom charges is provided in the Zenodo repository.[68]

For a subset of over 10k TMCs, we computed properties with other density functional approximations (DFAs) in addition to PBE0, all with the def2-TZVP basis in a subset we refer to as the manyDFA set. This set of over 10k TMCs consists of the smallest unique TMCs in our set by electron count, such that we preserve the distribution of transition metal centers in the overall dataset (Supporting Information Text S7 and Figures S2–S4). The 11 other functionals we chose are: PBE[90], M06-L[91], r$^2$-SCAN[92], B3LYP[93-95], B3LYP*[52,96], M06-2X[97], LRC-ωPBEh[98], ωB97X-D[99], ωB97M-V[100], PWPB95[101], and PBE0-DH[102]. Calculations for all 11 of these additional functionals were run using the same settings as the basis set projections in Psi4 and are initialized from the Psi4 PBE0/def2-TZVP molecular orbital coefficients.

**2d. Property-based filtering for outliers.**

Successfully converged calculations were filtered in two ways. First, any non-singlet states were tested for deviation of the $S^2$ operator from the expected $S(S+1)$ value by more than 1.0, and those were flagged as spin contaminated and removed from analysis as noted. For the atomic energies necessary for computing atomization energies, we noted a few cases where deviation of the $S^2$ operator from the expected $S(S+1)$ value was very slightly above 1.0, and these cases were noted but not removed from analysis to ensure that atomization energies could still be computed (Supporting Information Tables S6–S7). Secondly, after all data for a given spin state were determined, a generalized extreme student deviate (GESD) test[103] with an alpha value of 0.05 was used to determine outliers individually for each spin state. These tests were run with an upper bound for the number of outliers to search for at 500, and in all cases, the number of detected outliers was less than the specified upper bound. GESD tests were run on both TeraChem (i.e., PBE0/def2-SV(P)) and Psi4 (i.e., PBE0/def2-TZVP) data for spin-splitting energies, HOMO-LUMO gap, metal Mulliken charge, and the magnitude of the dipole moment. For PBE0/def2-TZVP data only, outliers were also reported for metal Löwdin charges, as well as the relative atomization energy, defined as the atomization energy scaled by the molecular weight. Even if they were not identified as outliers using the GESD test, all TMCs with a positive atomization energy were also marked as outliers in the CSV files containing TMCs and their properties. For the manyDFA set, outliers were determined individually for each DFT functional following the same approach for all of the same properties for which PBE0/def2-TZVP outliers were computed.

**3. Results and Discussion.**

We first calculated the electronic properties of all entries in BOS-TMC in their lowest-spin state (i.e., singlet or doublet) in their crystal structure geometry with optimized hydrogen positions and properties evaluated at the PBE0/def2-TZVP level of theory (see Methods). We report an initial set of properties including the electronic energy, HOMO energy, LUMO energy, HOMO-LUMO gap, Mulliken partial charges, Löwdin partial charges, and dipole moment magnitude for each structure. We describe spin-state dependent properties as well as relative energies of spin states in Sec. 3b, we extend properties to atomization energies in Sec. 3c, and we explore xc functional sensitivity in Sec. 3d.

**3a. Closed-Shell TMC Properties.**

From the initial set of 162.1k structures (128,357 unique molecular graphs) for which we assigned charge and spin, we obtained H-optimized properties of 159,014 complexes (i.e., 116,975 singlets and 42,039 doublets) with charges ranging from -8 to 8 (Figure 1 and Supporting Information Table S8). We also make note of doublets that have substantial spin contamination (3,377), which we eliminate from further analysis, although this does not affect the range of charges converged (Supporting Information Table S9). While the majority of successfully converged H-optimized calculations (97,185) are neutral, 26,443 TMCs have absolute charges, $|q|$ > 1, which had typically been excluded in prior sets[29]. Overall, the success rates of the singlets were slightly higher than those of the doublets, but both were relatively high at around 98% for the initial H-optimization (96% after removing spin contaminated cases) and 97% for the triple-zeta single-point energies (96% after removing spin contaminated cases, Supporting Information Table S8). This corresponds to a final set of 122,960 unique molecular graphs for which we obtained converged calculations without spin contamination (Supporting Information Table S9). For the vast majority of these TMCs, hydrogen optimization also preserves the molecular graph (see Methods and Supporting Information Text S4). For the typically omitted[29] $|q|$ > 1 set, we observe comparable or improved convergence rates compared to the full set (Supporting Information Table S9).

In total, the number of atoms in the TMCs with converged triple-zeta single-point energies (i.e., without substantial spin contamination) range from 2 to 245, with a mean around 60 and a standard deviation (std. dev.) of 28 (Figure 1 and Supporting Information Table S10). These sizes are slightly reduced from the H-optimized results (mean of 61 and complexes as large as 292 atoms), indicating an expected propensity of larger structures to fail to converge or to hit walltime limits with our larger basis set protocol (Supporting Information Table S10). While this difference is modest, it does suggest future work could focus more on converging very large structures, if they are the focus of a targeted discovery campaign. Additionally, analysis reveals a diverse range of metal identities and coordination environments, with overlapping but distinct distributions for 3d, 4d, or 5d metal centers (Figure 1).

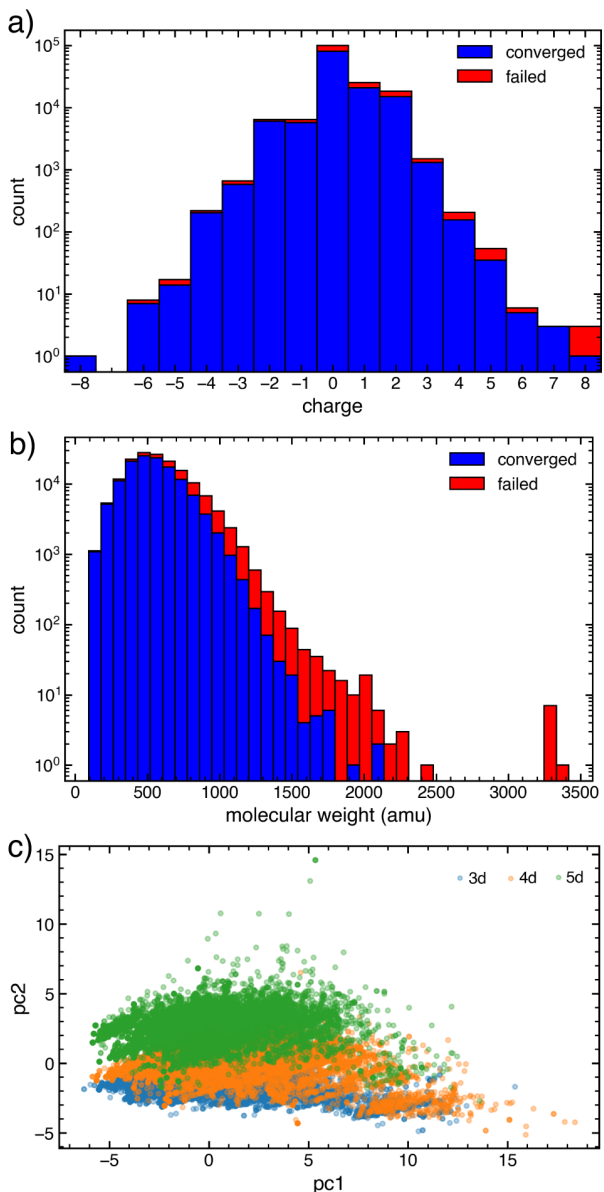

**Figure 1.** (a) Stacked histogram (i.e., grouped by converged and failed) of calculations by net complex charge. (b) Stacked histogram (i.e., grouped by converged and failed) of calculations by molecular weight (in amu). (c) Principal component analysis in metal-centered depth-two revised autocorrelations[104] of the full dataset grouped by period (i.e., 3d, 4d, or 5d metals).

We next evaluated the properties of converged TMCs after further filtering them to remove any outlier property values (see Methods). Over the full set of singlets and doublets, the triple-zeta-computed frontier orbital properties (i.e., HOMO-LUMO gap and individual HOMO or LUMO values) have a slight, expected trend with molecular weight or number of atoms in a TMC. As expected, the HOMO-LUMO gap decreases with increasing atom count (Pearson's $r$ = -0.16) or molecular weight (Pearson's $r$ = -0.19, Figure 2 and Supporting Information Figure S5). Unsurprisingly, individual HOMO and LUMO values are correlated with both each other (Pearson's $r$ = 0.94), and negatively with total charge (Pearson's $r \sim$ -0.95, i.e., more positive charge leads to deeper HOMO or LUMO values, Supporting Information Figure S5). The dipole moment magnitude does not correlate with molecular weight (Pearson's $r$ = 0.02, Figure 2). The metal partial charges computed with Mulliken or Löwdin analysis span a wide range that is relatively insensitive to overall charge of the TMC, but interestingly the two schemes have opposite trends with charge, molecular weight, or number of atoms (Supporting Information Figure S5). They also correlate only moderately with each other (Pearson's $r$ = 0.38, Supporting Information Figure S5). Specifically, the Löwdin scheme has the expected trend that increased positive charge of the complex leads to more positive metal partial charge (Pearson's $r$ 0.19), while the Mulliken scheme has the reversed trend (Pearson's $r$ -0.26, Supporting Information Figure S5). Given the expected better performance of Löwdin charges, we thus focus all subsequent analysis to only this partial charge scheme. Comparison of pairs of the three types of properties (i.e., frontier orbitals, charges, and dipole moments) also reveals either extremely limited (i.e., metal Löwdin charge versus dipole, Pearson's $r$ = 0.04) or weakly correlated trends (i.e., dipole and HOMO-LUMO gap, Pearson's $r$ = -0.23, Supporting Information Figure S5).

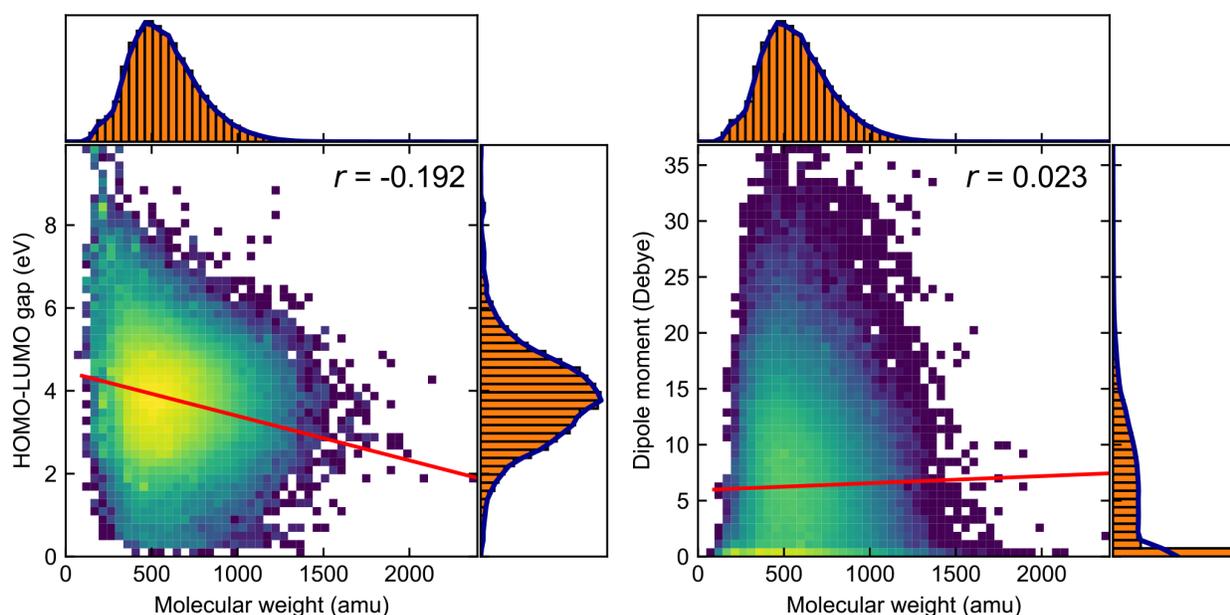

**Figure 2.** Molecular weight (amu) vs. HOMO-LUMO gap (eV, left) and center of mass dipole moment (Debye, right). Marginal 1D histograms are shown at top and right for each of the two quantities. The data is colored by KDE density in 100 bins for each dimension with a shared color bar ranging from $10^{-7}$ (dark purple) to 0.0009 (yellow) density.

BOS-TMC is not the first dataset of experimentally characterized structures of transition metal complexes from the CSD. We thus aimed to make direct comparisons to the TMCs in our set that overlap with the tmQMg[29] dataset reported in 2023. We note that in our set, no heavy atoms are relaxed, whereas they were relaxed in tmQMg with DFT in the gas phase. Furthermore, due to differences in curation strategies between the two sets, we only identified 33,448 TMCs in our set that are in common with the 60,799 present in tmQMg (Supporting Information Table S11). The main reason for this limited overlap is that we required each TMC to have a known charge and oxidation state. Other differences in composition between the two sets are attributable to tmQMg requiring all complexes to have at least one H and C atom and only allowing limited other atom types (i.e., B, Si, N, P, As, O, S, Se, F, Cl, Br, and I), which are not requirements in our set. We also have additional unique complexes deposited in the 2020–2024 period after tmQM and tmQMg were curated. We also analyzed how highly charged species (i.e., $|q| > 1$) excluded from tmQM[29] and tmQMg[43] but included in our set differed from the subset that overlaps with those structures in tmQMg. For these 26,353 highly-charged TMCs (i.e., after excluding 760 outliers for any property), we observe HOMO-LUMO gap averages of 4.24 eV with a std. dev. of 1.49 eV that are greater than the 3.78 eV average (1.03 eV std. dev.) for the same sized set of randomly selected $|q| \leq 1$ TMCs, although the ranges of the two sets are similar (Supporting Information Table S12 and Figure S6). This increase in HOMO-LUMO gap average arises from the intermediately charged positive and negative complexes (i.e., $|q|$ from 2–4), as gaps do shrink for the most strongly charged complexes in the set when HOMO and LUMO values become more comparable, regardless of their sign (Supporting Information Figure S6 and Tables S13–S15). These TMCs also have shifted dipole moment distributions than their less charged counterparts, with mean values that are smaller by over 4 Debye (i.e., 6.98 Debye vs 2.94 Debye for $|q| > 1$, Supporting Information Tables S16 and S17 and Figures S6 and S7). For metal partial charge on the other hand, there is no discernible difference between the two data subsets, consistent with the limited correlation between metal partial charge and TMC net charge (Supporting Information Table S18).

To further assess the added diversity in our set versus both the 61k CSD TMCs in tmQMg[43] or 86k CSD TMCs in tmQM[29] and 70k COD TMCs in OMol25[35], we evaluated the effect of sequential data addition on the four reported properties. Comparison to OMol25 is less straight forward due to the fact that those structures were obtained from a different database, but we expect it to have the same diversity as a 70k subset of BOS-TMC, as the curation strategies were similar. Specifically, we started from the 33,448 TMCs for which we have successfully converged single-points in common with the 61k in tmQMg[43] (i.e., after removal of property outliers and spin-contaminated TMCs, Supporting Information Table S11). We then explored how the mean and range of HOMO, LUMO, HOMO-LUMO gap, metal partial charges, and dipole moments change as we increase the dataset. To test this, we randomly add TMCs in batches of 1k starting with the less charged species (i.e., $|q| \leq 1$) that are most similar to the charge of TMCs in the tmQMg set and then adding more highly charged TMCs only after we had exhausted the set of less charged TMCs. We repeated this process with five different random folds to also obtain uncertainties. We carried out this analysis to capture the increasing diversity of the data as we increase the size to our full set of 153,495 TMCs for which we have converged singlet/doublet single-point energy-derived properties using triple-zeta basis sets (i.e., after removing GESD property outliers).

As we add more neutral TMCs beyond the 33,448 that overlap with tmQMg, our mean HOMO-LUMO gap decreases at first from 4.1 eV to below 3.8 eV, but it then increases when we add the charged species to over 3.85 eV, highlighting two competing effects in the overall properties of our set (Figure 3). We also observe significant shifts in dipole moment magnitudes, which decrease from over 7.5 Debye on average to less than 7.0 Debye before addition of highly charged complexes, after which dipole moments decrease further to 6.3 Debye on average (Figure 3). For maximum and minimum HOMO-LUMO gap, we observe the expected behavior that the ranges sampled by these complexes increase (from 0.4–8.3 eV up to 0.0–10.0 eV), as both HOMO and LUMO ranges also get wider (Supporting Information Figure S8). Conversely, mean metal partial charges (ca. -0.15) are not strongly sensitive to dataset size, although the maximum charge does increase from around 1.05 to nearly 1.20 when starting from our initial 33k set and increasing to the full 153.5k set (Figure 3).

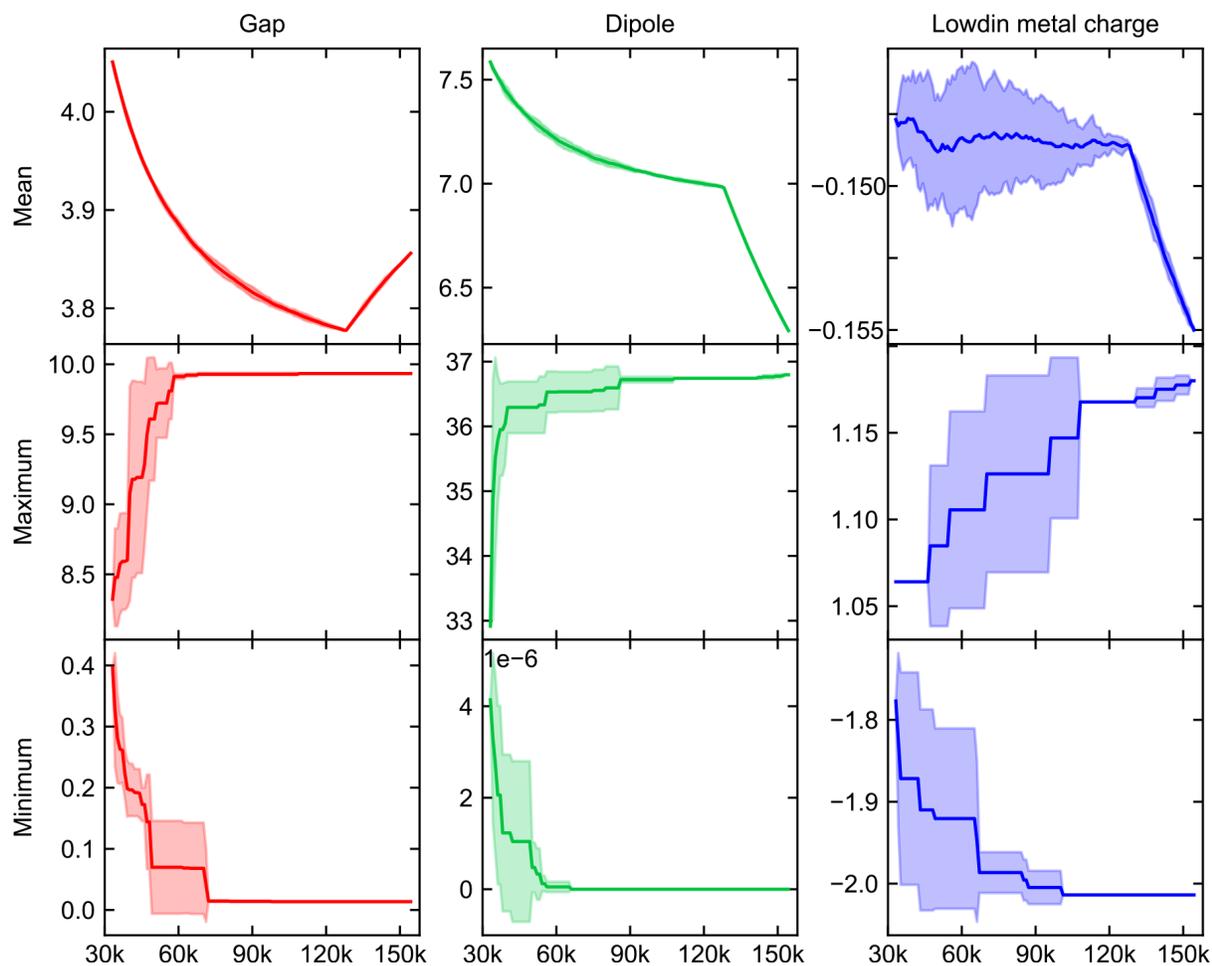

**Figure 3.** Effect of sequential addition of data on mean (top), max (middle), and min (bottom) of the energetic gap (eV, left), dipole moment magnitude (Debye, middle), and Löwdin metal charge (right) starting from a set of TMCs in common with tmQMg and adding ca. 30k $|q| > 1$ complexes last. The shaded regions correspond to the standard deviation (2x for the mean plot and 1x for max. and min.) observed when the process is repeated with five different folds of the data while preserving the ordering of addition of TMCs with respect to charge.

To find examples of TMCs with distinct properties not captured in the tmQMg overlap set, we identified outliers with respect to the mean and standard deviation (i.e., 3.77 ± 1.03 eV) of HOMO-LUMO gap of a typical set (e.g., tmQMg) consisting of the first 65k TMCs in our dataset with $|q| \leq 1$. We refer to this set of 65k TMCs as the "weakly charged" set because it excludes $|q| > 1$ TMCs. Notably, these are not GESD outliers with respect to our own dataset distributions in the larger full dataset (see Methods). Our complete set leads to 2,734 additional TMCs (or 3% of the additional set) that are in the tails of that initial distribution (i.e., mean ± three standard deviations) but are not GESD property outliers. Similar observations hold for structures in the tail of the distribution for the dipole moment, while individual HOMO or LUMO values vary even more significantly due to their differences in values with charge (Supporting Information Table S19). Of the TMCs that are in the tails of the distribution for all four properties (i.e., dipole moment, HOMO-LUMO gap, and individual HOMO or LUMO values), we identify ten TMCs that all have relatively high positive (ca. +2–5) charges (Supporting Information Table S20). These complexes all have very high dipole moments (> ca. 25 Debye), and most have small HOMO-LUMO gaps (ca. 0.4-0.6 eV for all but one 6.99 eV gap TMC), despite having strongly negative (i.e., with bound electrons) HOMO and LUMO levels (Supporting Information Table S20). Most such TMCs were not included in the tmQMg set due to their higher but still plausible charges. Thus, although experimental tmQM, tmQMg, and OMol25 datasets represent valuable resources, it is evident that an updated CSD set that includes more diverse chemistry, even in low-spin states, will provide a valuable resource for both machine learning model development and electronic structure investigations. Further consideration of spin state influence on properties is described next.

## 3b. Spin-state Dependent TMC Properties.

We next calculated properties of all BOS-TMC structures in multiplicities above their low-spin states (i.e., singlet or doublet) at the PBE0/def2-TZVP level of theory. This corresponded to intermediate-spin (IS) states (i.e., triplet or quartet) for all $d^2$–$d^9$ TMCs and high-spin (HS) states (i.e., quintet or sextet) for the $d^2s^2$–$d^7s^1$ 3d TMCs in the set. Approximately 31% of initial BOS-TMC structures (i.e., 49,892 TMCs) are initially modeled in three spin states, 59% (i.e., 96,424 TMCs) are modeled in at least two spin states, and only a small minority (10% or 15,793) are only computed in a single LS state (Supporting Information Table S2). We successfully computed 140,169 IS and 46,824 HS states (i.e., after removing spin-contaminated cases) with overall success rates of around 97% for both types of spin states (Supporting Information Tables S21 and S22). Accounting for pairs of converged calculations, the BOS-TMC set consists of 137,219 IS-LS, 45,203 HS-IS, and 45,776 HS-LS vertical spin-splitting energies (Supporting Information Table S23). In addition, we report the same properties for the higher-spin states that we did for the LS states (i.e., HOMO, LUMO, HOMO-LUMO gap, dipole moment, and Mulliken and Löwdin partial charges). This set of 358,317 attempted structure/spin configurations (corresponding to 341,849 successful calculations without spin contamination) makes BOS-TMC by far the largest dataset of electronic structure properties on experimentally synthesized transition metal complexes in open-shell configurations.

To analyze the potential benefit of including IS and HS states in our dataset, we computed the ground state spin at the PBE0/def2-TZVP level of theory (xc dependence is considered in Sec. 3d). For the midrow 3d TMCs for which we computed all three spin states and were not PBE0/def2-TZVP outliers in any property, 9,945 of 44,070 TMCs (i.e., 22%) have HS ground states (Figure 4 and Supporting Information Tables S24 and S25). Over the entire set for which LS and IS states were computed, 17,415 of 135,598 TMCs (i.e., 13%) have IS ground states (Figure 4 and Supporting Information Tables S24 and S25). Most notably, 3d metals such as Cr, Mn, and Fe have more open-shell ground states than closed-shell ground states, in accordance with expectations about mid-row 3d TMCs (Figure 4).

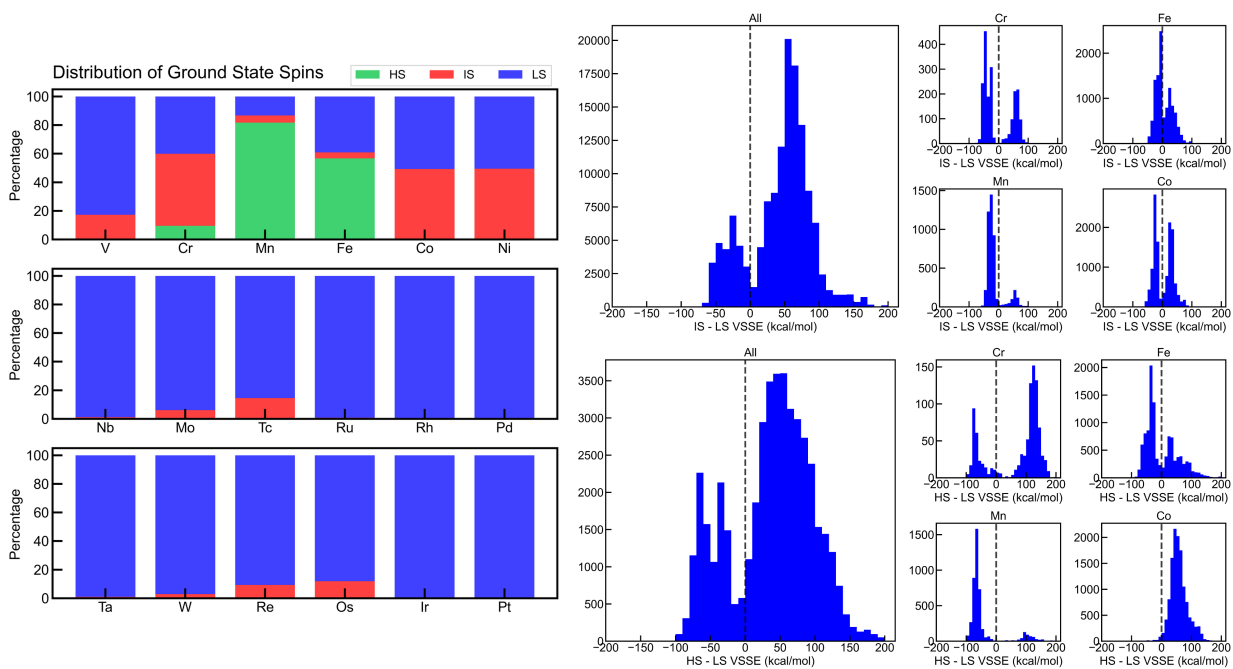

**Figure 4.** Distribution of ground state spins for transition metal complexes (left), grouped by the identity of the metal center. All transition metals omitted have >98% low-spin ground states. Distribution of vertical intermediate–low (top right) and high–low (bottom right) spin-splitting energies for both all transition metal complexes and for mid-row 3*d* metals. Only transition metal complexes which have converged and are not outliers in any property, for every spin state modeled for their corresponding metal and oxidation state, are shown.

In terms of quantitative HS-LS vertical spin-splitting energies, we observe a range of 396 kcal/mol, which exceeds that reported in our prior work[62] of around 90 kcal/mol by over 300 kcal/mol (Figure 4 and Supporting Information Table S26). One distinction with prior work is that those TMCs were all octahedral, whereas 45% of the TMCs for which we have both LS and HS states converged are not octahedral (Supporting Information Table S27). For example, the most HS-directing structures are both Mn complexes but one is a trigonal bipyramidal Mn(III)(Cl$^-$)$_5$ complex (HS-LS: -160 kcal/mol) while the other is an O-coordinated octahedral Mn(II) complex (HS-LS: -63 kcal/mol) with weak field ligands (Supporting Information Table S28). Conversely, strongly LS-directing structures include a square planar C-coordinated Cu(III) anionic complex (HS-LS 235 kcal/mol) and a Ni(III) monocationic octahedral complex with a combination of N- and Cl-coordinating ligands (HS-LS 217 kcal/mol, Supporting Information Table S28).

Over the full set of IS-LS VSSEs, (e.g., on early or late TMs as well as 4d and 5d metals), we observe an IS-LS range from -134 to 238 kcal/mol and mean of 44 kcal/mol that is nearly as large as that for HS-LS VSSEs (i.e., -160 to 235 kcal/mol and mean: 38 kcal/mol, Supporting Information Table S26). For the 4d and 5d metals for which we only computed IS-LS VSSEs, the extrema of the LS-directing and IS-directing TMCs are both non-octahedral (Supporting Information Table S28). These consist of a strongly LS tetrahedral Ir(III) cationic TMC with strong-field C-coordinating ligands (IS-LS: 239 kcal/mol) and a Pd(II) square planar neutral TMC with a mix of C- and Cl-coordinating ligands (IS-LS: -134 kcal/mol, Supporting Information Table S28).

For the subset where all three spin states are known, the IS-LS VSSE values have a range (-109 kcal/mol to 106 kcal/mol) and mean (-1.9 kcal/mol) that is reduced from the full HS-LS VSSE by about half (range: -160 to 235 kcal/mol and mean: 38 kcal/mol) and are reduced from the IS-LS computed over the whole set (Supporting Information Table S29). This is expected behavior for the prototypical case where IS states reside between LS and HS states when the three states are computed. For this subset, the HS-IS spin-splitting energies are more comparable to the HS-LS in terms of the mean (mean: 40 kcal/mol), but its range is reduced (-112 kcal/mol to 186 kcal/mol), consistent with IS states being more comparable to the LS state on average (Supporting Information Table S29). Overall, while the set contains more positive VSSEs (i.e., low-spin ground states), the large range suggests a substantial number of highly favored HS and IS states as well. This enhanced diversity in VSSEs is expected to offer opportunities for training more general spin-state prediction models beyond those that have primarily focused on octahedral TMCs.[26,61,105]

We next considered the extent to which studying TMCs in their revised (i.e., IS or HS) ground state would alter the properties we obtained on the LS states (in Sec. 3a). For the full set, 20% of TMCs (i.e., 27,360) computed in multiple spin states have IS or HS ground states (Supporting Information Table S25). Comparing to only the 33k TMCs our set has in common with the tmQMg set, we observe that 2,170 of these would have reassigned IS or HS (i.e., 1,353 IS and 817 HS) ground states at the PBE0 level of theory, which corresponds to a relatively small 7% of TMCs in this subset (Supporting Information Tables S11 and S30). This reduced rate of IS/HS structures in the set that overlaps with tmQMg relative to the full set could be due to a relatively low percentage of mid-row 3d metals in this set (ca. 6,622 or 20%) and the higher rate of oxidation states even among these metals that are typically LS (e.g., Co(III)).

We next returned to the full set to explore how studying properties in the GS instead of the LS state alters the reported values. For all 17,415 complexes in IS ground states or of 9,945 TMCs in HS ground states, we first compare the effect of changing spin state on dipole and metal partial charge as a reporter of change in the electronic state character (Supporting Information Tables S24 and S25). We observe absolute dipole moments to shift by 0.40 Debye on average (range: 0.00–23.34 Debye) on average from their LS values, and shifts in Löwdin partial charges are also substantial at around 0.l0-0.18 a.u. (range: 0.00–0.55 a.u., Figure 5 and Supporting Information Tables S31 and S32). These average shifts are consistent with a picture where HS and IS states populate successively more antibonding orbitals and therefore localize more charge on the metal. The largest deviation in dipoles from LS to the IS/HS GS in the set are observed for a LS-to-IS octahedral Co(III) TMC with N-coordinating ligands (shift = 23.3 Debye), and more moderate shifts of around 8-9 Debye are observed for LS to IS/HS Co(II/III) octahedral TMCs (Supporting Information Table S33). Nevertheless, there are also cases where changing the spin state has no effect (ca. 0 Debye shift) on a non-zero dipole moment, for example in anionic Co(I) or neutral Co(II) TMCs with mixed-field-strength linear N-coordinating or trigonal bipyramidal mixed Cl/N/P-coordinating ligands, respectively (Supporting Information Table S33). From these overall examples, we ascertain that intermediate field strength ligands and octahedral Co(II/III) structures in some but not all cases lead to greater shifts in dipole properties when revised to an IS or HS ground state.

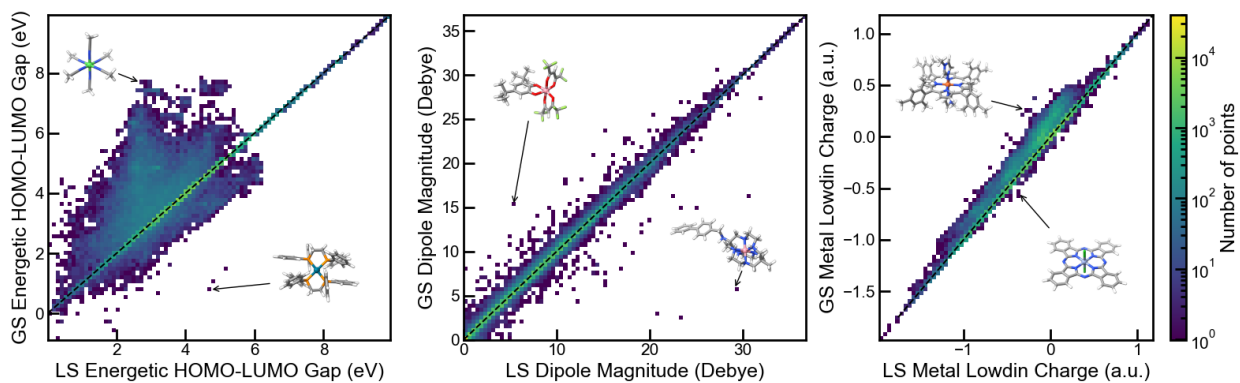

**Figure 5.** Parity between the ground state and low-spin state for the energetic HOMO-LUMO gap (left), magnitude of the dipole moment (middle), and metal Löwdin charge (right). Representative structures with large deviations from parity are shown. Only transition metal complexes which have converged and are not outliers in any property, for every spin state modeled for their corresponding metal and oxidation state, are shown. Atoms are colored as follows: hydrogen in white, carbon in gray, nitrogen in blue, oxygen in red, fluorine in green, phosphorous in orange, chromium in light blue, iron in dark orange, cobalt in pink, nickel in bright green, and palladium in dark blue.

For metal partial charges, we observe shifts as large as ca. 0.3–0.4 a.u. from the LS to the IS GS for an anionic Ni(II)($H_2O$)($Cl$)$_2$ or cationic Fe(III) octahedral complexes with weak field O/N-coordinating ligands (Supporting Information Table S34). Conversely, for cases where charges are essentially unchanged from LS to the IS GS, it is harder to make a general observation but the smallest shifts generally correspond to early (i.e., Cr(III)) or late (i.e., Zn(II)) metals with nearly full or empty d shells (Supporting Information Table S34). For charges, the largest shifts are larger from LS to HS GS (vs IS GSes), and the largest values are again observed in Ni(II) or Fe(III) octahedral TMCs with intermediate field N-coordinating ligands (Supporting Information Table S34). There are nevertheless also cases with minimal shifts in charge occur from the LS state to the HS GS, in this case again corresponding to Cr(II) or Co(II) TMCs with relatively empty or filled d-shells among the metals for which we computed HS states (Supporting Information Tables S33 and S34). Regardless of specific chemical trends, electronic properties of TMCs are often shifted when considering their open-shell ground states in comparison to the closed-shell, higher-energy spin state.

For the HOMO, LUMO, and HOMO-LUMO gap properties, we would expect even more substantial shifts with the change in electronic state because the characters of the frontier orbitals shift upon revision of the ground state spin. For example, a $d^6$ singlet octahedral TMC has three doubly occupied orbitals with a large gap to the first unoccupied orbital while the equivalent quintet should have a much smaller gap between the highest occupied minority spin orbital and the next frontier orbital. Comparing LS HOMO-LUMO gaps to revised HOMO-LUMO gaps corresponding to IS or HS ground states, we observe an absolute shift on average of 1.28 eV (range: 0.00–5.00 eV) for IS ground states and a smaller average shift for HS ground states of 0.65 eV (range: 0.00–4.01 eV, Figure 5 and Table S35). Analyzing HOMO level shifts and LUMO level shifts individually, the HOMO generally decreases with the reassigned higher-spin ground state while the LUMO increases. This explains why the majority of TMCs exhibit an increase in HOMO-LUMO gap for the LS versus upon revision of the ground spin state (Figure 5 and Supporting Information Figure S9). The largest deviations in individual HOMO or LUMO values from LS to IS or HS GS in the set are around 2.5–4.5 eV (Supporting Information Table S35).

For the TMCs with the largest shift in HOMO-LUMO gap, HOMO level, or LUMO level from LS to IS/HS GS, we observe many of the largest shifts (e.g., 5.0 eV in HOMO-LUMO gap) to be in very small (i.e., with monatomic ligands) tetrahedral structures as well as a number of cases in larger octahedral structures (Supporting Information Table S36). There are also counter examples where the shifts in these properties are minimal, typically in octahedral TMCs, although that may be due to the fact that they are well represented in our dataset (Supporting Information Table S36). The wide range of shifts in frontier orbital energy properties across the TMC sets highlights the challenges associated with aiming to generalize trends across a diverse dataset regarding when properties will be strongly spin-state-dependent.

Overall, adding IS and HS states to the set of computed properties shifts their mean values but tends not to increase the maximum reported value over that for the LS states alone. Considering only the ground state IS and HS states, the average HOMO-LUMO gap stays comparable or increases (LS: 3.86 eV vs ca. 3.64–4.08 eV for IS and HS GSs, Supporting Information Table S37). If we instead consider all IS and HS structures, the average gap of those structures is lower at around 1.87 eV (Supporting Information Table S37). Even after eliminating GESD outliers, we observe a number of negative HOMO-LUMO gap structures in the open-shell cases because our outlier removal does not disallow these structures, but the minimum value for the gap is much closer to zero for IS and HS GS TMCs (Supporting Information Table S37). This trend in the gap is consistent with lower average HOMO and LUMO values (by ca. 1.4–2.0 eV) for IS and HS TMCs as well (Supporting Information Table S37). Mean metal partial charges are relatively insensitive to spin state in comparison to dipole moment or frontier orbital energies, but the range of values is narrower for HS states (-1.73 to 1.03 a.u.) than it is for the LS (-1.97 to 1.18 a.u.) or IS states (-2.03 to 1.12 a.u., Supporting Information Table S37). Dipole moments instead show more substantial differences. They are, on average, lower for the IS states (5.61 Debye vs. LS 6.28 Debye) and lowest for HS states (5.04 Debye), although this effect is again smaller, or even reversed, when considering only higher-spin ground states (5.64–6.59 Debye, Supporting Information Table S37). Given the greater diversity in property values with the same molecular graph but different spin, we anticipate the spin-state-specific properties reported in the BOS-TMC set to be useful for developing spin-aware data-driven models.

In our analysis of ground state reassignment, it is worth considering sensitivity to xc functional. We consider this in more detail in Sec. 3d. Nevertheless, because we compute vertical spin-splitting energies (VSSEs), we are computing multiple spin states at a single geometry that corresponds to the ground state structure in the crystal for a single spin state. As a result, the VSSE gap should be larger than if we computed an adiabatic one that involved relaxation, and therefore we would expect that the sensitivity to the xc functional of the ground state assignment (i.e., not necessarily the magnitude of the gap) to be reduced relative to the adiabatic case.

**3c. Atomization Energies of TMCs.**

We next computed atomization energies (AEs) of all TMCs in the dataset in each spin state. For TMCs with net charges that differ from those of the metal ion oxidation state, we introduced an algorithm dependent on individual atoms' electron affinities and ionization potentials to assign the excess charge (Supporting Information Text S6). We computed these atomization energies as the difference in the total electronic energy from the constituent atoms (i.e., without any zero-point energy corrections). Over the dataset of singlet or doublet structures, atomization energies span a wide range, from –1272.88 to –1.96 eV, which expectedly tracks with molecular weight (Figure 6). As the metal center is just one atom in a relatively large TMC, we do not expect open-shell atomization energies to differ substantially from the closed-shell counterparts. Nevertheless, for 4.6% of the 138,773 complexes where we have computed more than one spin state, the open-shell atomization energy differs by more than 5% from the closed-shell counterpart (Supporting Information Figure S10). We observe discrepancies up to 55.2% differences in the atomization energy (Supporting Information Figure S10). Analyzing the subset with the most extreme differences, they all correspond to cases that have relatively small atomization energies (e.g., Cu or Zn with Cl and Br ligands), explaining the drastic differences when the metal spin state is altered (Supporting Information Table S38). Over the full set, the metals with the greatest average shift in atomization energies are primarily late transition metals (i.e., Cd, Zn, Au), with some exceptions (i.e., Cr), although a number of other metals (e.g., Mn, Fe, Cu) have several hundred TMCs with large spin-state-dependent shifts in the atomization energy (Supporting Information Table S39). Besides these extreme cases, the dependence of the atomization energy on spin state exhibits metal- and complex-charge-specific trends (Figure 6). To make analysis less sensitive to the molecular weight of the complex, we also report the relative atomization energy (relAE), which is the atomization energy divided by the molecular weight. In particular, elevated median values of the absolute spin-state-dependent shift in relAEs is highest for late transition metals Zn, Cd, Pd, and Ag. A further stratification by the charge of the complex indicates that these effects are not determined solely by the metal but depend on the metal-charge combination, most notably for trianionic Co TMCs that exhibit a high spin-state dependent shift (Figure 6). We expect that some metal- and charge-specific trends in atomization energies cannot be fully deconvoluted from dependence on the ligands, as large shifts can be expected for complexes with the smallest number of atoms, but this analysis broadly suggests that late transition metals are more prone to having strongly spin-state dependent atomization energies.

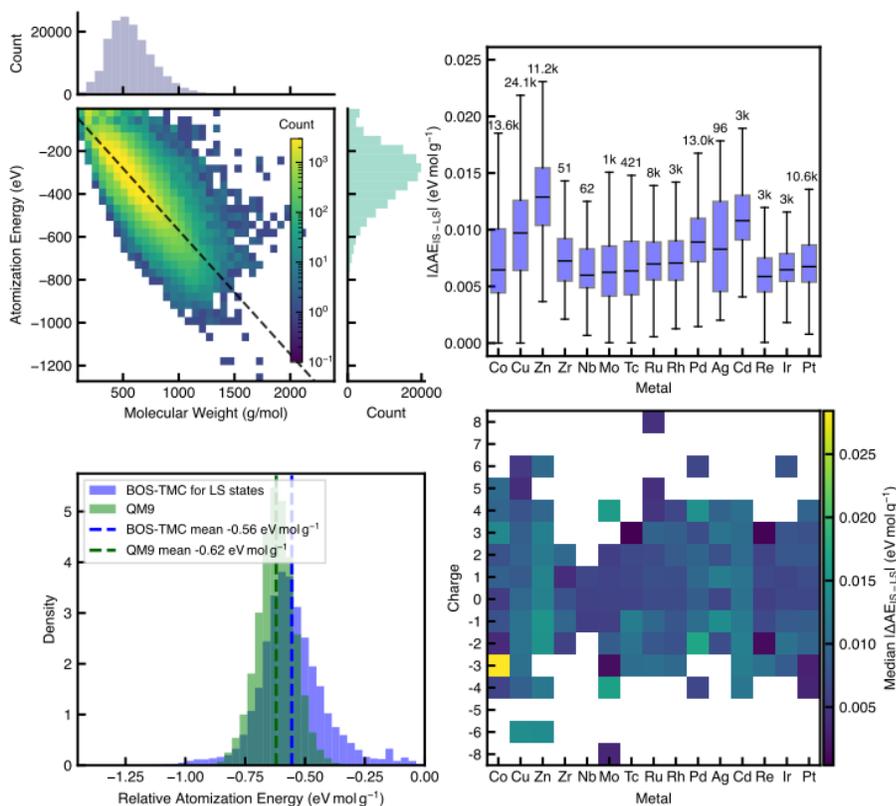

**Figure 6.** Atomization energy statistics for BOS-TMC. The top-left panel shows the joint distribution of atomization energy and molecular weight for low-spin (singlets or doublets) structures as a log-scaled 2D density map with marginal histograms and a linear best-fit trend (dashed line). The bottom-left panel compares the distribution of the relative atomization energies with respect to molecular weight (eV mol g$^{-1}$) for low-spin BOS-TMC structures against QM9, with dashed lines indicating the mean for each dataset. The top-right panel summarizes per-metal low-to-intermediate spin splittings in absolute relative atomization energies (eV mol g$^{-1}$) for the 15 metals with the largest median splitting, ordered by periodic table and annotated with the number of complexes for each distribution. The bottom-right panel supplements the previous panel by showing the heatmap of the median absolute relative atomization energy spin differences as a function of metal and charge.

In order to make comparisons to other datasets, we focus our analysis on relAE values. We first compare to the relAE values reported for the 134k complexes in the QM9 dataset[5], which were obtained with the B3LYP level of theory and 6-31G(2df,p) basis set. In comparison to QM9, the mean relAE in our dataset is less negative (-0.56 eV mol g$^{-1}$ vs. -0.62 eV mol g$^{-1}$), suggesting that our TMCs are generally less stable than the organic molecules in QM9 (Supporting Information Table S40). Notably, our dataset also spans a wider range of both more and less stable molecules from –1.286 to –0.003 eV mol g$^{-1}$ vs –1.07 to –0.236 eV mol g$^{-1}$ in QM9 (Figure 6). We also analyzed the atomization energies in terms of the more neutral TMCs representative of those included in tmQMg[43] (i.e., which did not report atomization energies) as well as the more highly charged TMCs only in our set. For the subset of TMCs we identified with high net charges (i.e., $|q| > 1$) that had been excluded from tmQMg[43], their higher net charge leads to lower-magnitude AE or relAEs (Supporting Information Figure S11). We also quantitatively compared only closed-shell singlet atomization energies or relative atomization energies between the 26,219 non-outlier, singlet cases in our set with $|q| > 1$ to a similarly sized set of $|q| \leq 1$ TMCs more representative of tmQMg[43]. Over these two sets, we observe the expected behavior that the average stability of more highly charged TMCs is lower (mean: -0.519 eV mol g$^{-1}$ vs -0.563 eV mol g$^{-1}$), although both sets have similar minimum values (i.e., -1.269 eV mol g$^{-1}$ for $|q|>1$ vs -1.267 eV mol g$^{-1}$ for $|q|\leq1$, Supporting Information Table S41). As expected, the range for $|q|\leq1$ cases expands when considering all data and not just an equally-sized subset, highlighting the added diversity that comes simply from increasing the number of TMCs considered (Supporting Information Table S41).

Focusing specifically on the most extreme cases in the relAE and AE in our full set, we observe that early TMs (e.g., Mo or Cr) metals with large O/N-coordinating organic ligands have the most negative atomization energies (Figure 6 and Supporting Information Table S42). Conversely, the species with atomization energies closest to zero generally have Cu or Ag metals with small monoatomic halide (i.e., Cl, Br, I) ligands (Figure 6 and Supporting Information Table S42). Although it is challenging to decouple ligand preferences of metals from trends in metal-specific atomization energies, we observe that early (e.g., Ti, V, Mo) and mid-row (e.g., Fe, Co) metals have the most negative relAEs, while late TMs (e.g., Cu, Zn, Pt, Au, Hg), consistent with expectations based on their electron-configuration-dependent bonding preferences, have the least negative relAEs (Supporting Information Table S43). Overall, analysis of the atomization energies in the BOS-TMC dataset suggests that our computed atomization energies should be a rich target for assessing ML model architectures with a diversity not present in other sets.

**3d. Method-dependence of Properties.**

We investigated the sensitivity of reported properties to DFT functional choice by evaluating properties with multiple exchange correlation functionals on a subset of the BOS-TMC set. Our full set of twelve exchange-correlation functionals span the rungs of "Jacob's ladder" from semi-local to double hybrid functionals (see Methods), and thus we selected over 12k TMCs (referred to as the manyDFA set) for their small size while preserving adequate representation of all metals in the full set in order to reduce computational cost particularly for double hybrids (Figure 7 and Supporting Information Table S44). From this initial set, we successfully computed properties for over 10k TMCs in up to three spin states (Supporting Information Table S44 and Figure S12). Some properties in the BOS-TMC set, such as HOMO, LUMO, or HOMO-LUMO gap, would be expected to be strongly sensitive to how well the exchange correlation functional recovers piecewise linearity[106] (i.e., eliminates delocalization error[107-109]). Thus, we emphasize comparison of properties that are expected to be more comparable across functionals: vertical spin-splitting energies, dipole moment magnitude, metal partial charges, and atomization energies.

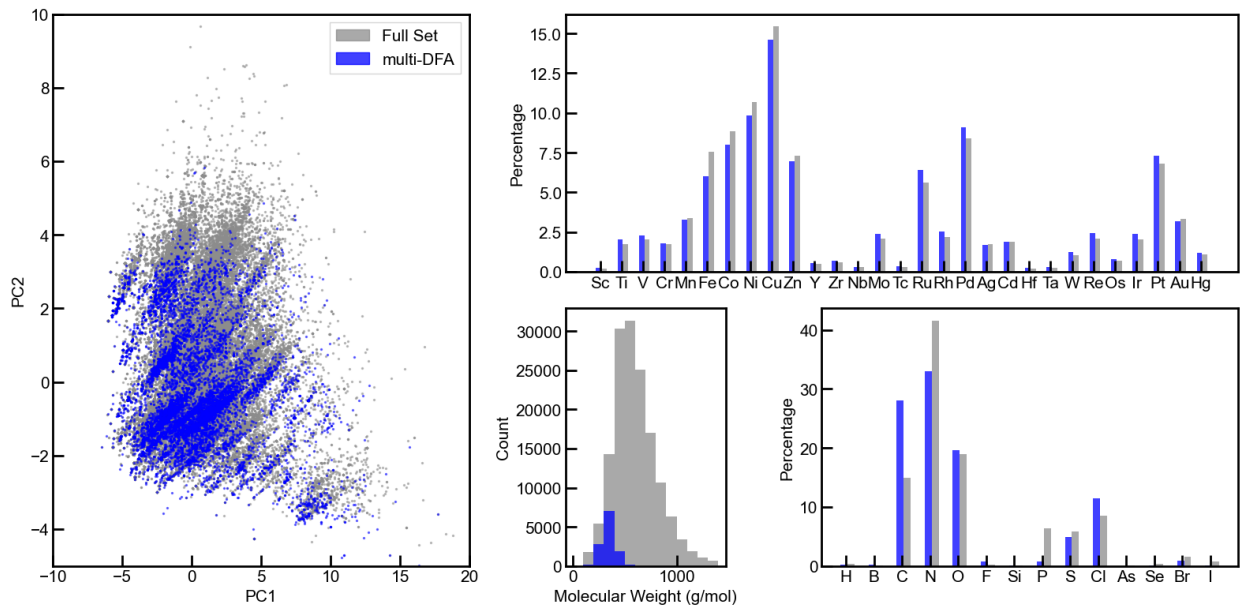

**Figure 7.** Distributions of properties in the manyDFA set compared to the full distribution in BOS-TMC. Principal component analysis of depth-2 metal-centered revised autocorrelations of the manyDFA set plotted over the full BOS-TMC set (left). Comparison how frequently different metal centers occur in the sets (top right). Distribution of molecular weights in the manyDFA set overlaid on the distribution in the full set (bottom middle). Comparison of how frequently different coordinating atoms occur in the manyDFA vs. full set (bottom right).

Spin-splitting energies are known to be strongly sensitive to exchange correlation (xc) functional, especially the Hartree–Fock (HF) exchange fraction.[50-60] For the 9.8k TMCs in our manyDFA set that have properties computed at multiple spin states, we compute standard deviations (std. dev.) over the 12 functionals in the VSSE as high as 25 kcal/mol for IS-LS splittings and 40 kcal/mol for HS-LS spin splittings (Supporting Information Table S45). This means that 141 TMCs would have disagreements for the ground state assignment (Supporting Information Table S46). Comparing trends in metals and coordinating atoms among the most extreme sets with high and low std. dev. in VSSE, we observe the most extreme variation in spin splitting occurs in a range of field strengths in both octahedral and tetrahedral/square planar geometries (Figure 8 and Supporting Information Table S45). Thus challenge cases are well represented across coordination chemistry. A range of metals is also observed in these extreme cases, including Fe or Cu for the IS-LS cases and Fe or Ni for the HS-LS cases, all of which are in M(II) oxidation states but with a range of net charges (-2 to 2) on the TMC (Supporting Information Table S45). Nevertheless, for 23.8% of the data, standard deviations across all xcs in the set are as low as 3 kcal/mol. These less variable spin splitting energies (std. dev. of 1–3 kcal/mol) correspond to a range of TMCs that are generally four-coordinate or lower coordination with late 3d TMs or 4d/5d metals (Figure 8 and Supporting Information Table S45).

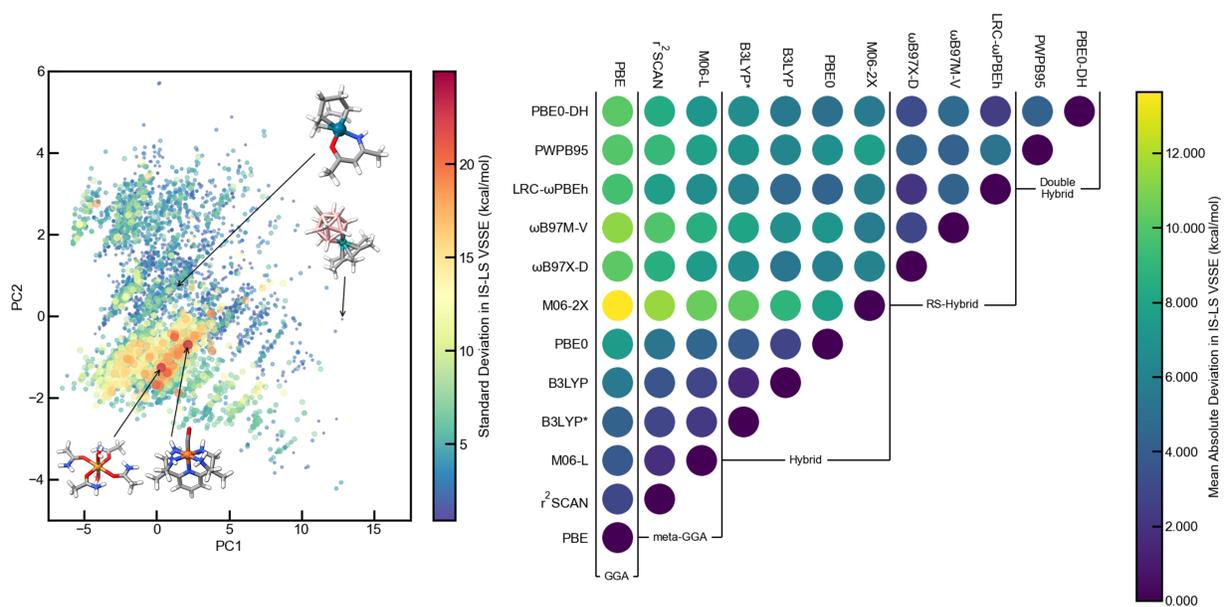

**Figure 8.** Principal component analysis of depth-2, metal-centered revised autocorrelations in the manyDFA set, colored and sized by the standard deviation in the intermediate-low vertical spin-splitting energies among 12 functionals (left). Only structures which converged and were not spin contaminated for all 12 functionals and were not outliers for any property for any functional are included. Selected structures with high and low standard deviations have been highlighted. Atoms are colored as follows: hydrogen in white, boron in pink, carbon in gray, nitrogen in blue, oxygen in red, iron in dark orange, copper in light orange, ruthenium in light teal, and palladium in dark teal. Triangle plot colored by the mean absolute deviation between pairs of functionals (right). Functionals are grouped by which rung of Jacob's ladder they belong to.

Over the full set, range-separated hybrids (i.e., LRC-ωPBEh and ωB97M-V or ωB97X-D) and double hybrids (i.e., PBE0-DH and PWPB95) tend to agree most closely in terms of the correlation of IS-LS VSSE values (Figure 8). Wider disagreement is observed among the larger set of global hybrids (e.g., M06-2X and B3LYP*), which have a mean absolute deviation of around 10 kcal/mol, Figure 8). Of all the functionals, PBE and M06-2X disagree most strongly with other functionals, although it is noteworthy that $r^2$-SCAN exhibits substantive disagreement with higher-rung functionals as well (Figure 8). Interestingly, M06-L agreement with other functionals is better than that of $r^2$-SCAN, despite the fact that M06-2X from the same family disagrees with other xcs, suggesting the overriding effect of HF exchange (Figure 8).

As HF exchange (HFX) is one of the greatest factors in tuning spin splitting, we created four groupings: semilocal functionals with low or no HF exchange (PBE, M06-L, and $r^2$-SCAN) those with moderate HF exchange (B3LYP*, B3LYP, PBE0), range-separated hybrids (RSHs: LRC-ωPBEh, ωB97X-D, ωB97M-V), and those with high HF exchange (M06-2X, PBE0-DH, PWPB95). These groupings do not necessarily correspond to "rungs" of Jacob's ladder as they mix meta-GGA hybrids and double hybrids, for example, when the HF exchange fraction coincides or meta-GGAs and PBE when the functionals lack HF exchange. When we compute std. devs. only among functionals in each HFX-grouped subset, they decrease markedly in most cases (i.e., by 50–60%) compared to the std. dev. across the full set but still remain high for high-HFX calculation of HS-LS VSSE (Supporting Information Table S47). This unusual behavior of the high-HFX group can likely be attributed to the uncorrelated behaviors of M06-2X with the double hybrids (Supporting Information Table S47).

Complexes with the greatest degree of disagreement for the IS-LS or HS-LS VSSE even when grouped by HF exchange fraction are distinct for each grouping (Supporting Information Table S48). Range-separated hybrids and moderate exchange groupings exhibit the highest sensitivity for prototypical Fe(II/III) complexes (both octahedral and linear), although other TMCs are the most sensitive for other HF exchange groupings (Supporting Information Table S48). The greatest std. devs. for HS-LS VSSEs grouped by HF exchange are more uniform because they are predominantly computed for mid-row 3d metals and thus correspond to Mn(II) or Fe(II/III) metal centers and one Ni(II) complex (Supporting Information Table S48). The associated ligand fields with the greatest disagreement are strong-field C-coordinating linear haptic for the RSH case (std. dev. 25.5 kcal/mol) whereas semi-local and moderate groupings (semilocal: 17.3 kcal/mol, moderate: 34.1 kcal/mol) functionals have weaker field N- or N/-Cl coordinating ligands in higher coordination numbers (Supporting Information Table S48). Thus, the highest sensitivities to xc choice in evaluating spin splitting energy are strongly method- and ligand-dependent and cannot be isolated to a few coordination numbers, types, or metal oxidation states. We have provided a ranked list in the Zenodo repository[68] of complexes by exchange correlation sensitivities and anticipate that the most sensitive TMCs could be a useful set of molecules for further benchmarking methods.

Dipole moments and metal partial charges are both reporters of the overall electron density and provide insight into whether discrepancies between exchange-correlation functionals are evident not just in properties but also in the quality of their densities[110-113]. Although we now compare dipole moments and metal partial charges across the 10k subset, in comparison to spin splitting energies, the range of properties sampled for both dipole moments and metal charges is considerably lower. Dipole moments have moderate standard deviations as large as 3.9 Debye across the set (Supporting Information Table S47). Representative structures with these greatest differences are characterized by Cu(II) or Cd(II) metals in square planar, square pyramidal, and pentagonal bipyramidal geometries with a mix of predominantly N- and O-coordinating ligands, which bears some similarity to the same complexes that had high IS-LS VSSE deviations (Supporting Information Tables S45 and S49). The TMCs with the highest standard deviations in dipole moments also correspond to cases with high average dipole magnitude values, but there are a number of high dipole cases with low std. dev. as well (Supporting Information Figure S13). In this analysis, we exclude cases with trivially zero dipole moments (i.e., < 0.1 Debye) that would be expected to drive down the standard deviation (Supporting Information Table S49). Complexes with a dipole of significant magnitude but little variation among functionals include a Ag(I)(Cl$^-$)$_4$ TMC as well as two larger Pd(II) and Sc(III) TMCs with intermediate ligand field strengths (Supporting Information Table S49). These TMCs with low dipole moment variation are more distinct from the set that had low VSSE standard deviations, but we note that the lowest sensitivities for the dipole correspond to $d^0$ or $d^{10}$ metals (Supporting Information Table S49). Grouping by HFX strength, we observe that the dipole moment standard deviations decrease markedly within a subclass with respect to the whole set of functionals but slightly less so for the semilocal functionals, likely due to the uncorrelated behavior of PBE in comparison to M06-L or r$^2$-SCAN (Supporting Information Table S47). Returning to "rung"-based analysis, we also observe improved agreement within each "rung" subgrouping for dipole moment magnitude prediction in comparison to other properties such as spin-splitting, with the greatest discrepancy remaining among PBE or M06-2X especially in comparison to double hybrids (Supporting Information Figure S14).

Analyzing metal partial charges, we observe a relatively low deviation among functionals overall (Supporting Information Figure S15). The greatest deviations (ca. 0.1 a.u.) are for PBE compared to double hybrids. Consistent with observations of anomalous density behavior of the Minnesota functionals[111], M06-L and M06-2X have more distinct partial charges versus the VSSE or dipole moments (Supporting Information Figure S15). We initially expected that behavior for charges and dipoles should be consistent, but we observe no correlation between the cases with large standard deviations in dipole moment and metal partial charge properties (Supporting Information Figure S16). Nevertheless, trends in grouping by functional HF exchange fraction remain consistent, with this grouping lowering the standard deviation substantially for all classes but the semilocal set, for which the std. dev. instead increases (Supporting Information Table S47). Overall, we observe greater xc agreement in terms of correlations of dipole moments or metal charges computed across functionals than for spin splitting energies (Figure 8 and Supporting Information Figures S14 and S15). This emphasizes that deviations in xc predictions of spin splitting can occur even when densities and density-derived properties are consistent.

Next, we compared atomization energies and the relative atomization energies computed with each of the DFAs. First, to determine if any of the DFAs bind TMCs more strongly on average than others, we computed the mean relAE for each xc functional. In line with our expectations of delocalization error, the semi-local functionals (e.g., PBE) had the most negative average relAE (Supporting Information Table S50). Unnormalized AE values had standard deviations as large as 3.45 eV across the set. The structures with these greatest standard deviations in atomization energies are characterized by high-valent early transition metal (i.e., Ti(IV), V(IV), V(V)) anionic octahedral TMCs with azide ligands (Supporting Information Table S51). Similar trends are observed for relAEs, but the outliers shift slightly to include not just two V(V) complexes but also a Co(III)(CO)$_4$ TMC (Supporting Information Table S52). The TMCs with largest variations in atomization energies are necessarily distinct from those that had the largest spin splitting variation or metal partial charge or dipole variation because most of these TMCs have either d$^0$ or d$^1$ formal electron configurations. Nevertheless, we generally observe a large grouping of TMCs with relatively high deviations in both dipole and atomization energy in the PCAs for the two quantities (Supporting Information Figures S13 and S17). Several complexes have AEs or relAEs that are essentially unchanged with functional choice, and these generally correspond to $d^{10}$ (i.e., Cu, Ag, Cd, or Zn) metals coordinated to halides (Supporting Information Tables S51 and S52). Next, we compared the correlation of both AE and relAE between pairs of functionals (Supporting Information Figures S18 and S19). We observed the greatest deviations of AE or relAE for PBE with any other functional, and all other functionals agreed quite closely in comparison (Supporting Information Figures S18 and S19). Here, M06-2X is no longer an outlier by comparison, and trends hold regardless of whether AE or relAE are being compared.

As with the other properties, we also investigated if grouping different functionals separately based on HF exchange fraction would lead to smaller ranges in properties for AEs or relAEs. As expected, the AE or relAE standard deviations are substantially lower on average for all other rungs besides the semilocal grouping due to the unusual behavior of PBE, although maximum and average values are also slightly above the values observed over the whole set in the high-HFX subset as well (Supporting Information Table S47). In comparison to spin splitting energies where grouping by HF exchange fraction class reduced average standard deviations by around 50–60%, a similar or larger reduction is observed for AE/relAE only in the case of range-separated hybrids or high-HFX cases but less substantial reductions are observed in the moderate HFX or semilocal functional categories (Supporting Information Table S47). Thus, HF exchange is likely not the sole parameter that influences atomization energies, which are likely also sensitive to differences in treatment of correlation.

Finally, we examined and ranked TMCs by their greatest deviation in all four key properties that we compared (i.e., IS-LS VSSE, dipole moment, Löwdin metal charge, and relative atomization energy). From this set, primarily neutral, square planar Cu(II) TMCs have the uniformly highest standard deviations, but the list also includes a neutral Ni(II) and a cationic Co(III) complex (Supporting Information Table S53). The ligand chemistry is relatively similar among these compounds, with primarily a combination of intermediate field N- and weaker field O/S-coordinating ligand atoms (Supporting Information Table S53). These observations highlight Cu(II) centers as a potential target for data-driven or fundamental electronic structure method development and emphasize that more attention should be paid to later 3d transition metals. Returning to comparisons among the functionals studied, PBE and M06-2X most consistently disagree with each other and among other functionals (Supporting Information Table S54). In comparison, the range separated hybrids ωB97X-D and LRC-ωPBEh have very high correlations with each other (Supporting Information Table S54). For properties such as atomization energy, we even see very high correlations among the global PBE0 hybrid and LRC-ωPBEh, which we expect may occur due to the narrow range of molecular sizes we study in the current set (Supporting Information Table S54). Finally, to aid future method development and validation, we provide in our Zenodo repository this cumulative ranking of transition metal complexes.[68] We also provide the PBE0 hydrogen-optimized structures of the 100 most challenging TMCs for DFT, as judged by their highest standard deviation across all functionals and properties considered.[68]

## 4. Conclusions

In summary, we curated the BOS-TMC dataset of DFT properties of 159k experimentally characterized mononuclear transition metal complexes (TMCs) corresponding to 126k unique molecular graphs spanning multiple charge and spin states. We first employed a text-parsing and iterative charge assignment scheme to assign metal oxidation states and overall charges to TMCs. From these charges and oxidation states, we then generated up to three spin states per complex depending on both metal electron configuration and row in the periodic table. We computed electronic properties using single-point PBE0/def2-TZVP calculations on structures with their experimental, crystallographically-resolved heavy-atom positions preserved but hydrogen atom positions optimized. By preserving these atomic coordinates, BOS-TMC eliminates potential deviations in properties from those in their crystal environment that have been introduced by optimizing structures at the semi-empirical (tmQM) or DFT level in the gas phase (e.g., tmQM[29], tmQMg[43], and OMol25[35]). Thus, we expect our dataset to be a useful complement to these other datasets. Relative to existing datasets such as OMol25[35] and tmQM[29], BOS-TMC contains a larger number of synthesized TMCs, especially those with higher absolute charges and a greater diversity of spin states. As a result, BOS-TMC spans a distinct property distribution, including for HOMO–LUMO gaps, individual HOMO or LUMO levels, dipole moments.

We also reported TMC atomization energies for the first time and showed that in comparison to QM9 atomization energies, the per-atom atomization energies of TMCs are lower than those in organic molecule sets but span a wider range. In contrast to existing data sets that calculated predominantly the low-spin state or high-spin states of only 3d metals, we explored spin-state dependence and found our set to include 27,360 mid-row 3d and 4d/5d TMCs (i.e., 18%) with non–low-spin ground states. We showed how recalculating properties with the appropriate open-shell ground state led to shifts in HOMO-LUMO gaps, partial charges, dipole moments, and atomization energies. These results demonstrate that by sampling open-shell configurations, we are more likely to capture the realistic electronic structure diversity characteristic of experimentally realized coordination compounds.

Finally, we assessed density functional dependence using a >10k complex subset evaluated with a total of twelve density functionals spanning multiple rungs of "Jacob's ladder". We observed large sensitivities in HS-LS vertical spin-splitting energies, where functional choice led to variations of up to 40 kcal/mol in computed properties for a single complex. We also observed substantial discrepancies in dipole moments, atomization energies, and partial charges, especially for PBE and M06-2X or r$^2$-SCAN in comparison to alternative functionals. Our analysis further identified Cu(I/II) TMCs as the most prone to high variations among xc functionals regardless of property considered. These trends highlight the continuing challenges of functional selection for open-shell transition-metal chemistry, including for training machine learning models, and underscore the value of BOS-TMC as a benchmark for future methodological development.

Together, we provide 2.9 M properties for the full dataset and 2.8 M property/xc functional combinations for the subset of TMCs for which we explored xc functional dependence. These results establish BOS-TMC as a comprehensive DFT dataset of experimentally synthesized transition metal complexes that we anticipate will serve as a useful resource for machine-learning model development for property prediction, electronic structure method benchmarking, and large-scale exploration of charge, spin, and bonding trends in transition-metal chemistry.

AUTHOR INFORMATION

**Corresponding Author**

*email:hjkulik@mit.edu

**Author Contributions**

CRediT: Aaron G Garrison: Data curation, Formal Analysis, Investigation, Methodology, Validation, Visualization, Writing – original draft, Writing – review & editing; Jacob W Toney: Data curation, Formal Analysis, Investigation, Methodology, Validation, Visualization, Writing – original draft, Writing – review & editing; Tatiana Nikolaeva: Data curation, Formal Analysis, Investigation, Methodology, Visualization, Writing – original draft, Writing – review & editing; Roland G St. Michel: Data curation, Methodology, Visualization, Writing – review & editing; Christopher J Stein: Funding acquisition, Project administration, Supervision, Validation, Writing

– review & editing; Heather J Kulik: Conceptualization, Formal Analysis, Funding acquisition, Investigation, Methodology, Project administration, Supervision, Validation, Visualization, Writing – original draft, Writing – review & editing

**Notes**

The authors declare no competing financial interest.

ASSOCIATED CONTENT

**Supporting Information**. Workflow for curating the initial transition metal complex (TMC) dataset; procedures for hydrogen addition, charge assignment, and treatment of odd-electron ligands; statistics on molecular graph changes during preprocessing; allowed spin multiplicities for metals and atomic spin states used in atomization energy calculations; description of single-atom energy calculations and algorithm for determining atomization energies; examples of electron additions or removals used in atomization energy evaluation; scheme for selecting the manyDFA subset of ~10k TMCs; distributions of metal centers, electron counts, and structural properties in the dataset; statistics on spin contamination and calculation success rates across spin states, basis sets, and charges; structural and compositional statistics including atom counts and molecular weights; correlations and distributions of electronic and structural properties; comparison with existing datasets and overlap analysis; distributions and statistics of HOMO, LUMO, HOMO–LUMO gaps, dipole moments, and partial charges by molecular charge; summary statistics and identification of property outliers; analysis of ground-state spin distributions by metal and oxidation state; statistics and examples of vertical spin-state energy differences (VSSEs); symmetry statistics for complexes; comparisons of intermediate- and high-spin properties relative to low-spin states; shifts in dipole moments, charges, and frontier orbital energies across spin states; sensitivity of atomization energies to spin states and analysis of deviations; relative atomization energy comparisons with small-molecule datasets and across charges; identification of extreme cases in atomization energies and metal-dependent trends; curation of the manyDFA dataset and statistics of calculations across exchange–correlation functionals; functional-dependent failure rates and ground-state reassignments; statistical variation of properties across functionals including dipole moments, charges, and atomization energies; principal component analyses and mean absolute deviations across functionals; correlations and variability of properties across exchange–correlation methods; identification of complexes with unusually large functional sensitivity across multiple properties. (PDF)

This material is available free of charge via the Internet at http://pubs.acs.org.

**Data and Software Availability statement**

All data required to reproduce this work is provided either in the Supporting Information PDF file

or in the Zenodo repository.[68]

ACKNOWLEDGMENT

The work on DFT functional sensitivity was supported by the U.S. Department of Energy under grant number DE-SC0024174 (for A.G.G. and H.J.K.). The broader workflows were supported by a UPI from the Dow Chemical Company (for J.W.T., R.G.S.M., and H.J.K.). This work was also supported by the Technical University of Munich – Institute for Advanced Study (for T.N., C.J.S., and H.J.K.). A.G.G. and J.W.T. were partially supported by Chemical Engineering MathWorks Engineering Fellowships. J.W.T. was partially supported by a Leslye Miller Fraser and Darryl M. Fraser Fellowship from the MIT School of Engineering. H.J.K. was partially supported by a Simon Family Faculty Research Innovation Fund. The authors acknowledge the MIT SuperCloud and Lincoln Laboratory Supercomputing Center for providing HPC resources that have contributed to the research results reported in this work. The computational work in part made use of Expanse at San Diego Supercomputer Center through allocation CHE140073 from the Advanced Cyberinfrastructure Coordination Ecosystem: Services & Support (ACCESS) program, which is supported by National Science Foundation grants #2138259, #2138286, #2138307, #2137603, and #2138296. The authors thank Adam H. Steeves for providing a critical reading of the manuscript.

REFERENCES

(1) Keith, J. A.; Vassilev-Galindo, V.; Cheng, B.; Chmiela, S.; Gastegger, M.; Muller, K.-R.; Tkatchenko, A. Combining Machine Learning and Computational Chemistry for Predictive Insights into Chemical Systems. *Chemical reviews* **2021,** *121*, 9816-9872.

(2) Aldossary, A.; Campos-Gonzalez-Angulo, J. A.; Pablo-García, S.; Leong, S. X.; Rajaonson, E. M.; Thiede, L.; Tom, G.; Wang, A.; Avagliano, D.; Aspuru-Guzik, A. In Silico Chemical Experiments in the Age of Ai: From Quantum Chemistry to Machine Learning and Back. *Advanced Materials* **2024,** *36*, 2402369.

(3) Deringer, V. L.; Caro, M. A.; Csányi, G. Machine Learning Interatomic Potentials as Emerging Tools for Materials Science. *Advanced Materials* **2019,** *31*, 1902765.

(4) Kulichenko, M.; Nebgen, B.; Lubbers, N.; Smith, J. S.; Barros, K.; Allen, A. E.; Habib, A.; Shinkle, E.; Fedik, N.; Li, Y. W. Data Generation for Machine Learning Interatomic Potentials and Beyond. *Chemical Reviews* **2024,** *124*, 13681-13714.

(5) Ramakrishnan, R.; Dral, P. O.; Rupp, M.; von Lilienfeld, O. A. Quantum Chemistry Structures and Properties of 134 Kilo Molecules. *Scientific Data* **2014,** *1*, 140022.

(6) Schreiner, M.; Bhowmik, A.; Vegge, T.; Busk, J.; Winther, O. Transition1x-a Dataset for Building Generalizable Reactive Machine Learning Potentials. *Scientific Data* **2022,** *9*, 779.

(7) Zhao, Q.; Vaddadi, S. M.; Woulfe, M.; Ogunfowora, L. A.; Garimella, S. S.; Isayev, O.; Savoie, B. M. Comprehensive Exploration of Graphically Defined Reaction Spaces. *Scientific Data* **2023,** *10*, 145.

(8) Smith, J. S.; Zubatyuk, R.; Nebgen, B.; Lubbers, N.; Barros, K.; Roitberg, A. E.; Isayev, O.; Tretiak, S. The Ani-1ccx and Ani-1x Data Sets, Coupled-Cluster and Density Functional Theory Properties for Molecules. *Scientific data* **2020,** *7*, 134.

(9) Groom, C. R.; Bruno, I. J.; Lightfoot, M. P.; Ward, S. C. The Cambridge Structural Database. *Acta Crystallogr B Struct Sci Cryst Eng Mater* **2016,** *72*, 171-9.

(10) Nandy, A.; Taylor, M. G.; Kulik, H. J. Identifying Underexplored and Untapped Regions in the Chemical Space of Transition Metal Complexes. *The Journal of Physical Chemistry Letters* **2023,** *14*, 5798-5804.

(11) Orpen, A. G.; Connelly, N. G. Structural Systematics: The Role of P-A .Sigma.* Orbitals in Metal-Phosphorus .Pi.-Bonding in Redox-Related Pairs of M-Pa3 Complexes (a = R, Ar, or; R = Alkyl). *Organometallics* **1990**, *9*, 1206-1210.

(12) Dunne, B. J.; Morris, R. B.; Orpen, A. G. Structural Systematics. Part 3. Geometry Deformations in Triphenylphosphine Fragments: A Test of Bonding Theories in Phosphine Complexes. *J. Chem. Soc., Dalton Trans.* **1991**, 653-661.

(13) Orpen, A. G.; Salter, I. D. Structural Systematics. 2. Metal Framework Rearrangements in Cluster Compounds Containing the Au2ru3 Fragment. *Organometallics* **1991,** *10*, 111-117.

(14) Morton, D. A. V.; Orpen, A. G. Structural Systematics. Part 4. Conformations of the Diphosphine Ligands in M2(M-Ph2pch2pph2) and M(Ph2pch2ch2pph2) Complexes. *J. Chem. Soc., Dalton Trans.* **1992**, 641-653.

(15) Garner, S. E.; Orpen, A. G. Structural Systematics. Part 5. Conformation and Bonding in the Chiral Metal Complexes [M(H5-C5r5)(Xo)Z(Pph3)]. *J. Chem. Soc., Dalton Trans.* **1993**, 533-541.

(16) Martín, A.; Orpen, A. G. Structural Systematics. 6.1 Apparent Flexibility of Metal Complexes in Crystals. *Journal of the American Chemical Society* **1996**, *118*, 1464-1470.

(17) Anderson, K. M.; Orpen, A. G. On the Relative Magnitudes of Cis and Trans Influences in Metal Complexes. *Chemical Communications* **2001**, 2682-2683.

(18) Vela, S.; Laplaza, R.; Cho, Y.; Corminboeuf, C. Cell2mol: Encoding Chemistry to Interpret Crystallographic Data. *npj Comput. Mater.* **2022**, *8*, 188.

(19) Cundari, T. R. Computational Studies of Transition Metal−Main Group Multiple Bonding. *Chemical Reviews* **2000**, *100*, 807-818.

(20) Sinha, V.; Laan, J. J.; Pidko, E. A. Accurate and Rapid Prediction of $Pk_a$ of Transition Metal Complexes: Semiempirical Quantum Chemistry with a Data-Augmented Approach. *Physical Chemistry Chemical Physics* **2021,** *23*, 2557-2567.

(21) Jover, J.; Fey, N.; Harvey, J. N.; Lloyd-Jones, G. C.; Orpen, A. G.; Owen-Smith, G. J. J.; Murray, P.; Hose, D. R. J.; Osborne, R.; Purdie, M. Expansion of the Ligand Knowledge Base for Monodentate P-Donor Ligands (Lkb-P). *Organometallics* **2010,** *29*, 6245-6258.

(22) Jover, J.; Fey, N.; Harvey, J. N.; Lloyd-Jones, G. C.; Orpen, A. G.; Owen-Smith, G. J. J.; Murray, P.; Hose, D. R. J.; Osborne, R.; Purdie, M. Expansion of the Ligand Knowledge Base for Chelating P,P-Donor Ligands (Lkb-Pp). *Organometallics* **2012,** *31*, 5302-5306.

(23) Fey, N.; Tsipis, A. C.; Harris, S. E.; Harvey, J. N.; Orpen, A. G.; Mansson, R. A. Development of a Ligand Knowledge Base, Part 1: Computational Descriptors for Phosphorus Donor Ligands. *Chemistry - A European Journal* **2006**, *12*, 291-302.

(24) Fey, N.; Harvey, J. N.; Lloyd-Jones, G. C.; Murray, P.; Orpen, A. G.; Osborne, R.; Purdie, M. Computational Descriptors for Chelating P,P- and P,N-Donor Ligands[1]. *Organometallics* **2008,** *27*, 1372-1383.

(25) Gensch, T.; dos Passos Gomes, G.; Friederich, P.; Peters, E.; Gaudin, T.; Pollice, R.; Jorner, K.; Nigam, A.; Lindner-D'Addario, M.; Sigman, M. S.; Aspuru-Guzik, A. A Comprehensive Discovery Platform for Organophosphorus Ligands for Catalysis. *Journal of the American Chemical Society* **2022**, *144*, 1205-1217.

(26) Taylor, M. G.; Yang, T.; Lin, S.; Nandy, A.; Janet, J. P.; Duan, C.; Kulik, H. J. Seeing Is Believing: Experimental Spin States from Machine Learning Model Structure Predictions. *J. Phys. Chem. A* **2020**, *124*, 3286-3299.

(27) Taylor, M. G.; Nandy, A.; Lu, C. C.; Kulik, H. J. Deciphering Cryptic Behavior in Bimetallic Transition-Metal Complexes with Machine Learning. *The Journal of Physical Chemistry Letters* **2021,** *12*, 9812-9820.

(28) Jablonka, K. M.; Ongari, D.; Moosavi, S. M.; Smit, B. Using Collective Knowledge to Assign Oxidation States of Metal Cations in Metal–Organic Frameworks. *Nature Chemistry* **2021,** *13*, 771-777.

(29) Balcells, D.; Skjelstad, B. B. Tmqm Dataset—Quantum Geometries and Properties of 86k Transition Metal Complexes. *Journal of Chemical Information and Modeling* **2020,** *60*, 6135-6146.

(30) Cho, Y.; Laplaza, R.; Vela, S.; Corminboeuf, C. Automated Prediction of Ground State Spin for Transition Metal Complexes. *Digit Discov* **2024,** *3*, 1638-1647.

(31) Arunachalam, N.; Gugler, S.; Taylor, M. G.; Duan, C.; Nandy, A.; Janet, J. P.; Meyer, R.; Oldenstaedt, J.; Chu, D. B. K.; Kulik, H. J. Ligand Additivity Relationships Enable Efficient Exploration of Transition Metal Chemical Space *Journal of Chemical Physics* **2022,** *157*, 184112.

(32) Duan, C.; Chu, D. B. K.; Nandy, A.; Kulik, H. J. Detection of Multi-Reference Character Imbalances Enables a Transfer Learning Approach for Chemical Discovery with Coupled Cluster Accuracy at Dft Cost. *Chemical Science* **2022,** *13*, 4962-4971.

(33) Liu, F.; Duan, C.; Kulik, H. J. Rapid Detection of Strong Correlation with Machine Learning for Transition-Metal Complex High-Throughput Screening. *The Journal of Physical Chemistry Letters* **2020,** *11*, 8067-8076.

(34) Chen, S.; Nielson, T.; Zalit, E.; Skjelstad, B. B.; Borough, B.; Hirschi, W. J.; Yu, S.; Balcells, D.; Ess, D. H. Automated Construction and Optimization Combined with Machine Learning to Generate Pt (Ii) Methane C–H Activation Transition States. *Topics in Catalysis* **2022,** *65*, 312-324.

(35) Levine, D. S.; Shuaibi, M.; Spotte-Smith, E. W. C.; Taylor, M. G.; Hasyim, M. R.; Michel, K.; Batatia, I.; Csányi, G.; Dzamba, M.; Eastman, P.; Frey, N. C.; Fu, X.; Gharakhanyan, V.; Krishnapriyan, A. S.; Rackers, J. A.; Raja, S.; Rizvi, A.; Rosen, A. S.; Ulissi, Z.; Vargas, S.; Zitnick, C. L.; Blau, S. M.; M.Wood, B. The Open Molecules 2025 (Omol25) Dataset, Evaluations, and Models. *arXiv* **2025**, DOI:https://doi.org/10.48550/arXiv.2505.08762 https://doi.org/10.48550/arXiv.2505.08762.

(36) Duan, C.; Nandy, A.; Terrones, G.; Kastner, D. W.; Kulik, H. J. Active Learning Exploration of Transition Metal Complexes to Discover Method-Insensitive and Synthetically Accessible Chromophores. *JACS Au* **2023,** *3*, 391-401.

(37) Kneiding, H.; Nova, A.; Balcells, D. Directional Multiobjective Optimization of Metal Complexes at the Billion-System Scale. *Nature computational science* **2024,** *4*, 263-273.

(38) Glerup, J.; Moensted, O.; Schaeffer, C. E. Nonadditive and Additive Ligand Fields and Spectrochemical Series Arising from Ligand Field Parameterization Schemes.  Pyridine as a Nonlinearly Ligating .Pi.-Back-Bonding Ligand toward Chromium(Iii). *Inorganic Chemistry* **1976,** *15*, 1399-1407.

(39) Vanquickenborne, L.; Hendrickx, M.; Hyla-Kryspin, I. Ab Initio Study of Progressive Ligand Substitution in Octahedral Transition-Metal Complexes. *Inorganic Chemistry* **1989,** *28*, 770-773.

(40) Morris, R. H. Ligand Additivity Effects and Periodic Trends in the Stability and Acidity of Octahedral. Eta. 2-Dihydrogen Complexes of D6 Transition Metal Ions. *Inorganic Chemistry* **1992,** *31*, 1471-1478.
(41) Bursten, B. E. Ligand Additivity: Applications to the Electrochemistry and Photoelectron Spectroscopy of D6 Octahedral Complexes. *Journal of the American Chemical Society* **1982,** *104*, 1299-1304.
(42) Chu, D. B. K.; González-Narváez, D. A.; Meyer, R.; Nandy, A.; Kulik, H. J. Ligand Many-Body Expansion as a General Approach for Accelerating Transition Metal Complex Discovery. *Journal of Chemical Information and Modeling* **2024,** *64*, 9397-9412.
(43) Kneiding, H.; Lukin, R.; Lang, L.; Reine, S.; Pedersen, T. B.; De Bin, R.; Balcells, D. Deep Learning Metal Complex Properties with Natural Quantum Graphs. *Digital Discovery* **2023,** *2*, 618-633.
(44) Garrison, A. G.; Heras-Domingo, J.; Kitchin, J. R.; Dos Passos Gomes, G.; Ulissi, Z. W.; Blau, S. M. Applying Large Graph Neural Networks to Predict Transition Metal Complex Energies Using the Tmqm_Wb97mv Data Set. *J Chem Inf Model* **2023,** *63*, 7642-7654.
(45) Kneiding, H.; Balcells, D. Tmqmg* Dataset: Excited State Properties of 74k Transition Metal Complexes. *ChemRxiv* **2025**, DOI:10.26434/chemrxiv-2025-pdd8p 10.26434/chemrxiv-2025-pdd8p.
(46) Gražulis, S.; Chateigner, D.; Downs, R. T.; Yokochi, A. F.; Quirós, M.; Lutterotti, L.; Manakova, E.; Butkus, J.; Moeck, P.; Le Bail, A. Crystallography Open Database–an Open-Access Collection of Crystal Structures. *Applied Crystallography* **2009,** *42*, 726-729.
(47) Bühl, M.; Kabrede, H. Geometries of Transition-Metal Complexes from Density-Functional Theory. *Journal of chemical theory and computation* **2006,** *2*, 1282-1290.
(48) Waller, M. P.; Braun, H.; Hojdis, N.; Bühl, M. Geometries of Second-Row Transition-Metal Complexes from Density-Functional Theory. *Journal of chemical theory and computation* **2007,** *3*, 2234-2242.
(49) Buhl, M.; Reimann, C.; Pantazis, D. A.; Bredow, T.; Neese, F. Geometries of Third-Row Transition-Metal Complexes from Density-Functional Theory. *Journal of chemical theory and computation* **2008,** *4*, 1449-1459.
(50) Mortensen, S. R.; Kepp, K. P. Spin Propensities of Octahedral Complexes from Density Functional Theory. *The Journal of Physical Chemistry A* **2015,** *119*, 4041-4050.
(51) Ioannidis, E. I.; Kulik, H. J. Towards Quantifying the Role of Exact Exchange in Predictions of Transition Metal Complex Properties. *The Journal of Chemical Physics* **2015,** *143*, 034104.
(52) Reiher, M.; Salomon, O.; Artur Hess, B. Reparameterization of Hybrid Functionals Based on Energy Differences of States of Different Multiplicity. *Theoretical Chemistry Accounts: Theory, Computation, and Modeling (Theoretica Chimica Acta)* **2001,** *107*, 48-55.
(53) Reiher, M. Theoretical Study of the Fe(Phen)2(Ncs)2 Spin-Crossover Complex with Reparametrized Density Functionals. *Inorganic Chemistry* **2002,** *41*, 6928-6935.
(54) Droghetti, A.; Alfè, D.; Sanvito, S. Assessment of Density Functional Theory for Iron(Ii) Molecules across the Spin-Crossover Transition. *The Journal of Chemical Physics* **2012,** *137*, 124303.

(55) Ganzenmüller, G.; Berkaïne, N.; Fouqueau, A.; Casida, M. E.; Reiher, M. Comparison of Density Functionals for Differences between the High- (T2g5) and Low- (A1g1) Spin States of Iron(Ii) Compounds. Iv. Results for the Ferrous Complexes [Fe(L)('Nhs4')]. *The Journal of Chemical Physics* **2005,** *122*, 234321.

(56) Kepp, K. P. Theoretical Study of Spin Crossover in 30 Iron Complexes. *Inorganic Chemistry* **2016,** *55*, 2717-2727.

(57) Zein, S.; Borshch, S. A.; Fleurat-Lessard, P.; Casida, M. E.; Chermette, H. Assessment of the Exchange-Correlation Functionals for the Physical Description of Spin Transition Phenomena by Density Functional Theory Methods: All the Same? *The Journal of Chemical Physics* **2007,** *126*, 014105.

(58) Boguslawski, K.; Jacob, C. R.; Reiher, M. Can Dft Accurately Predict Spin Densities? Analysis of Discrepancies in Iron Nitrosyl Complexes. *Journal of Chemical Theory and Computation* **2011,** *7*, 2740-2752.

(59) Bleda, E. A.; Trindle, C.; Altun, Z. Studies on Spin State Preferences in Fe(Ii) Complexes. *Computational and Theoretical Chemistry* **2015,** *1073*, 139-148.

(60) Milko, P.; Iron, M. A. On the Innocence of Bipyridine Ligands: How Well Do Dft Functionals Fare for These Challenging Spin Systems? *Journal of Chemical Theory and Computation* **2014,** *10*, 220-235.

(61) Duan, C.; Chen, S.; Taylor, M. G.; Liu, F.; Kulik, H. J. Machine Learning to Tame Divergent Density Functional Approximations: A New Path to Consensus Materials Design Principles. *Chem Sci* **2021,** *12*, 13021-13036.

(62) Duan, C.; Nandy, A.; Meyer, R.; Arunachalam, N.; Kulik, H. J. A Transferable Recommender Approach for Selecting the Best Density Functional Approximations in Chemical Discovery. *Nature Computational Science* **2023,** *3*, 38-47.

(63) Vennelakanti, V.; Taylor, M. G.; Nandy, A.; Duan, C.; Kulik, H. J. Assessing the Performance of Approximate Density Functional Theory on 95 Experimentally Characterized Fe(Ii) Spin Crossover Complexes. *J Chem Phys* **2023,** *159*, 024120.

(64) Stein, T.; Autschbach, J.; Govind, N.; Kronik, L.; Baer, R. Curvature and Frontier Orbital Energies in Density Functional Theory. *The journal of physical chemistry letters* **2012,** *3*, 3740-3744.

(65) Duan, C.; Ladera, A. J.; Liu, J. C.; Taylor, M. G.; Ariyarathna, I. R.; Kulik, H. J. Exploiting Ligand Additivity for Transferable Machine Learning of Multireference Character across Known Transition Metal Complex Ligands. *J Chem Theory Comput* **2022,** *18*, 4836-4845.

(66) St. Michel, R. G.; Jang, R. J.; Garrison, A. G.; Kevlishvili, I.; Kulik, H. J. The Bos-Lig Dataset: Accurate Ligand Charges from a Consensus Approach for 66,810 Experimentally Synthesized Ligands *arXiv* http://arxiv.org/abs/2604.06043 **2026**, DOI:http://arxiv.org/abs/2604.06043 http://arxiv.org/abs/2604.06043.

(67) Kramida, A., Ralchenko, Yu., Reader, J., and NIST ASD Team. Nist Atomic Spectra Database (Ver. 5.12). **2024**, DOI:https://dx.doi.org/10.18434/T4W30F https://dx.doi.org/10.18434/T4W30F.

(68) Garrison, A. G.; Toney, J. W.; Nikolaeva, T.; St. Michel, R. G.; Stein, C. J.; Kulik, H. J. Zenodo Dataset for "the Bos-Tmc Dataset: Dft Properties of 159k Experimentally Characterized Transition Metal Complexes Spanning Multiple Charge and Spin States". 2026. https://dx.doi.org/10.5281/zenodo.19410245. (Accessed April 4, 2026).

(69) Seritan, S.; Bannwarth, C.; Fales, B. S.; Hohenstein, E. G.; Isborn, C. M.; Kokkila-Schumacher, S. I. L.; Li, X.; Liu, F.; Luehr, N.; Snyder, J. W.; Song, C.; Titov, A. V.; Ufimtsev, I. S.; Wang, L. P.; Martínez, T. J. Terachem: A Graphical Processing Unit-Accelerated Electronic Structure Package for Large-Scale Ab Initio Molecular Dynamics. *WIREs Computational Molecular Science* **2020**, *11*.
(70) Titov, A. V.; Ufimtsev, I. S.; Luehr, N.; Martinez, T. J. Generating Efficient Quantum Chemistry Codes for Novel Architectures. *J Chem Theory Comput* **2013**, *9*, 213-21.
(71) Adamo, C.; Barone, V. Toward Reliable Density Functional Methods without Adjustable Parameters: The Pbe0 Model. *The Journal of Chemical Physics* **1999**, *110*, 6158-6170.
(72) Weigend, F.; Ahlrichs, R. Balanced Basis Sets of Split Valence, Triple Zeta Valence and Quadruple Zeta Valence Quality for H to Rn: Design and Assessment of Accuracy. *Phys Chem Chem Phys* **2005**, *7*, 3297-305.
(73) Grimme, S.; Antony, J.; Ehrlich, S.; Krieg, H. A Consistent and Accurate Ab Initio Parametrization of Density Functional Dispersion Correction (Dft-D) for the 94 Elements H-Pu. *J Chem Phys* **2010**, *132*, 154104.
(74) Becke, A. D.; Johnson, E. R. A Density-Functional Model of the Dispersion Interaction. *J Chem Phys* **2005**, *123*, 154101.
(75) Johnson, E. R.; Becke, A. D. A Post-Hartree-Fock Model of Intermolecular Interactions. *J Chem Phys* **2005**, *123*, 24101.
(76) Johnson, E. R.; Becke, A. D. A Post-Hartree-Fock Model of Intermolecular Interactions: Inclusion of Higher-Order Corrections. *J Chem Phys* **2006**, *124*, 174104.
(77) Grimme, S.; Ehrlich, S.; Goerigk, L. Effect of the Damping Function in Dispersion Corrected Density Functional Theory. *J Comput Chem* **2011**, *32*, 1456-65.
(78) Wang, L. P.; Song, C. Geometry Optimization Made Simple with Translation and Rotation Coordinates. *J Chem Phys* **2016**, *144*, 214108.
(79) Rajpurohit, S.; Vennelakanti, V.; Kulik, H. J. Improving Predictions of Spin-Crossover Complex Properties through Dft Calculations with a Local Hybrid Functional. *J. Phys. Chem. A* **2024**, *128*, 9082-9089.
(80) Liu, F.; Luehr, N.; Kulik, H. J.; Martinez, T. J. Quantum Chemistry for Solvated Molecules on Graphical Processing Units Using Polarizable Continuum Models. *J Chem Theory Comput* **2015**, *11*, 3131-44.
(81) Barone, V.; Cossi, M. Quantum Calculation of Molecular Energies and Energy Gradients in Solution by a Conductor Solvent Model. *J Phys Chem A* **1998**, *102*, 1995-2001.
(82) Lange, A. W.; Herbert, J. M. A Smooth, Nonsingular, and Faithful Discretization Scheme for Polarizable Continuum Models: The Switching/Gaussian Approach. *J Chem Phys* **2010**, *133*, 244111.
(83) York, D. M.; Karplus, M. A Smooth Solvation Potential Based on the Conductor-Like Screening Model. *J Phys Chem A* **1999**, *103*, 11060-11079.
(84) Alvarez, S. A Cartography of the Van Der Waals Territories. *Dalton Trans* **2013**, *42*, 8617-36.
(85) Pulay, P. Improved Scf Convergence Acceleration. *Journal of Computational Chemistry* **2004**, *3*, 556-560.
(86) Hu, X.; Yang, W. Accelerating Self-Consistent Field Convergence with the Augmented Roothaan-Hall Energy Function. *J Chem Phys* **2010**, *132*, 054109.

(87) Ioannidis, E. I.; Gani, T. Z. H.; Kulik, H. J. Molsimplify: A Toolkit for Automating Discovery in Inorganic Chemistry. *Journal of Computational Chemistry* **2016,** *37*, 2106-2117.

(88) Terrones, G. G.; St. Michel II, R. G.; Toney, J. W.; Ball, A. K.; Wang, Y.; Garrison, A. G.; Nandy, A.; Meyer, R.; Edholm, F.; Oh, C.; Pujet, S. G.; Chu, D. B. K.; Muhammetgulyyev, D.; Kulik, H. J. Molsimplify 2.0: Improved Structure Generation for Automating Discovery in Inorganic Molecular and Reticular Chemistry. *Journal of Chemical Information and Modeling* **2026,** *66*, 2753-2767.

(89) Smith, D. G. A.; Burns, L. A.; Simmonett, A. C.; Parrish, R. M.; Schieber, M. C.; Galvelis, R.; Kraus, P.; Kruse, H.; Di Remigio, R.; Alenaizan, A.; James, A. M.; Lehtola, S.; Misiewicz, J. P.; Scheurer, M.; Shaw, R. A.; Schriber, J. B.; Xie, Y.; Glick, Z. L.; Sirianni, D. A.; O'Brien, J. S.; Waldrop, J. M.; Kumar, A.; Hohenstein, E. G.; Pritchard, B. P.; Brooks, B. R.; Schaefer, H. F., 3rd; Sokolov, A. Y.; Patkowski, K.; DePrince, A. E., 3rd; Bozkaya, U.; King, R. A.; Evangelista, F. A.; Turney, J. M.; Crawford, T. D.; Sherrill, C. D. Psi4 1.4: Open-Source Software for High-Throughput Quantum Chemistry. *J Chem Phys* **2020,** *152*, 184108.

(90) Perdew, J. P.; Burke, K.; Ernzerhof, M. Generalized Gradient Approximation Made Simple. *Phys Rev Lett* **1996,** *77*, 3865-3868.

(91) Zhao, Y.; Truhlar, D. G. A New Local Density Functional for Main-Group Thermochemistry, Transition Metal Bonding, Thermochemical Kinetics, and Noncovalent Interactions. *J Chem Phys* **2006,** *125*, 194101.

(92) Furness, J. W.; Kaplan, A. D.; Ning, J.; Perdew, J. P.; Sun, J. Accurate and Numerically Efficient R(2)Scan Meta-Generalized Gradient Approximation. *J Phys Chem Lett* **2020,** *11*, 8208-8215.

(93) Becke, A. D. Density-Functional Exchange-Energy Approximation with Correct Asymptotic Behavior. *Phys Rev A Gen Phys* **1988,** *38*, 3098-3100.

(94) Lee, C.; Yang, W.; Parr, R. G. Development of the Colle-Salvetti Correlation-Energy Formula into a Functional of the Electron Density. *Phys Rev B Condens Matter* **1988,** *37*, 785-789.

(95) Stephens, P. J.; Devlin, F. J.; Chabalowski, C. F.; Frisch, M. J. Ab Initio Calculation of Vibrational Absorption and Circular Dichroism Spectra Using Density Functional Force Fields. *The Journal of Physical Chemistry* **2002,** *98*, 11623-11627.

(96) Salomon, O.; Reiher, M.; Hess, B. A. Assertion and Validation of the Performance of the B3lyp⋆ Functional for the First Transition Metal Row and the G2 Test Set. *The Journal of Chemical Physics* **2002,** *117*, 4729-4737.

(97) Zhao, Y.; Truhlar, D. G. The M06 Suite of Density Functionals for Main Group Thermochemistry, Thermochemical Kinetics, Noncovalent Interactions, Excited States, and Transition Elements: Two New Functionals and Systematic Testing of Four M06-Class Functionals and 12 Other Functionals. *Theoretical Chemistry Accounts* **2007,** *120*, 215-241.

(98) Rohrdanz, M. A.; Martins, K. M.; Herbert, J. M. A Long-Range-Corrected Density Functional That Performs Well for Both Ground-State Properties and Time-Dependent Density Functional Theory Excitation Energies, Including Charge-Transfer Excited States. *J Chem Phys* **2009,** *130*, 054112.

(99) Chai, J. D.; Head-Gordon, M. Long-Range Corrected Hybrid Density Functionals with Damped Atom-Atom Dispersion Corrections. *Phys Chem Chem Phys* **2008,** *10*, 6615-20.

(100) Mardirossian, N.; Head-Gordon, M. Omegab97m-V: A Combinatorially Optimized, Range-Separated Hybrid, Meta-Gga Density Functional with Vv10 Nonlocal Correlation. *J Chem Phys* **2016,** *144*, 214110.

(101) Goerigk, L.; Grimme, S. Efficient and Accurate Double-Hybrid-Meta-Gga Density Functionals-Evaluation with the Extended Gmtkn30 Database for General Main Group Thermochemistry, Kinetics, and Noncovalent Interactions. *J Chem Theory Comput* **2011,** *7*, 291-309.

(102) Bremond, E.; Adamo, C. Seeking for Parameter-Free Double-Hybrid Functionals: The Pbe0-Dh Model. *J Chem Phys* **2011,** *135*, 024106.

(103) Rosner, B. Percentage Points for a Generalized Esd Many-Outlier Procedure. *Technometrics* **1983,** *25*, 165-172.

(104) Janet, J. P.; Kulik, H. J. Resolving Transition Metal Chemical Space: Feature Selection for Machine Learning and Structure-Property Relationships. *J. Phys. Chem. A* **2017,** *121*, 8939-8954.

(105) Jin, H.; Merz Jr, K. M. Modeling Fe (Ii) Complexes Using Neural Networks. *Journal of Chemical Theory and Computation* **2024,** *20*, 2551-2558.

(106) Perdew, J. P.; Parr, R. G.; Levy, M.; Balduz, J. L. Density-Functional Theory for Fractional Particle Number: Derivative Discontinuities of the Energy. *Physical Review Letters* **1982,** *49*, 1691-1694.

(107) Mori-Sánchez, P.; Cohen, A. J.; Yang, W. Many-Electron Self-Interaction Error in Approximate Density Functionals. *The Journal of chemical physics* **2006,** *125*.

(108) Johnson, E. R.; Mori-Sánchez, P.; Cohen, A. J.; Yang, W. Delocalization Errors in Density Functionals and Implications for Main-Group Thermochemistry. *The Journal of chemical physics* **2008,** *129*.

(109) Cohen, A. J.; Mori-Sánchez, P.; Yang, W. Insights into Current Limitations of Density Functional Theory. *Science* **2008,** *321*, 792-794.

(110) Gani, T. Z. H.; Kulik, H. J. Where Does the Density Localize? Convergent Behavior for Global Hybrids, Range Separation, and Dft+U *Journal of Chemical Theory and Computation* **2016,** *12*, 5931-5945.

(111) Medvedev, M. G.; Bushmarinov, I. S.; Sun, J.; Perdew, J. P.; Lyssenko, K. A. Density Functional Theory Is Straying from the Path toward the Exact Functional. *Science* **2017,** *355*, 49-52.

(112) Sim, E.; Song, S.; Burke, K. Quantifying Density Errors in Dft. *The journal of physical chemistry letters* **2018,** *9*, 6385-6392.

(113) Kim, M.-C.; Sim, E.; Burke, K. Understanding and Reducing Errors in Density Functional Calculations. *Physical review letters* **2013,** *111*, 073003.

**For Table of Contents Use Only**

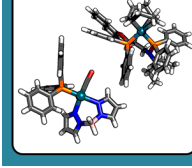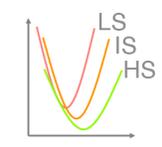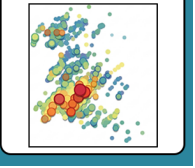

**Supporting Information for**
*The BOS-TMC Dataset: DFT Properties of 159k Experimentally Characterized Transition Metal Complexes Spanning Multiple Charge and Spin States*

Aaron G. Garrison[1,#], Jacob W. Toney[1,#], Tatiana Nikolaeva[1,2], Roland G. St. Michel[1,5], Christopher J. Stein[2,3,4], and Heather J. Kulik[1,6,7]*

[1]*Department of Chemical Engineering, Massachusetts Institute of Technology, Cambridge, MA 02139, USA*

[2]*Department of Chemistry, TUM School of Natural Sciences, Technical University of Munich, Garching, Germany*

[3]*Institute for Advanced Study, Technical University of Munich, Lichtenbergstrasse 2 a, D-85748 Garching, Germany*

[4]*Department of Materials Science and Engineering, Massachusetts Institute of Technology, Cambridge, MA 02139, USA*

[5]*Catalysis Research Center, Technical University of Munich, Garching, Germany*

[6]*Atomistic Modeling Center, Technical University of Munich, Garching, Germany*

[7]*Department of Chemistry, Massachusetts Institute of Technology, Cambridge, MA 02139, USA*

[#]These authors contributed equally
*corresponding author email: hjkulik@mit.edu

**Contents**

**Table S1.** Steps used[a,b] to curate the set of TMCs in BOS-TMC. Uniqueness was determined via Weisfeiler-Lehman graph hashes after converting the 3D structures into 2D graphs.

| Filter | Count |
| --- | --- |
| Containing A Transition Metal | 604,262 |
| Single Transition Metal | 324,061 |
| At Least One Heavy Atom Bonded to Metal | 307,676 |
| Single Coordination Center (i.e., one metal and only at most one species with more than four bonds) | 299,035 |
| Trustworthy Hydrogen | 270,953 |
| Valid Oxidation State | 165,305 |
| Atomic Filters[a] | 163,875 |
| Formula Filters[b] | 162,109 |
| Unique molecular graphs | 128,357 |

[a]The step labeled "Atomic Filters" incorporates filters for complexes containing lanthanides, actinides, post-transition metal clusters, and deuteriums. Post-transition metal clusters are defined as structures containing more than five atoms from a predefined set of metals and metalloids (Al, Ga, Ge, As, In, Sn, Sb, Te, Tl, Pb, Bi, Po, At).

[b]We removed inconsistent structures in which the component-level structures did not agree with the reported entry-level composition. This was typically due to duplicate atoms or missing atomic coordinates (especially hydrogens), which led to mismatches between the expected structure, deposited structure, or exported structure. The step labeled "Formula Filters" removes any structures where the atoms in the .xyz file do not match the molecular formula of the corresponding CSD entry, or the CSD molecular formula is polymeric.

**Text S1.** Method for hydrogen addition and subsequent addition quality check.

Before performing any charge inference or ligand analysis, each accepted CSD component was passed through a standardized procedure to ensure its molecular representation was chemically complete. All steps were executed using the CCDC Python API.

1. **Initial atom accounting and ghost-atom screening**
   For every component, we first recorded the total number of atomic sites and counted how many had valid, properly indexed atom labels. Any site lacking a numerical index (e.g., labeled simply "X" or left blank) was treated as a ghost atom. We then computed the number of hydrogens that should be present to satisfy standard valence requirements based on the existing bonding network.
2. **Hydrogen completion with fallback handling**
   Each component was duplicated, and hydrogens were added using CCDC's valence-based hydrogen-completion method (add_hydrogens('missing')). If this process failed, typically due to undefined or ambiguous bond orders, the script attempted to infer bond orders and then repeated hydrogen addition.
3. **Post-processing checks for structural consistency**
   After hydrogen addition, every complex was assigned a binary reliability flag (h_trustworthy) based on the outcome:

- **Case A – No change required**
  If no atoms were added, the original structure already contained a complete hydrogen representation. These components were accepted as trustworthy without modification.
- **Case B – Expected additions only**
  If the algorithm added exactly the number of missing hydrogens predicted from valence analysis—with no ghost atoms or unexpected sites introduced—the hydrogen-completed version replaced the original component and was marked trustworthy.
- **Case C – Excess or inconsistent additions**
  If more atoms were added than expected, suggesting spurious or ambiguous insertions, the hydrogen-modified version was saved only for reference (csd_mol2string_suspect_h_added). The workflow reverted to the unmodified structure, and the entry was flagged as non-trustworthy.
- **Case D – Complete failure**
  If both hydrogen-addition attempts failed, the component was marked non-trustworthy.

**Text S2.** Workflow for determining net charge of transition metal complexes.

Because many complexes lack reliable net charge annotations, we inferred component charges through an iterative unit-cell neutrality procedure. We began by assigning charges to frequently occurring, non-metal species, then propagated charges to unknown components by balancing the unit cell. Conflicts across identical hashes were resolved by majority consensus.

**Table S2**. Summary of allowed spin multiplicities, corresponding nominal electron configurations, associated metals (with oxidation states), and number of TMCs corresponding to each set of spin states, all grouped by 3d, 4d, and 5d metals. The total number of calculations is the number of spin multiplicities multiplied by the number of TMCs.

| Row | Spin mult. (2S+1) | Nominal electron configurations | Metal(oxidation) state | Number of TMCs | Number of calcs. |
|---|---|---|---|---|---|
| 3d | 1,3,5 | $d^2s^2$-$d^7s^1$ | Ti(0), V(1), Cr(0), Cr(2), Mn(1), Mn(3), Fe(0), Fe(2), Fe(4), Co(1), Co(3), Co(5), Ni(2), Ni(4), Cu(3) | 32747 | 98241 |
| 3d | 2,4,6 | $d^3s^2$-$d^6s^1$ or $d^7$ | V(0), Cr(1), Mn(0), Mn(2), Fe(1), Fe(3), Co(2), Co(4), Ni(3), Cu(4) | 17145 | 51435 |
| 3d | 1,3 | $d^2s^0, d^8s^2$-$d^{10}$ | Sc(1), Ti(2), V(3), Cr(4), Mn(5), Ni(0), Cu(1), Zn(2) | 17767 | 35534 |
| 3d | 2,4 | $d^3$ or $d^2s^1$, $d^7s^2$ or $d^8s^1$ or $d^9$ | Ti(1), V(2), Cr(3), Mn(4), Fe(5), Co(0), Ni(1), Cu(2) | 22139 | 44278 |
| 3d | 1 | $d^0s^2, d^{10}s^2$ | Sc(3), Ti(4), V(5), Cr(6), Mn(7), Zn(0) | 4050 | 4050 |
| 3d | 2 | $d^1$ or $s^1$, $d^{10}s^1$ or $d^9s^2$ | Sc(2), Ti(3), V(4), Cr(5), Mn(6), Cu(0), Zn(1) | 1664 | 1664 |
| 4d | 1,3 | $d^2$-$d^9s^1$ | Zr(0), Zr(2), Nb(1), Nb(3), Mo(4), Mo(0), Mo(2), Tc(1), Tc(3), Tc(5), Ru(0), Ru(2), Ru(4), Ru(6), Rh(1), Rh(3), Rh(5), Pd(0), Pd(2), Pd(4), Ag(3), Cd(2) | 30226 | 60452 |
| 4d | 2,4 | $d^3$-$d^9$ or $d^8s^1$ | Nb(0), Nb(2), Mo(1), Mo(3), Tc(0), Tc(2), Tc(4), Ru(1), Ru(3), Ru(5), Rh(0), Rh(2), Rh(4), Pd(1), Pd(3), Ag(2) | 1188 | 2376 |
| 4d | 1 | $d^0s^2, d^{10}s^2$ | Y(3), Zr(4), Nb(5), Mo(6), Tc(7), Ag(1) | 6354 | 6354 |
| 4d | 2 | $d^1$ or $s^1$, $d^{10}s^1$ or $d^9s^2$ | Y(2), Y(4), Zr(3), Nb(4), Mo(5), Tc(6), Cd(1) | 383 | 383 |
| 5d | 1,3 | $d^2$-$d^9s^1$ | Hf(0), Hf(2), Ta(1), Ta(3), W(0), W(2), W(4), Re(1), Re(3), Re(5), Os(0), Os(2), Os(4), Os(6), Ir(1), Ir(3), Ir(5), Pt(0), Pt(2), Pt(4), Pt(6), Au(1), Au(3), Au(5) | 24452 | 48904 |
| 5d | 2,4 | $d^3$-$d^9$ or $d^8s^1$ | Ta(0), Ta(2), W(1), W(3), Re(0), Re(2), Re(4), Os(1), Os(3), Os(5), Ir(0), Ir(2), Ir(4), Pt(1), Pt(3), Au(2), Au(4) | 652 | 1304 |
| 5d | 1 | $d^0s^2, d^{10}s^2$ | Hf(4), Ta(5), W(6), Re(7), Os(8), Hg(2) | 3158 | 3158 |
| 5d | 2 | $d^1$ or $s^1$, $d^{10}s^1$ or $d^9s^2$ | Hf(3), Ta(4), W(5), Re(6), Au(0), Hg(1) | 184 | 184 |
| **3 spin states** | | | | **49892** | **149676** |
| **2 spin states** | | | | **96424** | **192848** |
| **1 spin state** | | | | **15793** | **15793** |
| **Totals** | | | | **162109** | **358317** |

**Text S3.** Scheme for identifying transition metal complexes with noninnocent ligands.

For each transition metal complex, we used the total complex charge to determine compatible total TMC spin multiplicities (i.e., singlet/triplet/quintet for an even number of electrons or doublet/quartet/sextet for an odd number of electrons). We also determined the compatible spin states for the metal based on its identity and oxidation state (Table S2). If the multiplicities compatible with the metal center were different from those compatible with the overall complex based on its charge, this suggested that there are an odd overall number of electrons on the ligands, i.e., there must be at least one noninnocent ligand. We then incremented up the total spin multiplicity of the complex by 1 to account for this discrepancy.

**Text S4.** Summary of TMCs in the set at each step of the calculation workflow.
We note the following statistics in terms of changed molecular graphs and whether the structures had $<S^2>$ values that deviated by more than 1 for $S(S+1)$ (i.e., were spin contaminated):

1. **All attempted TeraChem calcs (total//unique): 162,109//128,357**
2. **Calcs which completed but the graph changed: 1,026**
    a. 22 of these were problematic (i.e., hydrogen atoms flew away from the complex, identified by any hydrogen atom being over 1.5 Å away from the nearest heavy atom) and marked as failures
    b. All others (1,004) were kept in (see explanation below)
3. **All successful TeraChem calcs (total//unique): 159,014//125,830**
4. **All attempted Psi4 calcs (total//unique): 159,014//125,830**
5. **All successful Psi4 calcs (total//unique): 156,522//123,777**
    a. Successful **and** graph preserved (total//unique): 155,529//122,845
    b. Successful **and** not spin contaminated (total//unique): 154,856//122,960
    c. Successful **and** graph preserved **and** not spin contaminated (total//unique): 153,872//122,037

Explanation of changed molecular graphs:

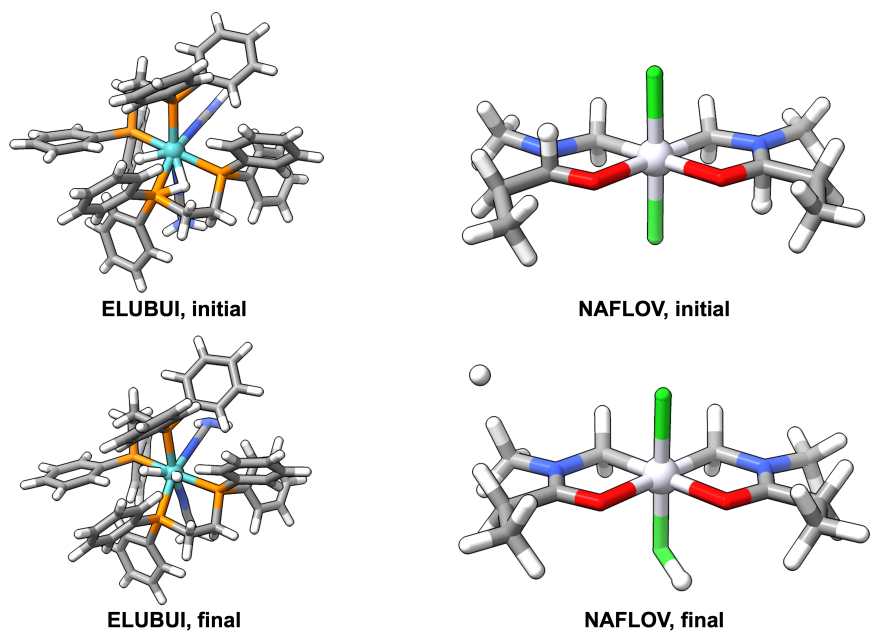

**ELUBUI, initial**    **NAFLOV, initial**

**ELUBUI, final**    **NAFLOV, final**

Above depicts representative structures with changed molecular graphs during optimization. The top row corresponds to structures before optimization (i.e., from the original crystal structure), while the bottom row depicts structures after optimizing the positions of all hydrogen atoms at the PBE0-D3BJ/def2-SV(P) level of theory. Both structures were identified to have a change in molecular graph before and after optimization, and included at least one hydrogen atom greater than 1.5 Å away from the nearest heavy atom. The left structure depicts a metal hydride with poorly resolved metal-hydrogen bonding before optimization, while the right structure depicts a hydrogen atom dissociating from the complex. The left structure is chemically meaningful and retained in the dataset, while the right structure is discarded. Atoms are colored as follows: hydrogen in white, carbon in gray, nitrogen in blue, oxygen in red, phosphorus in orange, chlorine in green, molybdenum in light blue, and platinum in light gray. Metal atoms and hydrogen atoms of interest are depicted as spheres.

**Table S3.** Spin states of neutral non-metal atoms used for atomization energy calculations, as determined from PBE0/def2-TZVP.

| Element | Ground state spin |
|---|---|
| H | 2 |
| Li | 2 |
| Be | 1 |
| B | 2 |
| C | 3 |
| N | 4 |
| O | 3 |
| F | 2 |
| Na | 2 |
| Mg | 1 |
| Al | 2 |
| Si | 3 |
| P | 4 |
| S | 3 |
| Cl | 2 |
| K | 2 |
| Ca | 1 |
| Ga | 2 |
| Ge | 3 |
| As | 4 |
| Se | 3 |
| Br | 2 |
| Rb | 2 |
| Sr | 1 |
| In | 2 |
| Sn | 3 |
| Sb | 4 |
| Te | 3 |
| I | 2 |
| Cs | 2 |
| Ba | 1 |
| Tl | 2 |
| Pb | 3 |
| Bi | 4 |

**Text S5.** Description of the methodology used to obtain single-atom energies.

For each single-atom energy required for the workflow in Text S6, the following convergence schemes were attempted, where subsequent steps were only attempted if earlier calculation attempts failed:
1. In Psi4, converge a PBE0/def2-SV(P) calculation starting from a superposition of atomic densities (SAD) guess. Then, project this to PBE0/def2-TZVP.
2. In ORCA v6.1.0, converge a PBE0/def2-TZVP calculation from a polarized atomic densities guess.

In the case of atomization energies computed for the manyDFA set, the following schemes were attempted in the following order:
1. From a converged PBE0 result, perform a DFT calculation using the selected functional and def2-TZVP starting from the PBE0 wave function. If this fails, start from a SAD guess.
2. In ORCA v6.1.0, perform a DFT calculation with the selected functional and def2-TZVP from a polarized atomic densities guess.

3. In Psi4, attempt to converge the DFA with def2-TZVP starting from a functional that is not PBE0.
    a. In the case of semilocal DFAs, take a converged PBE result and initialize the calculations for other semilocal DFAs from PBE.
        i. If the PBE calculation did not converge, successively converge PBE+x% calculations, where x is the amount of Hartree-Fock exchange (HFX) and ranges between 0% and 20%, as was done in prior work.[1]
    b. In the case of M06-2X, PBE0-DH, and PWPB95, obtain a PBE+50% HFX result and initialize the calculations from these. The PBE+50% HFX are again obtained by successively converging PBE+x% HFX calculations, where x ranges from 20% to 50%.
    c. In the case of a range-separated hybrid (RSH), initialize the calculation from one of the other RSH functionals that converged.
4. For semilocal DFAs, if the strategy in (3.a.i) did not work, attempt to converge the semilocal DFA+20% HFX from a PBE0 result, and step down in 1% HFX intervals. If the semilocal DFA converges, utilize that result. Otherwise, utilize a linear extrapolation of the lowest two HFX% calculations to converge to estimate what the energy would be at 0% HFX. This extrapolation should be reasonable for low HFX% regimes, as established in prior work.[1] Calculations requiring HFX extrapolation are detailed in Figure S1.

Atomization energies were computed even if the required single-atom calculations were spin contaminated. All atoms with a spin contamination metric, $\langle S^2 \rangle - S(S+1) > 1$, are reported in Tables S6 and S7.

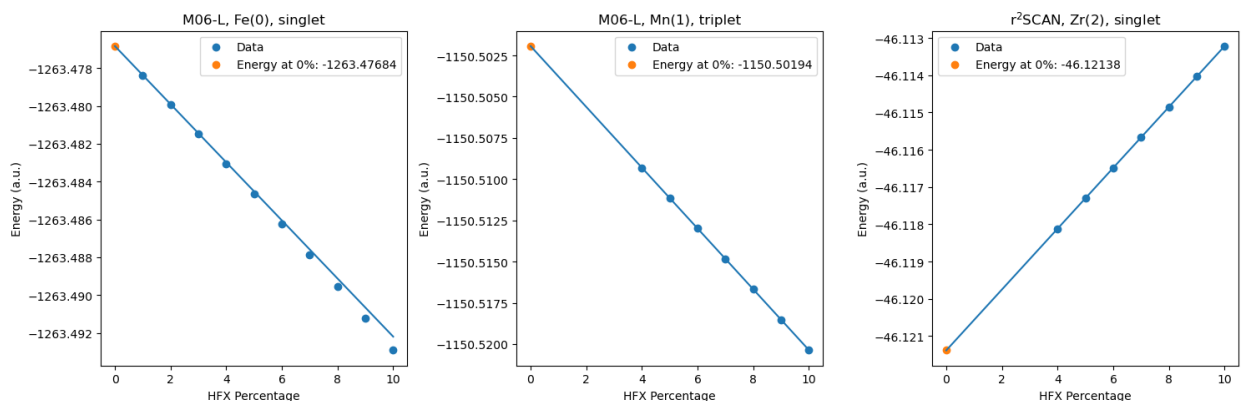

**Figure S1.** Atoms requiring HFX extrapolation to obtain energies, along with the values used for extrapolation. All energies are in hartree.[a]

**Text S6.** Algorithm for determining the atomization energies of transition metal complexes.
1. Given the oxidation state of the metal and the total charge of the transition metal complex (TMC), determine the total charge of all ligands ($N$) by subtracting the metal oxidation state from the total charge of the TMC. Then, -$N$ gives the number of excess electrons to be distributed amongst the ligand atoms.
2. Calculate the energy of the metal center, maintaining the same oxidation and spin state as it had in the transition metal complex, using the same functional and basis set as the TMC.
    a. In the case of non-innocent ligands, i.e., if the total electron count of all ligands was odd, the metal spin was incremented by one such that the metal oxidation and spin

state were compatible while maintaining the classification of low-, intermediate-, or high-spin (e.g., singlet to doublet, or quartet to triplet). This shortcut avoided computing non-metal atoms in non-ground states and impacted fewer than 2,200 complexes.
3. Compile a list of all nonmetal atoms present in the transition metal complex, $\{A_i\}$, in their neutral charge states and the lowest-energy spin state, as determined from PBE0/def2-TZVP calculations on the isolated atoms for the first five accessible spin multiplicities (i.e., 1, 3, 5, 7, 9 or 2, 4, 6, 8, 10). More details on the single-atom PBE0/def2-TZVP results are found in Text S5 and Table S6.
4. If $N < 0$, then there are $-N$ additional electrons that need to be assigned. For each of these electrons:
   a. Find the atom type in $\{A_i\}$ that has the lowest electron attachment energy, calculated as $E(A_i^{-1}) - E(A_i)$, where superscripts denote the charge of the atom, and all atom energies are in their ground state spin.
   b. Remove one atom of type $A_i$ from the list of atoms, and replace it with $A_i^{-1}$ in its ground state spin. Then, go back to step 4a and repeat until all $-N$ electrons have been assigned.
5. If $N > 0$, then there are $N$ electrons that need to be removed. For each of these electrons:
   a. Find the atom type in $\{A_i\}$ that has the lowest electron detachment energy, calculated as $E(A_i^{+1}) - E(A_i)$, where superscripts denote the charge of the atom, and all atom energies are in their ground state spin.
   b. Remove one atom of type $A_i$ from the list of atoms, and replace it with $A_i^{+1}$ in its ground state spin. Then, go back to step 5a and repeat until $N$ electrons have been removed.
6. Determine the sum of the energies of the atoms in the transition metal complex by adding the metal atom energy to the sum of nonmetal atom energies, where the additional/missing electrons have been incorporated into the nonmetal atoms as to minimize the resultant sum of atom energies.
7. Calculate the atomization energy as the energy of the transition metal complex minus the sum of the energies of all atoms calculated in step 6.

In the case of non-PBE0 atomization energies (i.e., for the manyDFA set), atom energies are computed with the same functional used to obtain the complex energy, again with the def2-TZVP basis set. More details on these calculations are found in Text S5 and Table S7.

**Table S4.** Elements with electrons added to them during the atomization energy procedure (i.e., those with high electron affinities), alongside the spin state used to compute the energies of the anionic species, as determined from PBE0/def2-TZVP.

| Element | Charge | Spin |
|---------|--------|------|
| H | -1 | 1 |
| B | -1 | 3 |
| C | -1 | 4 |
| C | -2 | 3 |
| N | -1 | 3 |
| N | -2 | 2 |
| N | -3 | 1 |
| O | -1 | 2 |
| O | -2 | 1 |
| F | -1 | 1 |
| Si | -1 | 4 |
| S | -1 | 2 |
| S | -2 | 1 |
| Cl | -1 | 1 |
| Cl | -2 | 2 |
| Se | -1 | 2 |
| Se | -2 | 1 |
| Br | -1 | 1 |
| Br | -2 | 2 |
| Sn | -1 | 4 |
| Te | -1 | 2 |
| Te | -2 | 1 |
| I | -1 | 1 |
| I | -2 | 2 |
| Pb | -1 | 4 |

**Table S5.** Elements with electrons removed from them during the atomization energy procedure (i.e., those with low ionization potentials), alongside the spin state used to compute the energies of the cationic species, as determined from PBE0/def2-TZVP.

| Element | Charge | Spin |
|---------|--------|------|
| Li | +1 | 1 |
| B | +1 | 1 |
| C | +1 | 2 |
| Na | +1 | 1 |
| Si | +1 | 2 |
| P | +1 | 3 |
| S | +1 | 4 |
| Ga | +1 | 1 |
| Ge | +1 | 2 |
| Sn | +1 | 2 |
| Sb | +1 | 3 |
| Tl | +1 | 1 |

**Text S7.** Scheme for determining the set of >10k TMCs studied with multiple functionals.
1. From the set of all TMCs sampled in BOS-TMC, determine the distribution of metal centers (Figure S2).
2. For each metal center:
    a. Determine the total number of structures, $N$, needed in the sample as to preserve the overall distribution of metal centers:

$$\text{\# to include} = N = 12{,}000 * \frac{\text{\# unique structures containing metal}}{\text{\# unique structures in BOS} - \text{TMC}}$$

We selected 12,000 TMCs so that, after attrition of some TMCs due to failed calculations or spin contamination, the resultant dataset would still contain over 10,000 data points where all twelve functionals converge.

b. Obtain the $N$ smallest unique TMCs containing that metal center (rounded up). Uniqueness is determined by unique graph hashes and coordination formulae (i.e., metal identity and ligand molecular formulae). In the case of duplicates, the lowest R-factor structure (i.e., the higher-quality crystal structure) was used. These uniqueness checks were carried out independent of charge, oxidation, and spin state, meaning that transition metal complexes that only varied by the addition/removal of electrons were not treated as distinct. This differs from the uniqueness definition for the overall set, where uniqueness is determined by i) the graph hash, which assesses the connectivity of atoms and identity of bonded atoms in a molecule, and ii) whether it was run as a singlet or doublet. That is, TMCs that would be treated as identical in this smaller set may be viewed as distinct for the full set.

This scheme allows for the subset studied with multiple functionals to not be biased towards early transition metals, as would have been done if only the smallest 10k structures (by electron count) were selected. (Figure S3). This does lead to slightly larger structures being selected on average but gives a more representative sample of chemical space (Figure S4). The metal center with the fewest TMCs included in this set is scandium, with 28 structures.

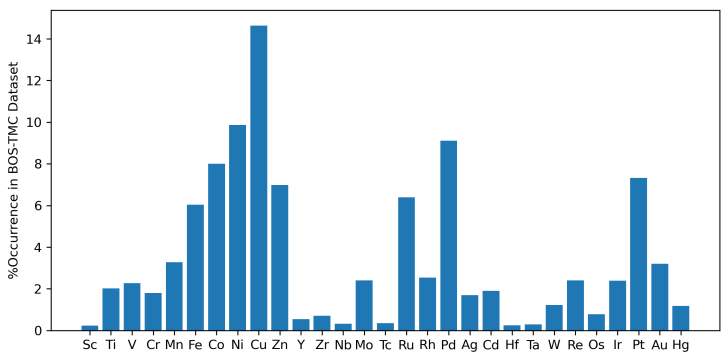

**Figure S2.** Distribution of metal centers among unique (i.e., different graph hashes and coordination formulae) structures in the BOS-TMC dataset.

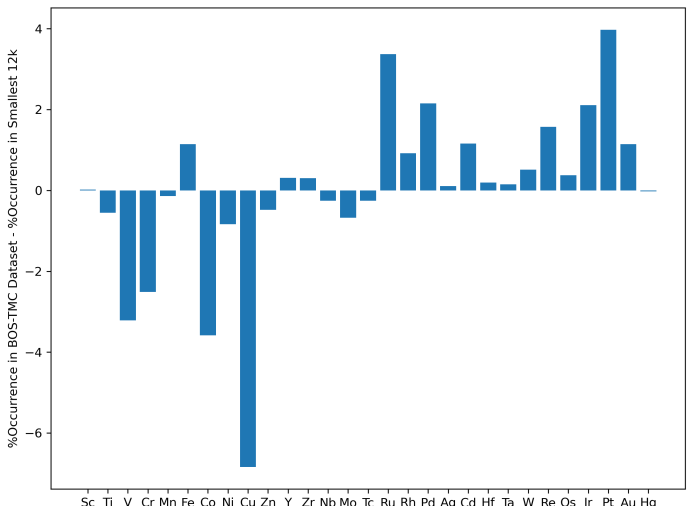

**Figure S3.** Difference in the distributions of metal centers among the entirety of unique structures in BOS-TMC (i.e., those with different graph hashes and coordination formulae) and the set of 12,000 TMCs with the smallest atom counts. Negative numbers indicate higher occurrence in the smallest 12k structures, while positive numbers indicate higher occurrence in the overall dataset.

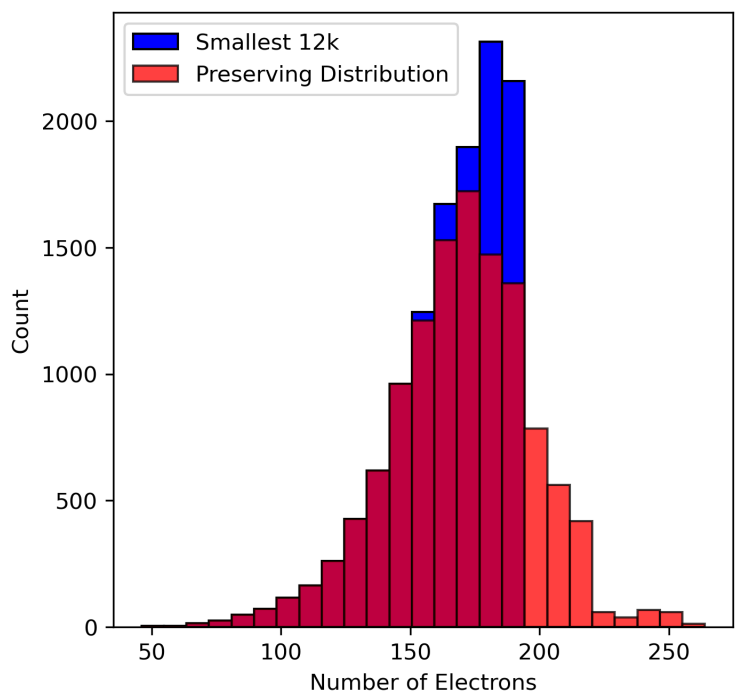

**Figure S4.** Distribution of electron counts in the unique TMCs in the smallest 12k unique structures by electron count (orange) and in the smallest 12k structures such that we preserve the distribution of metal centers among unique structures in BOS-TMC (blue).

**Table S6.** All atom energies used for computing PBE0 atomization energies that had a deviation of the $\langle S^2 \rangle$ operator from its expected $S(S+1)$ value over 1, along with the magnitude of that deviation.

| Element and Charge | Spin | Deviation from $S(S+1)$ |
|---|---|---|
| Cr(0) | 5 | 1.0002 |
| Mn(0) | 2 | 1.9978 |
| Co(0) | 2 | 1.0109 |
| Pd(3) | 2 | 1.0015 |
| Pd(4) | 3 | 1.0018 |
| Os(0) | 3 | 1.0059 |
| Pt(3) | 2 | 1.0011 |

**Table S7.** All atom energies used for computing manyDFA atomization energies that had a deviation [Sdev] of the $\langle S^2 \rangle$ operator from its expected $S(S+1)$ value over 1, along with the magnitude of that deviation. The reported format is Metal(ox), [Sdev].

| Functional | LS spin contaminated atoms | IS spin contaminated atoms | HS spin contaminated atoms |
|---|---|---|---|
| PBE | V(0), [1.664]; V(2), [1.001]; Cr(1), [2.001]; Mn(2), [2.001]; Mn(4), [1.001]; Fe(1), [1.143]; Fe(3), [2.001]; Mo(1), [2.001]; Mo(3), [1.001]; Tc(2), [2.001]; Tc(4), [1.001]; Ru(3), [2.001]; Ru(5), [1.001]; Pd(3), [1.001]; W(3), [1.001]; Re(2), [2.001]; Re(4), [1.001]; Os(3), [2.001]; Os(5), [1.001]; Ir(4), [1.003]; Pt(3), [1.004] | Cr(0), [1.876]; Cr(1), [1.001]; Cr(2), [1.001]; Mn(1), [1.459]; Mn(2), [1.001]; Mn(3), [1.001]; Fe(0), [1.014]; Fe(3), [1.001]; Fe(4), [1.000]; Nb(1), [1.001]; Mo(0), [1.930]; Tc(2), [1.001]; Tc(3), [1.001]; Ru(0), [1.004]; Ru(3), [1.001]; Rh(5), [1.001]; W(2), [1.001]; Re(1), [1.985]; Os(2), [1.080]; Os(3), [1.001]; Ir(4), [1.001] | |
| r²SCAN | V(0), [1.660]; V(2), [1.001]; Cr(3), [1.001]; Mn(4), [1.001]; Fe(1), [1.171]; Fe(3), [1.001]; Co(2), [1.001]; Ni(3), [1.001]; Mo(1), [2.003]; Tc(4), [1.002]; Ru(3), [2.003]; Ru(5), [1.002]; Rh(2), [1.005]; Pd(3), [1.002]; Re(4), [1.002]; Os(3), [2.003]; Os(5), [1.002]; Ir(4), [1.002]; Pt(3), [1.001] | V(1), [1.001]; Cr(1), [1.001]; Cr(2), [1.002]; Mn(2), [1.001]; Mn(3), [1.002]; Fe(0), [1.022]; Fe(2), [1.001]; Co(3), [1.001]; Ni(4), [1.001]; Ru(2), [1.007]; Rh(3), [1.003]; Rh(5), [1.002]; Pd(4), [1.002]; Re(1), [2.003]; Os(2), [1.186]; Ir(3), [1.012]; Pt(4), [1.004] | Cr(0), [1.000] |
| M06-L | V(0), [1.671]; V(2), [1.001]; Cr(1), [2.001]; Mn(4), [1.001]; Fe(3), [1.001]; Ru(3), [2.003]; Ru(5), [1.002]; Pd(3), [1.001]; Os(3), [2.002]; Ir(4), [2.002]; Pt(3), [1.001] | V(1), [1.001]; Cr(0), [1.806]; Cr(1), [1.001]; Cr(2), [1.001]; Mn(2), [1.001]; Mn(3), [1.001]; Fe(0), [1.023]; Fe(2), [1.003]; Co(3), [1.001]; Mo(0), [1.849]; Ru(0), [1.006]; Ru(2), [1.007]; Rh(3), [1.002]; Rh(5), [1.002]; Pd(4), [1.002]; Re(1), [1.912]; Os(2), [1.002]; Ir(3), [1.002]; Pt(4), [1.001] | Cr(0), [1.001] |
| B3LYP* | Pd(3), [1.001]; Pt(3), [1.001] | Cr(2), [1.001]; Fe(0), [1.012]; Pd(4), [1.001] | Cr(0), [1.000] |
| B3LYP | Pd(3), [1.001]; Pt(3), [1.001] | Fe(0), [1.012]; Pd(4), [1.001] | Cr(0), [1.000] |
| PBE0 | Pd(3), [1.001]; Pt(3), [1.001] | Pd(4), [1.002] | Cr(0), [1.000] |
| M06-2X | Pd(3), [1.001]; Pt(3), [1.001] | Cr(0), [1.015]; Pd(4), [1.002] | Cr(0), [1.000] |
| LRC-ωPBEh | Pd(3), [1.001]; Pt(3), [1.001] | Pd(4), [1.002] | Cr(0), [1.000] |
| ωB97X-D | Pd(3), [1.001]; Pt(3), [1.001] | Pd(4), [1.001] | Cr(0), [1.000] |
| ωB97M-V | | Fe(0), [1.010] | |
| PBE0-DH | Pd(3), [1.003]; Pt(3), [1.002] | Cr(0), [1.011]; Pd(4), [1.003] | Cr(0), [1.000] |
| PWPB95 | | Fe(0), [1.015] | Cr(0), [1.000] |

**Table S8.** Summary of attempted and converged calculations carried out with TeraChem (PBE0/def2-SV(P)) for H-optimized structures and with Psi4 (PBE0/def2-TZVP) for single-point energy calculations. All successful TeraChem H-optimizations (including those above the spin contamination threshold) were attempted with Psi4 except for 193 singlet and 429 doublet calculations that exceeded the walltime limit. The reported success rates are all out of the initial pool of candidate structures. The spin contamination success rate (spin cont. OK) excludes doublets where the expectation value of $S^2$ deviated from $S(S+1)$ by more than 1. Lower spin contamination is observed for the Psi4 calculations, leading to a higher success rate for the Psi4 than the TeraChem calculations. The total failed column indicates how many failed out of those attempted (i.e., versus Initial for the TeraChem H-opt but versus H-opt successful for the Psi4 Single point lines).

|  | Singlet |  | Doublet |  | Total successful |  | Total failed |
|---|---|---|---|---|---|---|---|
| Step | Count | Success rate | Count | Success rate | Count | Success rate | Count |
| Initial | 119088 | -- | 43021 | -- | 162109 | -- | -- |
| H-opt successful | 116975 | 98% | 42039 | 98% | 159014 | 98% | 3095 |
| H-opt (spin cont. OK) | 116975 | 98% | 38662 | 90% | 155637 | 96% | 6472 |
| Single point successful | 115383 | 97% | 41139 | 96% | 156522 | 97% | 2492 |
| Single point (spin cont. OK) | 115383 | 97% | 39473 | 92% | 154856 | 96% | 4158 |

**Table S9.** Success rate of converged ('Conv') calculations with TeraChem or Psi4 grouped by singlet and doublet spin convention as well as being below the spin contamination threshold of 1.0 for the deviation of the expectation operator ('Spin OK'). Total successes and success rates across singlets and doublets are reported as well as grouped by cases where $|q| \leq 1$ and $|q| > 1$. Unique molecular graph counts are reported for the initial set and the final, non-spin contaminated Psi4 set. Any numbers not available or not computed are indicated by '--'.

|  | TeraChem | | | | | | Psi4 | | | | | |
|---|---|---|---|---|---|---|---|---|---|---|---|---|
|  | Init. all | Singlet Conv | Doublet Conv | Dbl. Conv & Spin OK | Tot. Success | Success rate | Init. all | Singlet Conv | Doublet Conv | Dbl. Conv & Spin OK | Tot. Succes | Success rate |
| q | | | | | | | | | | | | |
| -8 | 2 | 1 | 0 | 0 | 1 | 50% | 1 | 1 | 0 | 0 | 1 | 100% |
| -7 | 0 | 0 | 0 | 0 | 0 | -- | 0 | 0 | 0 | 0 | 0 | -- |
| -6 | 9 | 5 | 3 | 3 | 8 | 89% | 8 | 5 | 3 | 3 | 8 | 100% |
| -5 | 17 | 14 | 3 | 3 | 17 | 100% | 17 | 13 | 3 | 3 | 16 | 94% |
| -4 | 226 | 165 | 55 | 49 | 214 | 95% | 220 | 164 | 55 | 54 | 218 | 99% |
| -3 | 669 | 266 | 394 | 318 | 584 | 87% | 660 | 264 | 391 | 318 | 582 | 88% |
| -2 | 6467 | 3847 | 2584 | 2059 | 5906 | 91% | 6429 | 3840 | 2578 | 2124 | 5964 | 93% |
| -1 | 6724 | 4101 | 2520 | 2087 | 6188 | 92% | 6617 | 4071 | 2504 | 2277 | 6348 | 96% |
| 0 | 101762 | 75295 | 24322 | 22692 | 97987 | 96% | 99557 | 74184 | 23668 | 23001 | 97185 | 98% |
| 1 | 25660 | 19468 | 5857 | 5588 | 25056 | 98% | 25324 | 19208 | 5774 | 5672 | 24880 | 98% |
| 2 | 18740 | 12501 | 5838 | 5482 | 17983 | 96% | 18324 | 12337 | 5712 | 5630 | 17967 | 98% |
| 3 | 1540 | 1118 | 382 | 303 | 1421 | 92% | 1500 | 1110 | 375 | 317 | 1427 | 95% |
| 4 | 214 | 139 | 69 | 66 | 205 | 96% | 208 | 137 | 68 | 66 | 203 | 98% |
| 5 | 58 | 45 | 10 | 10 | 55 | 95% | 55 | 39 | 7 | 7 | 46 | 84% |
| 6 | 8 | 5 | 1 | 1 | 6 | 75% | 6 | 5 | 1 | 1 | 6 | 100% |
| 7 | 3 | 3 | 0 | 0 | 3 | 100% | 3 | 3 | 0 | 0 | 3 | 100% |
| 8 | 7 | 2 | 1 | 1 | 3 | 43% | 3 | 2 | 0 | 0 | 2 | 67% |
| 9 | 2 | 0 | 0 | 0 | 0 | 0% | 0 | 0 | 0 | 0 | 0 | -- |
| 10 | 0 | 0 | 0 | 0 | 0 | -- | 0 | 0 | 0 | 0 | 0 | -- |
| 11 | 0 | 0 | 0 | 0 | 0 | -- | 0 | 0 | 0 | 0 | 0 | -- |
| 12 | 1 | 0 | 0 | 0 | 0 | 0% | 0 | 0 | 0 | 0 | 0 | -- |
| all | 162109 | 116975 | 42039 | 38662 | 155637 | 96% | 158932 | 115383 | 41139 | 39473 | 154856 | 97% |
| $|q|>1$ | 27963 | 18111 | 9340 | 8295 | 27451 | 98% | 27434 | 17920 | 9193 | 8523 | 26443 | 96% |
| $|q|\leq1$ | 134146 | 98864 | 32699 | 30367 | 128186 | 96% | 131498 | 97463 | 31946 | 30950 | 128413 | 98% |
| unique mol. graph | 128,357 | -- | -- | -- | -- | | -- | -- | -- | -- | 122,960 | 96% |

**Table S10.** Number of atoms (left) and molecular weight (right) statistics: minimum value, maximum value, mean, standard deviation (std. dev.), and median value. The results reported included singlet and doublet converged calculations without substantial spin contamination for TeraChem and Psi4.

|  | # atoms | | | | | molecular weight | | | | |
|---|---|---|---|---|---|---|---|---|---|---|
|  | min | max | mean | std. dev. | median | min | max | mean | std. dev. | median |
| Initial set | 2 | 719 | 62.4 | 31.7 | 58 | 93 | 5051 | 588.5 | 234.5 | 554 |
| TeraChem converged | 2 | 292 | 61.2 | 28.4 | 58 | 93 | 3329 | 580.0 | 210.9 | 552 |
| Psi4 converged | 2 | 245 | 60.3 | 27.6 | 57 | 93 | 2395 | 573.8 | 205.0 | 548 |

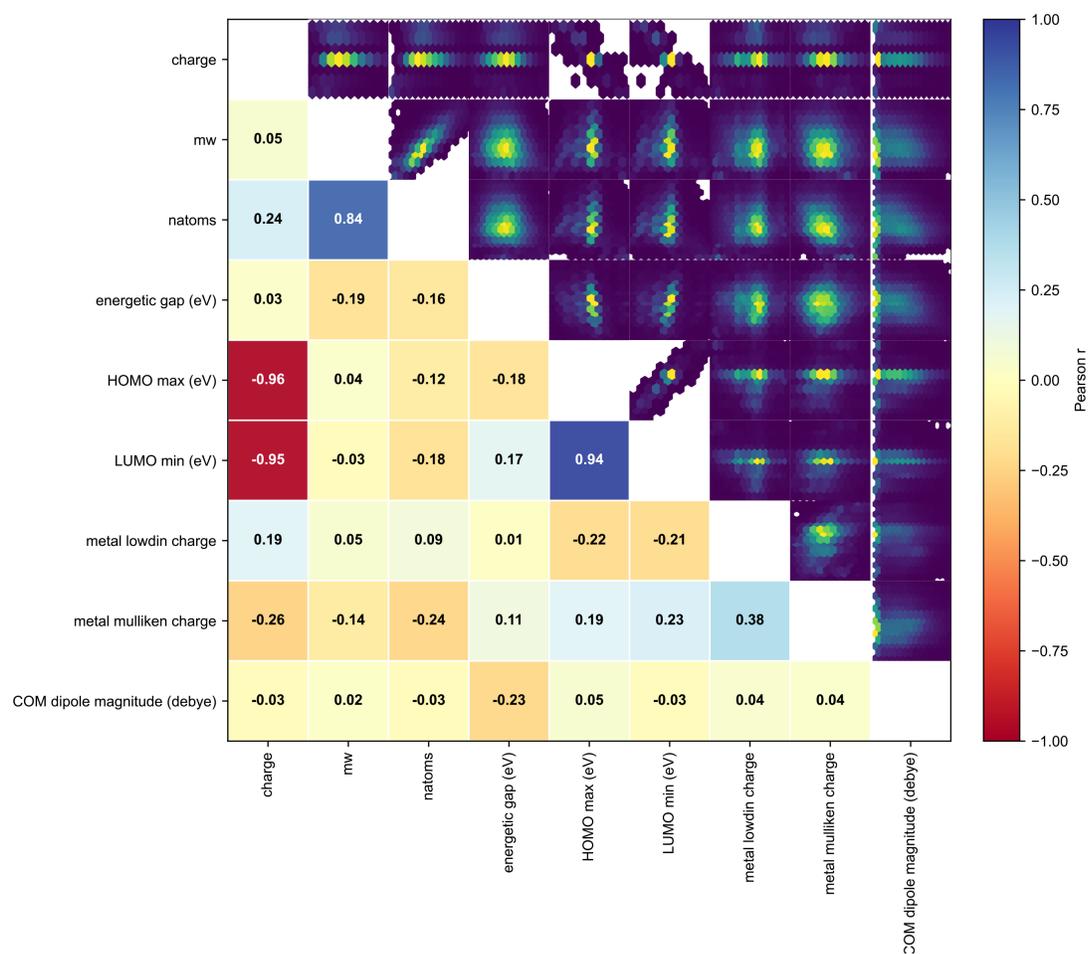

**Figure S5.** Correlation matrix of net charge (charge), molecular weight (mw), number of atoms (natoms), energetic HOMO-LUMO gap (eV), maximum HOMO value (eV), minimum LUMO value (eV), metal Löwdin charge, metal Mulliken charge, and COM dipole magnitude (Debye). The Pearson's r values are shown in the bottom left triangle of the matrix color coded according to the color bar, and KDE density plots of the individual correlations are shown in the top right triangle of the matrix.

**Table S11.** Summary of differences between the structures found in both BOS-TMC and tmQMg, where identical structures are determined solely by CSD refcode.[a]

|  | Count |
|---|---|
| Overlap by refcode | 33,893 |
| Structures have the same heavy atoms | 33,457 |
| Structures have the same molecular formula | 33,448 |
| Structure run as a singlet | 33,442 |
| Structure converged and was not an outlier for all available spin states for PBE0/def2-TZVP | 33,021 |

[a]We refer to all structures which contain the same refcode and molecular formula as the overlap between tmQMg and BOS-TMC, i.e., the size of the overlap is 33,448.

**Table S12.** Energetic gap (eV): min, max, mean, standard deviation, and median for 26,353 non-outlier complexes with |q| > 1, the full remaining set with |q| ≤1 and a representative 26,353-TMC sample for |q| ≤1.

|  | |q| > 1 | |q| ≤ 1 | |q| ≤1 sample |
|---|---|---|---|
| count | 26353 | 128042 | 26353 |
| min | 0.163 | 0.014 | 0.186 |
| max | 9.932 | 9.934 | 9.933 |
| mean | 4.239 | 3.777 | 3.777 |
| std | 1.489 | 1.034 | 1.035 |
| median | 4.043 | 3.792 | 3.797 |

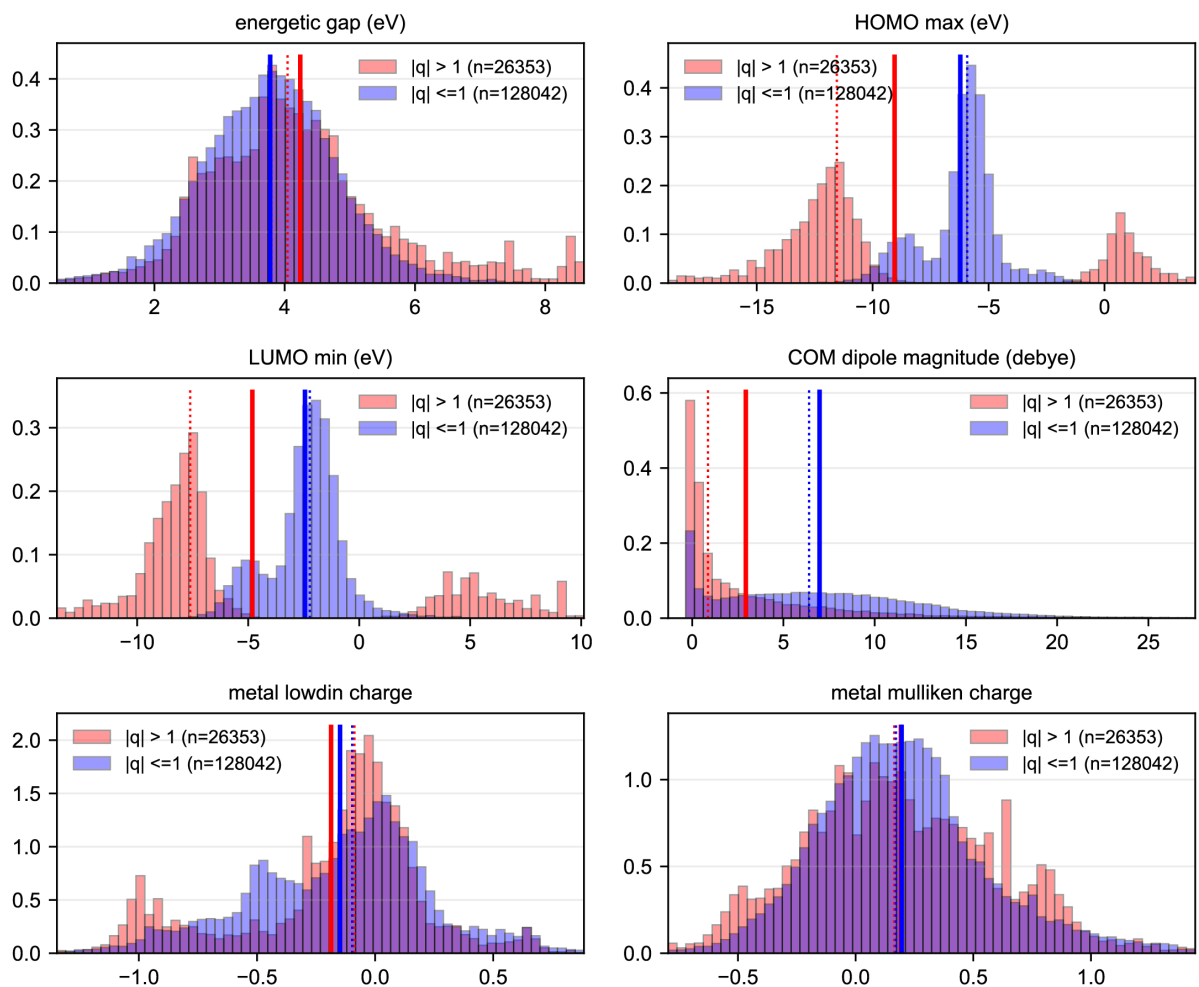

**Figure S6.** Normalized histograms of six quantities (left to right, top to bottom): energetic HOMO-LUMO gap (eV), maximum HOMO value (eV), minimum LUMO value (eV), COM dipole magnitude (Debye), metal Löwdin charge, metal Mulliken charge. Maximum HOMO / minimum LUMO refers to taking the energetically highest occupied (lowest unoccupied) level from either the majority or minority spin based on energetic criteria. Unlike the preceding figure, the quantities are shown for the |q| > 1 set of 26,353 complexes after eliminating outliers (red) as well as all 128,042 |q| ≤1 complexes (blue) with outliers removed. There are 60 bins shown for all quantities.

**Table S13.** HOMO-LUMO energetic gap (eV) grouped by net charge of the TMC statistics: mean, standard deviation (std), minimum (min), maximum (max), and median values.

| charge | count | mean | std | min | max | median |
|---|---|---|---|---|---|---|
| -8 | 1 | 3.098 | -- | 3.098 | 3.098 | 3.098 |
| -7 | 0 | -- | -- | -- | -- | -- |
| -6 | 8 | 3.711 | 0.759 | 2.511 | 4.424 | 4.016 |
| -5 | 16 | 4.373 | 0.785 | 3.455 | 6.052 | 4.298 |
| -4 | 218 | 4.421 | 1.266 | 0.760 | 7.391 | 4.533 |
| -3 | 582 | 4.350 | 1.518 | 0.415 | 8.120 | 4.207 |
| -2 | 5964 | 4.584 | 1.886 | 0.383 | 9.384 | 3.950 |
| -1 | 6337 | 3.596 | 1.450 | 0.014 | 9.934 | 3.417 |
| 0 | 97184 | 3.744 | 0.998 | 0.135 | 8.399 | 3.788 |
| 1 | 24880 | 3.913 | 1.077 | 0.016 | 8.493 | 3.853 |
| 2 | 17960 | 4.076 | 1.264 | 0.191 | 9.932 | 4.018 |
| 3 | 1415 | 4.770 | 1.825 | 0.163 | 9.929 | 4.640 |
| 4 | 203 | 3.724 | 1.597 | 0.231 | 7.641 | 3.593 |
| 5 | 46 | 3.900 | 2.360 | 0.426 | 7.089 | 3.767 |
| 6 | 6 | 4.072 | 2.041 | 0.385 | 6.272 | 4.756 |
| 7 | 3 | 2.416 | 1.777 | 0.365 | 3.501 | 3.381 |
| 8 | 2 | 2.120 | 2.340 | 0.466 | 3.775 | 2.120 |

**Table S14.** Maximum HOMO level (eV) grouped by net charge of the TMC statistics: mean, standard deviation (std), minimum (min), maximum (max), and median values.

| charge | count | mean | std | min | max | median |
|---|---|---|---|---|---|---|
| -8 | 1 | 13.794 | -- | 13.794 | 13.794 | 13.794 |
| -7 | 0 | -- | -- | -- | -- | -- |
| -6 | 8 | 14.380 | 1.249 | 12.702 | 15.532 | 15.014 |
| -5 | 16 | 10.158 | 1.521 | 8.043 | 12.865 | 9.659 |
| -4 | 218 | 6.880 | 2.094 | 0.828 | 10.396 | 6.956 |
| -3 | 582 | 3.692 | 1.569 | -0.723 | 11.880 | 3.477 |
| -2 | 5964 | 0.803 | 0.902 | -3.899 | 8.425 | 0.714 |
| -1 | 6337 | -2.829 | 0.881 | -9.903 | 1.272 | -2.784 |
| 0 | 97184 | -5.758 | 0.728 | -11.459 | 0.043 | -5.755 |
| 1 | 24880 | -8.867 | 0.853 | -14.090 | -4.165 | -8.790 |
| 2 | 17960 | -12.213 | 1.515 | -21.119 | -5.837 | -11.967 |
| 3 | 1415 | -16.988 | 2.843 | -23.956 | -9.558 | -16.590 |
| 4 | 203 | -17.123 | 2.693 | -24.639 | -11.358 | -16.645 |
| 5 | 46 | -20.069 | 3.392 | -26.198 | -15.621 | -18.946 |
| 6 | 6 | -20.025 | 1.126 | -21.322 | -18.774 | -20.048 |
| 7 | 3 | -21.428 | 3.075 | -23.235 | -17.877 | -23.171 |
| 8 | 2 | -21.742 | 1.661 | -22.917 | -20.567 | -21.742 |

**Table S15.** Minimum LUMO level (eV) grouped by net charge of the TMC statistics: mean, standard deviation (std), minimum (min), maximum (max), and median values.

| charge | count | mean | std | min | max | median |
| --- | --- | --- | --- | --- | --- | --- |
| -8 | 1 | 16.893 | -- | 16.893 | 16.893 | 16.893 |
| -7 | 0 | -- | -- | -- | -- | -- |
| -6 | 8 | 18.090 | 1.390 | 16.522 | 19.712 | 17.537 |
| -5 | 16 | 14.531 | 1.400 | 11.498 | 17.151 | 14.557 |
| -4 | 218 | 11.301 | 2.709 | 3.748 | 15.789 | 11.119 |
| -3 | 582 | 8.042 | 2.002 | 0.678 | 13.860 | 7.904 |
| -2 | 5964 | 5.388 | 1.885 | -3.498 | 9.716 | 5.039 |
| -1 | 6337 | 0.767 | 1.370 | -8.956 | 5.026 | 0.590 |
| 0 | 97184 | -2.014 | 0.861 | -10.325 | 1.052 | -2.016 |
| 1 | 24880 | -4.954 | 0.886 | -11.396 | -1.795 | -4.976 |
| 2 | 17960 | -8.137 | 1.200 | -17.270 | -4.052 | -7.969 |
| 3 | 1415 | -12.218 | 1.858 | -21.191 | -6.871 | -12.141 |
| 4 | 203 | -13.399 | 2.151 | -19.494 | -8.821 | -13.110 |
| 5 | 46 | -16.168 | 2.017 | -19.109 | -12.942 | -17.007 |
| 6 | 6 | -15.953 | 1.731 | -18.960 | -14.085 | -15.925 |
| 7 | 3 | -19.012 | 1.300 | -19.791 | -17.511 | -19.734 |
| 8 | 2 | -19.622 | 0.678 | -20.102 | -19.142 | -19.622 |

**Table S16.** COM magnitude dipole moment (Debye): min, max, mean, standard deviation, and median for 26,353 non-outlier complexes with $|q| > 1$, the full remaining set with $|q| \leq 1$ and a representative 26,353-TMC sample for $|q| \leq 1$.

|  | $|q| > 1$ | $|q| \leq 1$ | $|q| \leq 1$ sample |
| --- | --- | --- | --- |
| count | 26353 | 128042 | 26353 |
| min | 0.000 | 0.000 | 0.000 |
| max | 36.797 | 36.743 | 36.096 |
| mean | 2.938 | 6.984 | 6.997 |
| std | 4.605 | 5.502 | 5.490 |
| median | 0.866 | 6.408 | 6.404 |

**Table S17.** COM dipole magnitude (Debye) grouped by net charge of the TMC statistics: mean, standard deviation (std), minimum (min), maximum (max), and median values.

| charge | count | mean | std | min | max | median |
|---|---|---|---|---|---|---|
| -8 | 1 | 0.001 | -- | 0.001 | 0.001 | 0.001 |
| -7 | 0 | -- | -- | -- | -- | -- |
| -6 | 8 | 1.623 | 2.517 | 0.002 | 5.861 | 0.427 |
| -5 | 16 | 5.619 | 8.091 | 0.000 | 23.967 | 1.067 |
| -4 | 218 | 1.278 | 3.360 | 0.000 | 23.672 | 0.028 |
| -3 | 580 | 1.970 | 4.018 | 0.000 | 30.344 | 0.513 |
| -2 | 5958 | 1.511 | 3.331 | 0.000 | 35.138 | 0.335 |
| -1 | 6330 | 4.766 | 5.670 | 0.000 | 35.308 | 2.169 |
| 0 | 96885 | 7.144 | 5.559 | 0.000 | 36.743 | 6.594 |
| 1 | 24839 | 6.923 | 5.099 | 0.000 | 36.172 | 6.325 |
| 2 | 17919 | 3.348 | 4.676 | 0.000 | 36.473 | 1.529 |
| 3 | 1418 | 3.743 | 6.039 | 0.000 | 36.797 | 0.610 |
| 4 | 199 | 5.855 | 7.734 | 0.000 | 30.312 | 1.836 |
| 5 | 44 | 8.384 | 9.201 | 0.000 | 30.750 | 5.641 |
| 6 | 6 | 6.646 | 5.904 | 0.000 | 15.083 | 6.533 |
| 7 | 3 | 10.628 | 10.661 | 3.971 | 22.924 | 4.989 |
| 8 | 2 | 1.651 | 0.020 | 1.637 | 1.665 | 1.651 |

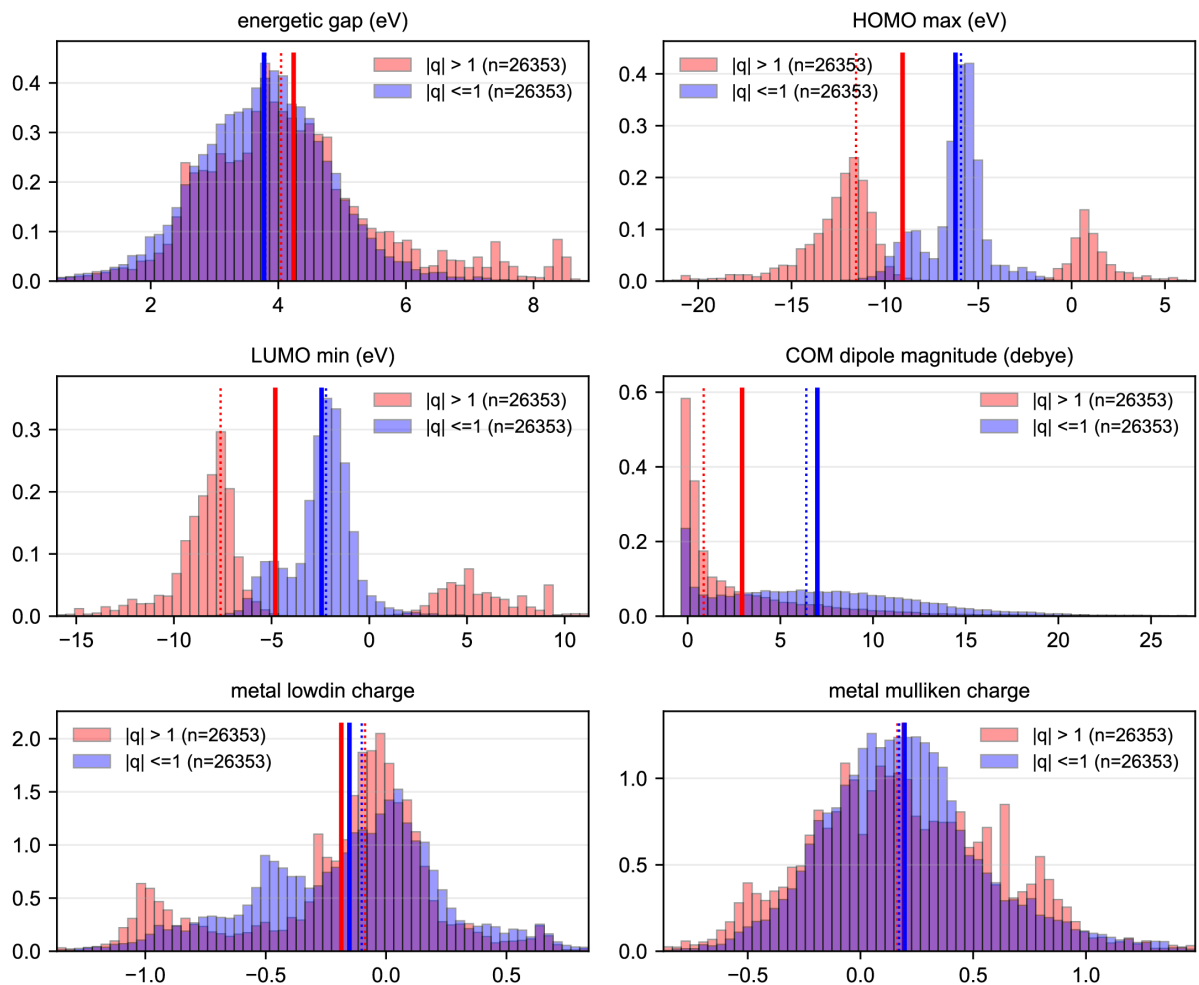

**Figure S7.** Normalized histograms of six quantities (left to right, top to bottom): energetic HOMO-LUMO gap (eV), maximum HOMO value (eV), minimum LUMO value (eV), COM dipole magnitude (Debye), metal Löwdin charge, metal Mulliken charge. The quantities are shown for the $|q| > 1$ set of 26,353 complexes after eliminating outliers (red) as well as a representative subset of 26,353 $|q| \leq 1$ complexes (blue) with outliers removed. There are 60 bins shown for all quantities.

**Table S18.** Metal Löwdin partial charge grouped by net charge of the TMC statistics: mean, standard deviation (std), minimum (min), maximum (max), and median values.

| charge | count | mean | std | min | max | median |
|---|---|---|---|---|---|---|
| -8 | 1 | -1.344 | -- | -1.344 | -1.344 | -1.344 |
| -7 | 0 | -- | -- | -- | -- | -- |
| -6 | 8 | 0.066 | 0.288 | -0.162 | 0.520 | -0.025 |
| -5 | 16 | 0.053 | 0.231 | -0.395 | 0.585 | 0.052 |
| -4 | 218 | -0.244 | 0.480 | -1.422 | 0.682 | -0.211 |
| -3 | 582 | -0.277 | 0.509 | -1.873 | 0.702 | -0.120 |
| -2 | 5964 | -0.560 | 0.504 | -1.666 | 1.107 | -0.807 |
| -1 | 6348 | -0.348 | 0.478 | -1.535 | 1.060 | -0.341 |
| 0 | 97185 | -0.147 | 0.388 | -2.014 | 1.168 | -0.093 |
| 1 | 24880 | -0.102 | 0.345 | -1.657 | 1.042 | -0.072 |
| 2 | 17967 | -0.064 | 0.276 | -1.617 | 1.180 | -0.049 |
| 3 | 1427 | -0.130 | 0.326 | -1.628 | 0.949 | -0.141 |
| 4 | 203 | -0.028 | 0.292 | -1.137 | 0.761 | -0.029 |
| 5 | 46 | -0.236 | 0.214 | -0.824 | 0.373 | -0.231 |
| 6 | 6 | -0.227 | 0.337 | -0.771 | 0.175 | -0.144 |
| 7 | 3 | -0.005 | 0.256 | -0.300 | 0.145 | 0.141 |
| 8 | 2 | 0.381 | 0.383 | 0.110 | 0.651 | 0.381 |

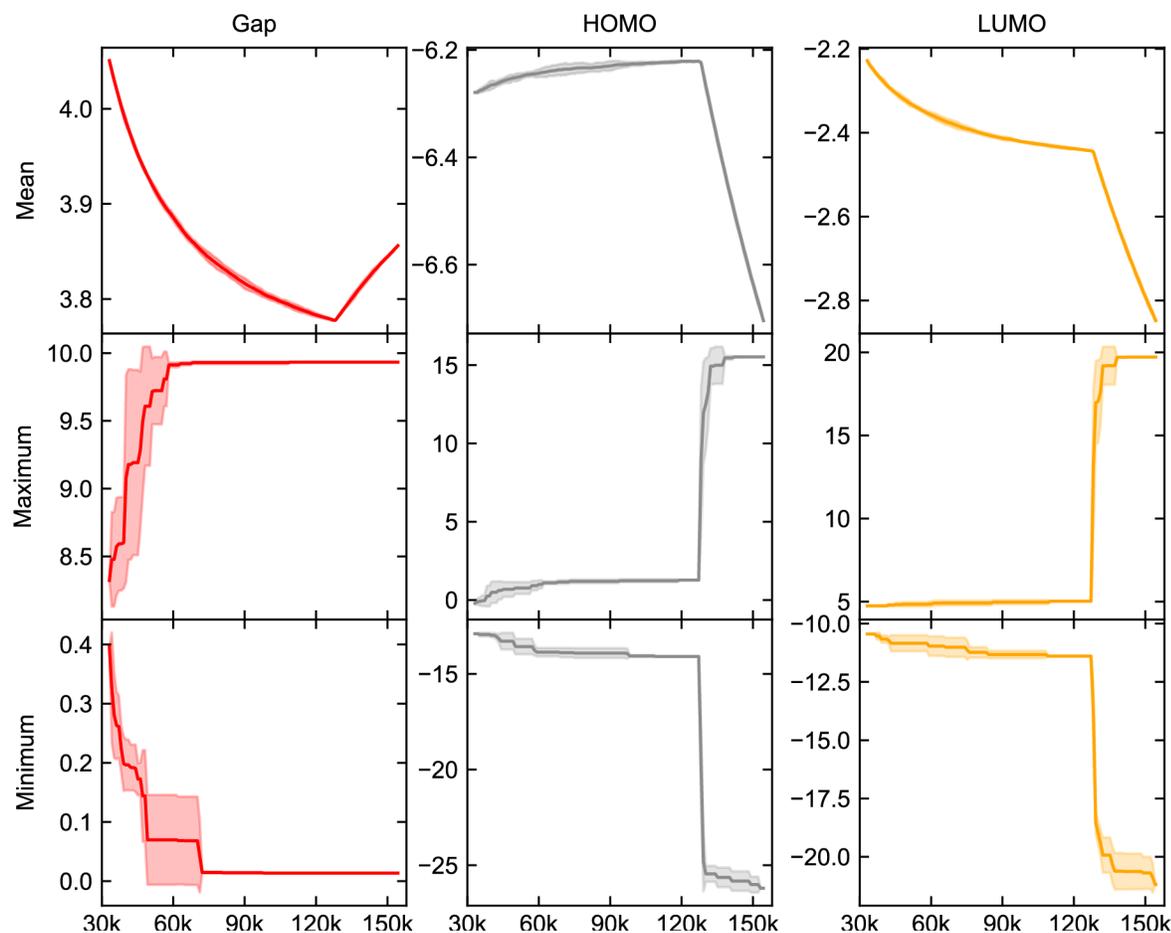

**Figure S8.** Effect of sequential addition of data on mean (top), max (middle), and min (bottom) of the energetic gap (eV, left), maximum HOMO level (eV, middle), and minimum LUMO level (eV, right) starting from a set of TMCs in common with tmQMg and adding ca. 30k |q| > 1 complexes last. The shaded regions correspond to the standard deviation (2x for the mean plot and 1x for max. and min.) observed when the process is repeated with five different folds of the data while preserving the ordering of addition of TMCs with respect to charge.

**Table S19.** Summary of statistics of properties in the 65k "weakly charged" set excluding |q| > 1 TMCs (this is representative of the size of the full tmQMg because it excludes |q| > 1): mean, min, max, std. dev., and median for energetic gap (eV), max HOMO (eV), min LUMO (eV), and dipole moment (Debye). The range of properties within 3 std. dev. of the mean ('lower' and 'upper') are also shown with the minimum dipole set to zero for the lower bound. The number of complexes in the full set that are outliers are indicated as well.

| Property | 65k "weakly charged" set | | | | | | | Full set | |
|---|---|---|---|---|---|---|---|---|---|
| | mean | min | max | std. dev. | median | lower | upper | # outliers | % additional |
| Energetic gap | 3.776 | 0.186 | 9.932 | 1.034 | 3.793 | 0.674 | 6.878 | 2734 | 3.1% |
| max HOMO | -6.212 | -14.090 | 1.060 | 1.631 | -5.919 | -11.106 | -1.318 | 23153 | 25.9% |
| min LUMO | -2.436 | -11.396 | 5.026 | 1.635 | -2.216 | -7.341 | 2.469 | 22759 | 25.5% |
| Dipole moment | 6.976 | 0.000 | 36.483 | 5.492 | 6.400 | 0.000 | 23.452 | 1439 | 1.6% |

**Table S20.** Properties of 10 outliers with respect to the "weakly charged" set in Table S19: refcode, net charge ($q$), metal species, chemical formula, number of atoms, dipole moment (Debye), max HOMO (eV), min LUMO (eV), and energetic gap (eV).

| refcode | q | spin | metal | formula | natoms | dipole (Debye) | HOMO max (eV) | LUMO min (eV) | HOMO-LUMO gap (eV) |
|---|---|---|---|---|---|---|---|---|---|
| **GEKKIQ** | 2 | 1 | Co | Co C6 H21 N6 O2 | 36 | 30.693 | -11.159 | -10.553 | 0.607 |
| **ROFBAP** | 3 | 1 | Co | Co C30 H42 N6 O | 80 | 29.325 | -11.905 | -11.423 | 0.482 |
| **SUWVAH** | 3 | 1 | Co | Co C28 H45 N7 | 81 | 29.113 | -11.640 | -11.255 | 0.385 |
| **LALFEN** | 3 | 1 | Co | Co C24 H45 N7 | 77 | 24.634 | -11.943 | -11.567 | 0.376 |
| **HUTYAX** | 3 | 2 | Cu | Cu C28 H62 N7 | 98 | 31.049 | -12.029 | -11.822 | 0.208 |
| **FEQVUV** | 4 | 2 | Cu | Cu C54 H50 N6 O15 | 126 | 24.402 | -11.358 | -11.127 | 0.231 |
| **SUWTUZ** | 5 | 1 | Co | Co C19 H40 N8 S | 69 | 24.776 | -18.000 | -17.534 | 0.466 |
| **TESSIT** | 5 | 1 | Co | Co C21 H42 N8 O | 73 | 23.500 | -17.598 | -17.147 | 0.451 |
| **TESSUF** | 5 | 1 | Co | Co C21 H42 N8 O | 73 | 24.198 | -17.816 | -17.343 | 0.472 |
| **ZEKNUY** | 5 | 1 | Co | Co C13 H38 N8 | 60 | 24.957 | -24.153 | -17.162 | 6.992 |

**Table S21.** Summary of attempted and converged intermediate-spin calculations carried out with TeraChem and Psi4 based on allowed spin multiplicities for the given metal and oxidation state (Table S2). The reported success rates are all out of the initial pool of candidate structures. The spin contamination success rate (spin cont. OK) excludes cases where the expectation value of $S^2$ deviated from $S(S+1)$ by more than 1. The total failed column indicates how many failed out of those attempted (i.e., versus Initial for the TeraChem calculations but versus TeraChem successful for the Psi4 calculations).

| | Triplet | | Quartet | | Total successful | | Total failed |
|---|---|---|---|---|---|---|---|
| Step | Count | Success rate | Count | Success rate | Count | Success rate | Count |
| Initial | 103710 | -- | 39885 | -- | 143595 | -- | -- |
| TeraChem single point successful | 103668 | 100% | 39861 | 100% | 143529 | 100% | 66 |
| TeraChem single point (spin cont. OK) | 103536 | 100% | 39622 | 99% | 143158 | 100% | 437 |
| Psi4 single point successful | 101502 | 98% | 38957 | 98% | 140459 | 98% | 3070 |
| Psi4 single point (spin cont. OK) | 101343 | 98% | 38826 | 97% | 140169 | 98% | 3360 |

**Table S22.** Summary of attempted and converged high-spin calculations (i.e., quintet or sextet) carried out with TeraChem and Psi4 based on allowed spin multiplicities for the given metal and oxidation state (Table S2). The reported success rates are all out of the initial pool of candidate structures. The spin contamination success rate (spin cont. OK) excludes cases where the expectation value of $S^2$ deviated from $S(S+1)$ by more than 1. The total failed column indicates how many failed out of those attempted (i.e., versus Initial for the TeraChem calculations but versus TeraChem successful for the Psi4 calculations).

|  | Quintet |  | Sextet |  | Total successful |  | Total failed |
|---|---|---|---|---|---|---|---|
| Step | Count | Success rate | Count | Success rate | Count | Success rate | Count |
| Initial | 32052 | -- | 16267 | -- | 48319 | -- | -- |
| TeraChem single point successful | 31854 | 99% | 16232 | 100% | 48086 | 100% | 233 |
| TeraChem single point (spin cont. OK) | 31833 | 99% | 16201 | 100% | 48034 | 99% | 285 |
| Psi4 single point successful | 31149 | 97% | 15701 | 97% | 46850 | 97% | 1236 |
| Psi4 single point (spin cont. OK) | 31137 | 97% | 15687 | 96% | 46824 | 97% | 1262 |

**Table S23.** Summary of attempted and converged spin-splitting energy calculations, computed by taking the difference in energy between a TMC in two different spin states. The reported success rates are all out of the initial pool of candidate structures. The spin contamination success rate (spin cont. OK) excludes cases where the expectation value of $S^2$ deviated from $S(S+1)$ by more than 1.

|  | IS-LS | | HS-LS | | HS-IS | |
|---|---|---|---|---|---|---|
| Step | Triplet-Singlet | Quartet-Doublet | Quintet-Singlet | Sextet-Doublet | Quintet-Triplet | Sextet-Quartet |
| Initial | 105,472 | 40,844 | 32,796 | 17,096 | 32,796 | 17,096 |
| TeraChem calculations successful | 103,668 | 39,861 | 31,854 | 16,232 | 31.854 | 16,232 |
| TeraChem calculations not spin contaminated | 103,536 | 36,429 | 31,833 | 13,659 | 31,740 | 16,006 |
| Psi4 single points successful | 100,569 | 38,504 | 30,752 | 15,440 | 30,610 | 15,426 |
| Psi4 single points not spin contaminated | 100,414 | 36,805 | 30,741 | 14,462 | 30,478 | 15,298 |
| Total by category | 137,219 | | 45,203 | | 45,776 | |

**Table S24.** Summary by metal and oxidation state of the counts of transition metal complexes modeled in each spin state. Each row represents a metal center and each column an oxidation state, and each entry is in the format (number of low-spin, number of intermediate-spin, number of high-spin), where a "-" denotes that no TMCs occurred with the corresponding metal and oxidation state, and shorter entries denote that the metal and oxidation were not run in all three spin states, as specified in Table S2. The only data contained in this table are structures which have converged and are not outliers as judged on properties evaluated from the PBE0/def2-TZVP calculations for all available spin states.

| Oxid. State | 0 | 1 | 2 | 3 | 4 | 5 | 6 | 7 | Total | Percent |
|---|---|---|---|---|---|---|---|---|---|---|
| Sc | - | - | (3) | (298) | - | - | - | - | (301) | (100) |
| Ti | (12, 1, 0) | (1, 0) | (65, 34) | (350) | (2263) | - | - | - | (2691, 35, 0) | (98.7, 1.3, 0) |
| V | (9, 1, 0) | (37, 20, 0) | (33, 95) | (35, 433) | (1194) | (1324) | - | - | (2632, 549, 0) | (82.7, 17.3, 0) |
| Cr | (620, 3, 0) | (34, 3, 1) | (34, 47, 196) | (20, 951) | (12, 32) | (67) | (37) | - | (824, 1036, 197) | (40.1, 50.4, 9.6) |
| Mn | (3, 0, 0) | (505, 0, 2) | (40, 11, 2633) | (6, 116, 1038) | (4, 98) | (34, 1) | (2) | - | (594, 226, 3673) | (13.2, 5.0, 81.7) |
| Fe | (243, 17, 2) | (54, 13, 3) | (2684, 165, 3687) | (1131, 211, 2316) | (27, 46, 5) | (2, 0) | - | - | (4141, 452, 6013) | (39.0, 4.3, 56.7) |
| Co | (21, 2) | (312, 103, 2) | (960, 5803, 23) | (4960, 138, 28) | (11, 2, 0) | (2, 0, 0) | - | - | (6266, 6048, 53) | (50.7, 48.9, 0.4) |
| Ni | (367, 5) | (174, 9) | (6835, 7950, 9) | (736, 7, 0) | (39, 1, 0) | - | - | - | (8151, 7972, 9) | (50.5, 49.4, 0.1) |
| Cu | (2) | (4610, 14) | (19211, 134) | (166, 7, 0) | - | - | - | - | (23989, 155, 0) | (99.4, 0.6, 0) |
| Zn | - | (5) | (11173, 42) | - | - | - | - | - | (11178, 42) | (99.6, 0.4) |
| Y | - | - | (5) | (569) | (1) | - | - | - | (575) | (100) |
| Zr | (3, 0) | - | (48, 0) | (8) | (838) | - | - | - | (897, 0) | (100, 0) |
| Nb | (2, 0) | (25, 0) | (1, 3) | (29, 2) | (62) | (304) | - | - | (423, 5) | (98.8, 1.2) |
| Mo | (365, 2) | (14, 0) | (291, 21) | (42, 94) | (489, 82) | (279) | (1585) | - | (3065, 199) | (93.9, 6.1) |
| Tc | - | (115, 0) | (21, 0) | (48, 48) | (4, 15) | (169, 1) | (9) | (10) | (376, 64) | (85.5, 14.5) |
| Ru | (101, 2) | (14, 2) | (7377, 31) | (704, 4) | (184, 29) | (3, 0) | (28, 2) | - | (8411, 70) | (99.2, 0.8) |
| Rh | (1, 0) | (1628, 2) | (87, 0) | (1614, 3) | (2, 0) | (5, 0) | - | - | (3337, 5) | (99.9, 0.1) |
| Pd | (204, 1) | (15, 0) | (12655, 21) | (35, 0) | (83, 1) | - | - | - | (12992, 23) | (99.8, 0.2) |
| Ag | - | (2670) | (40, 0) | (56, 0) | - | - | - | - | (2766, 0) | (100, 0) |
| Cd | - | (2) | (2924, 19) | - | - | - | - | - | (2926, 19) | (99.4, 0.6) |
| Hf | (2, 0) | - | (1, 0) | (3) | (307) | - | - | - | (313, 0) | (100, 0) |
| Ta | (1, 0) | (13, 0) | (3, 0) | (16, 4) | (33) | (331) | - | - | (397, 4) | (99.0, 1.0) |
| W | (312, 0) | (3, 0) | (336, 7) | (8, 14) | (259, 27) | (95) | (577) | - | (1590, 48) | (97.1, 2.9) |
| Re | (1, 0) | (1168, 1) | (72, 6) | (172, 138) | (11, 134) | (1298, 20) | (45) | (143) | (2910, 299) | (90.7, 9.3) |
| Os | (6, 0) | (1, 0) | (522, 4) | (145, 3) | (122, 115) | (10, 7) | (154, 1) | - | (960, 130) | (88.1, 11.9) |
| Ir | - | (522, 0) | (24, 0) | (2505, 5) | (37, 0) | (8, 2) | - | - | (3096, 7) | (99.8, 0.2) |
| Pt | (140, 0) | (9, 0) | (9049, 19) | (75, 0) | (1272, 6) | - | (2, 0) | - | (10547, 25) | (99.8, 0.2) |
| Au | (1) | (3364, 1) | (14, 0) | (1930, 1) | (2, 0) | (1, 0) | - | - | (5312, 2) | (100, 0) |
| Hg | - | (3) | (1726) | - | - | - | - | - | (1729) | (100) |
| Total | | | | | | | | | (123389, 17415, 9945) | (81.9, 11.6, 6.6) |

**Table S25.** Summary of calculations for which all available spin states converge and PBE0/def2-TZVP properties are not outliers for any property, computed for each spin state type and number in which the ground state was IS or HS.

|  | HS | IS | non-LS total |
|---|---|---|---|
| Number GS | 9945 | 17415 | 27360 |
| Total compatible TMCs | 44070 | 135598 | 135598 |
| Percent HS/IS GS | 22% | 13% | 20% |

**Table S26.** Statistics on the VSSEs of TMCs reported in BOS-TMC, reported in kcal/mol.

| VSSE Type | Mean | Median | Max | Min | St. Dev. |
|---|---|---|---|---|---|
| IS-LS | 43.56 | 53.34 | 238.54 | -134.33 | 45.30 |
| HS-LS | 37.55 | 45.43 | 235.30 | -160.41 | 59.12 |
| HS-IS | 40.04 | 47.80 | 186.44 | -112.35 | 51.16 |

**Table S27.** Summary of the geometry types of TMCs attempted and converged.

| Geometry type | Attempted | All available spin states converged and not spin contaminated | LS and HS states converge |
|---|---|---|---|
| octahedral | 54,761 | 50283 | 24,807 |
| square planar | 43,695 | 41684 | 8,235 |
| tetrahedral | 23,369 | 21619 | 4,877 |
| trigonal bipyramidal | 10,786 | 10164 | 2,607 |
| square pyramidal | 10,649 | 9959 | 2,034 |
| linear | 6,719 | 6478 | 732 |
| trigonal planar | 4,309 | 4063 | 659 |
| seesaw | 2,160 | 2034 | 402 |
| pentagonal bipyramidal | 1,331 | 1244 | 297 |
| trigonal prismatic | 1,215 | 1104 | 410 |
| T shape | 862 | 814 | 54 |
| bent | 653 | 624 | 20 |
| square antiprismatic | 631 | 565 | 12 |
| unknown | 485 | 469 | 8 |
| trigonal pyramidal | 309 | 295 | 32 |
| tricapped trigonal prismatic | 114 | 99 | 0 |
| pentagonal pyramidal | 49 | 46 | 14 |
| pentagonal planar | 12 | 10 | 3 |
| % non-octahedral | 66.2% | 66.8% | 45.1% |
| total | 162,109 | 151,554 | 45,203 |

**Table S28.** Details on structures in BOS-TMC that have the lowest/highest vertical spin-splitting energies. VSSEs are reported in kcal/mol. If no number is available, that is indicated as '--'.

| Refcode | Metal | Oxidation State | q | Formula | Coordinating atoms | Geometry | IS-LS VSSE | HS-LS VSSE | HS-IS VSSE |
|---|---|---|---|---|---|---|---|---|---|
| QEVZEX | Ir | 3 | 1 | Ir C40 H43 Cl N2 | C, C, C, C, C, Cl, N, N (haptic) | tetrahedral | 238.54 | -- | -- |
| DEKRET | Cu | 3 | -1 | Cu C4 H3 F9 | C, C, C, C | square planar | 87.49 | 235.30 | 147.82 |
| JADGOK | Ni | 3 | 1 | Ni C16 H36 Cl2 N4 | Cl, Cl, N, N, N, N | octahedral | 30.85 | 217.29 | 186.44 |
| ENUVUF | Pd | 2 | 0 | Pd C22 H26 Cl2 N6 | C, C, Cl, Cl | square planar | -134.33 | -- | -- |
| KANLIV | Mn | 3 | -2 | Mn Cl5 | Cl, Cl, Cl, Cl, Cl | trigonal bipyramidal | -109.20 | -160.41 | -51.21 |
| QOFYIU | Mn | 2 | 0 | Mn C36 H42 O6 S16 | O, O, O, O, O, O | octahedral | 48.85 | -63.51 | -112.35 |

**Table S29.** Statistics on the VSSEs of TMCs reported in BOS-TMC for which energies in all three spin states are available, reported in kcal/mol.

| VSSE Type | Mean | Median | Max | Min | St. Dev. |
|---|---|---|---|---|---|
| IS-LS | -1.94 | -7.70 | 106.11 | -109.20 | 35.04 |
| HS-LS | 38.10 | 45.81 | 235.30 | -160.41 | 58.92 |
| HS-IS | 40.04 | 47.80 | 186.44 | -112.35 | 51.16 |

**Table S30.** Number of TMCs (with percentages in parentheses) in BOS-TMC, where calculations for all available spin states converged, that also appear in tmQMg, sorted by metal and ground state spin, as determined by PBE0/def2-TZVP VSSE.[a]

| Metal | Singlet | Triplet | Quintet |
|---|---|---|---|
| Sc | 36 (100%) | 0 (0%) | 0 (0%) |
| Ti | 762 (99.7%) | 2 (0.3%) | 0 (0%) |
| V | 670 (93.1%) | 50 (6.9%) | 0 (0%) |
| Cr | 397 (94.1%) | 14 (3.3%) | 11 (2.6%) |
| Mn | 202 (70.9%) | 1 (0.4%) | 82 (28.8%) |
| Fe | 412 (35.8%) | 21 (1.8%) | 718 (62.4%) |
| Co | 1300 (96.6%) | 40 (3.0%) | 6 (0.4%) |
| Ni | 2277 (66.6%) | 1141 (33.4%) | 0 (0%) |
| Cu | 1101 (97.3%) | 31 (2.7%) | 0 (0%) |
| Zn | 3627 (99.9%) | 3 (0.1%) | 0 (0%) |
| Y | 91 (100%) | 0 (0%) | 0 (0%) |
| Zr | 238 (100%) | 0 (0%) | 0 (0%) |
| Nb | 85 (100%) | 0 (0%) | 0 (0%) |
| Mo | 1016 (99.1%) | 9 (0.9%) | 0 (0%) |
| Tc | 112 (99.1%) | 1 (0.9%) | 0 (0%) |
| Ru | 2046 (99.7%) | 7 (0.3%) | 0 (0%) |
| Rh | 963 (99.9%) | 1 (0.1%) | 0 (0%) |
| Pd | 5034 (99.9%) | 3 (0.1%) | 0 (0%) |
| Ag | 540 (100%) | 0 (0%) | 0 (0%) |
| Cd | 946 (99.9%) | 1 (0.1%) | 0 (0%) |
| Hf | 49 (100%) | 0 (0%) | 0 (0%) |
| Ta | 107 (99.1%) | 1 (0.9%) | 0 (0%) |
| W | 495 (99.0%) | 5 (1.0%) | 0 (0%) |
| Re | 1172 (99.0%) | 12 (1.0%) | 0 (0%) |
| Os | 203 (96.7%) | 7 (3.3%) | 0 (0%) |
| Ir | 795 (99.9%) | 1 (0.1%) | 0 (0%) |
| Pt | 3796 (99.9%) | 2 (0.1%) | 0 (0%) |
| Au | 1707 (100%) | 0 (0%) | 0 (0%) |
| Hg | 672 (100%) | 0 (0%) | 0 (0%) |
| Total | 30851 (93.4%) | 1353 (4.1%) | 817 (2.5%) |

[a]This set of 33,021 TMCs does not include any structures in tmQMg that were modeled as doublets in BOS-TMC, since those require a reassigned spin state, and also does not include any structures which failed to converge for a PBE0/def2-SV(P) or PBE0/def2-TZVP calculation in any available spin state, or were considered an outlier for PBE0/def2-TZVP. Table S11 describes the overlap between tmQMg and BOS-TMC in more detail.

**Table S31.** Statistics on the absolute difference in the dipole moment for structures with an IS or HS ground state compared to their LS values. All values in Debye. All structures with a ground state spin dipole moment less than 0.1 D were excluded from this analysis.

| Ground State | Mean | Median | Max | Min | St. Dev. | Signed Mean |
|---|---|---|---|---|---|---|
| IS | 0.398 | 0.258 | 23.34 | 0 | 0.592 | 0.122 |
| HS | 0.416 | 0.299 | 8.418 | 0 | 0.469 | 0.019 |

**Table S32.** Statistics on the absolute difference in the metal Löwdin charge for structures with an IS or HS ground state compared to their LS values. All values in a.u.

| Ground State | Mean | Median | Max | Min | St. Dev. | Signed Mean |
|---|---|---|---|---|---|---|
| IS | 0.096 | 0.100 | 0.380 | $2 \cdot 10^{-5}$ | 0.041 | 0.094 |
| HS | 0.184 | 0.185 | 0.554 | $3.9 \cdot 10^{-4}$ | 0.049 | 0.184 |

**Table S33.** Examples of structures with either a large dipole moment magnitude shift (top four) or negligible shift (bottom four) between different spin states, and do not have a low-spin ground state. For each spin transition (i.e., odd or even multiplicity, low- to intermediate- or high-spin), an example structure is given with a small shift and a large shift. All structures with a dipole moment less than 0.1 D in their ground state spin are excluded from this analysis.

| Absolute change in dipole magnitude (Debye) | Refcode | Metal | Oxidation State | Charge | Formula | Coordinating atoms | Geometry | Spin States |
|---|---|---|---|---|---|---|---|---|
| 23.34 | SUWVAH | Co | 3 | 3 | Co C28 H45 N7 | N, N, N, N, N, N | octahedral | Singlet-Triplet |
| 9.89 | VUSXEP | Co | 2 | 0 | Co C24 H22 F12 O6 | O, O, O, O, O, O | octahedral | Doublet-Quartet |
| 8.42 | BOGNEU01 | Co | 3 | 1 | Co C30 H44 N4 O2 | N, N, N, N, O, O | octahedral | Singlet-Quintet |
| 6.08 | ZOCRIS | Fe | 3 | 1 | Fe C22 H24 N4 O2 | N, N, N, N, O, O | octahedral | Doublet-Sextet |
| 0 | UYUKIL | Co | 1 | -1 | Co C40 H56 N2 Si2 | N, N | linear | Singlet-Triplet |
| 0 | MUSXON | Co | 2 | 0 | Co C43 H51 Cl2 N P2 | Cl, Cl, N, P, P | trigonal bipyramidal | Doublet-Quartet |
| 0 | HUSXUP | Cr | 2 | -2 | Cr C24 H40 S4 Si4 | S, S, S, S | square planar | Singlet-Quintet |
| 0 | PATWUD | Mn | 2 | 0 | Mn C28 H24 N6 O2 | N, N, N, N, O, O | octahedral | Doublet-Sextet |

**Table S34.** Examples of structures whose Löwdin partial charges on the metal center shifts between different spin states, and do not have a low-spin ground state (top four: large shift, bottom four: small shift). For each spin transition (i.e., odd or even multiplicity, low to intermediate or high spin), an example structure is given with a small shift and a large shift.

| Absolute change in Löwdin metal charge (a.u.) | Refcode | Metal | Oxidation State | Charge | Formula | Coordinating atoms | Geometry | Spin States |
|---|---|---|---|---|---|---|---|---|
| 0.33 | QIBPIA | Ni | 2 | -2 | Ni H8 Cl2 O4 | Cl, Cl, O, O, O, O | octahedral | Singlet-Triplet |
| 0.38 | GINLUN | Fe | 3 | 1 | Fe C44 H28 Cl4 N4 O2 | N, N, N, N, O, O | octahedral | Doublet-Quartet |
| 0.44 | FIBLEI | Ni | 2 | 0 | Ni C30 H33 N9 | N, N, N, N, N, N | octahedral | Singlet-Quintet |
| 0.55 | VOFMAE | Fe | 3 | 1 | Fe C64 H64 N8 | N, N, N, N, N, N | octahedral | Doublet-Sextet |
| $4.0 \cdot 10^{-5}$ | DULQUX | Zn | 2 | 0 | Zn C24 H21 N9 | N, N, N, N, N, N | octahedral | Singlet-Triplet |
| $2.0 \cdot 10^{-5}$ | ILUMEH | Cr | 3 | 0 | Cr C12 H30 P3 | C, C, C, C, C, C | octahedral | Doublet-Quartet |
| $3.9 \cdot 10^{-4}$ | SARSAG | Cr | 2 | -3 | Cr C5 N5 | C, C, C, C, C | square pyramidal | Singlet-Quintet |
| $1.1 \cdot 10^{-3}$ | BILBOQ | Co | 2 | 0 | Co C40 H48 N4 O4 | N, N, O, O, O, O | octahedral | Doublet-Sextet |

**Table S35.** Statistics on the absolute difference in the HOMO, LUMO, and HOMO-LUMO gap for structures with an IS or HS ground state compared to their LS values. All values in eV.

| Property | Ground State | Mean | Median | Max | Min | St. Dev. | Signed Mean |
|---|---|---|---|---|---|---|---|
| HOMO | IS | 0.604 | 0.460 | 2.503 | $2.72 \cdot 10^{-5}$ | 0.539 | -0.546 |
|  | HS | 0.379 | 0.283 | 1.776 | $2.72 \cdot 10^{-5}$ | 0.330 | -0.005 |
| LUMO | IS | 0.728 | 0.596 | 4.545 | 0.0 | 0.615 | 0.690 |
|  | HS | 0.362 | 0.218 | 2.451 | $2.72 \cdot 10^{-5}$ | 0.414 | 0.195 |
| Gap | IS | 1.277 | 1.060 | 5.004 | $5.44 \cdot 10^{-5}$ | 0.985 | 1.236 |
|  | HS | 0.653 | 0.479 | 4.008 | $8.88 \cdot 10^{-16}$ | 0.616 | 0.200 |

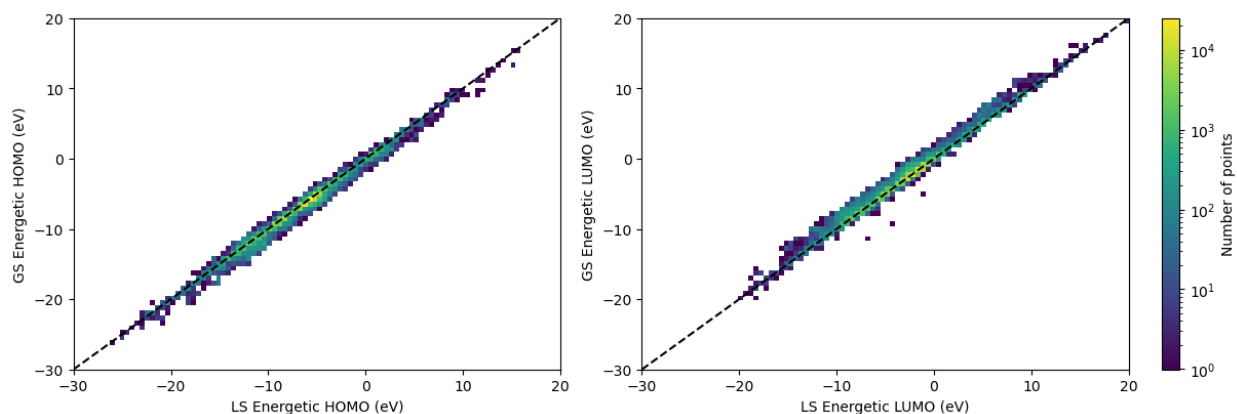

**Figure S9.** Ground state HOMO (left) and LUMO (right) against the low-spin (LS) frontier orbital energies, all in eV.

**Table S36.** Examples of structures whose HOMO, LUMO, or HOMO-LUMO gap shifted either by a large amount (top two in each grouping) or very little (bottom two in each grouping) upon a change in spin state, and do not have a low-spin ground state. For each property, an example is given for both a large and small shift for both low-spin to intermediate-spin and low-spin to high-spin transitions.

| Property | Absolute change in property (eV) | Refcode | Metal | Oxidation State | Charge | Formula | Coordinating atoms | Geometry | Spin States |
|---|---|---|---|---|---|---|---|---|---|
| HOMO | 2.50 | XENBIE | Ni | 2 | 2 | Ni C12 H18 N6 | N, N, N, N, N, N | octahedral | Singlet-Triplet |
| HOMO | 1.78 | IBIXIA01 | Mn | 2 | -2 | Mn Cl4 | Cl, Cl, Cl, Cl | tetrahedral | Doublet-Sextet |
| HOMO | $2.72 \cdot 10^{-5}$ | HUMPAI | Co | 2 | 1 | Co C26 H36 N7 O3 | N, N, N, N, O, O | octahedral | Doublet-Quartet |
| HOMO | $2.72 \cdot 10^{-5}$ | CARQUH01 | Fe | 2 | 0 | Fe C12 H16 N2 O8 | N, N, O, O, O, O | octahedral | Singlet-Quintet |
| LUMO | 4.55 | VACJUF | Rh | 1 | 1 | Rh C20 H30 | C, C, C, C, C, C (haptic) | square planar | Singlet-Triplet |
| LUMO | 2.45 | NOFKUO | Cr | 2 | 2 | Cr C8 H12 N4 | N, N, N, N | square planar | Singlet-Quintet |
| LUMO | 0 | SUYHUR | Ni | 2 | 0 | Ni C34 H26 N4 O4 | N, N, N, N, O, O | octahedral | Singlet-Triplet |
| LUMO | $2.72 \cdot 10^{-5}$ | FUJXEQ | Mn | 2 | 0 | Mn C56 H52 N2 O4 P4 S2 | N, N, O, O, O, O | octahedral | Doublet-Sextet |
| Gap | 5.00 | XENBIE | Ni | 2 | 2 | Ni C12 H18 N6 | N, N, N, N, N, N | octahedral | Singlet-Triplet |
| Gap | 4.01 | JUCJOJ | Mn | 2 | -2 | Mn Cl4 | Cl, Cl, Cl, Cl | tetrahedral | Doublet-Sextet |
| Gap | $5.44 \cdot 10^{-5}$ | XALLAY | Co | 2 | 0 | Co C30 H20 N6 O2 S2 | N, N, N, N | seesaw | Doublet-Quartet |
| Gap | $8.88 \cdot 10^{-16}$ | COZVIX | Fe | 2 | 0 | Fe C38 H36 N14 O2 S2 | N, N, N, N, O, O | octahedral | Singlet-Quintet |

**Table S37.** Statistics of the HOMO-LUMO gap (eV), HOMO (eV), LUMO (eV), dipole moment magnitude (Debye), and Löwdin metal partial charge (a.u.) for all TMCs in their LS, IS, and HS states as well as the full dataset all together and those IS states that correspond to ground states or HS states that correspond to ground states. GESD outliers of individual properties and spin-contaminated cases were removed prior to analysis.

|  | mean | std | min | max | median |
|---|---|---|---|---|---|
| | | | HOMO-LUMO gap | | |
| All LS | 3.86 | 1.14 | 0.01 | 9.93 | 3.83 |
| All IS | 0.21 | 2.43 | -7.53 | 7.74 | -0.48 |
| All HS | 0.28 | 2.56 | -6.50 | 7.81 | -0.49 |
| All | 1.87 | 2.68 | -7.53 | 9.93 | 2.74 |
| IS GS | 4.08 | 1.19 | -0.31 | 7.74 | 4.11 |
| HS GS | 3.64 | 1.10 | 0.17 | 7.81 | 3.52 |
| | | | HOMO | | |
| All LS | -6.71 | 3.21 | -26.20 | 15.53 | -6.04 |
| All IS | -4.89 | 3.59 | -23.70 | 18.23 | -4.32 |
| All HS | -5.24 | 4.03 | -24.34 | 16.52 | -4.68 |
| All | -5.76 | 3.60 | -26.20 | 18.23 | -5.57 |
| IS GS | -7.32 | 3.79 | -23.09 | 13.50 | -6.13 |
| HS GS | -6.51 | 3.45 | -23.84 | 11.61 | -5.71 |
| | | | LUMO | | |
| All LS | -2.85 | 3.21 | -21.19 | 19.71 | -2.37 |
| All IS | -4.68 | 3.35 | -22.34 | 17.49 | -4.43 |
| All HS | -4.96 | 3.91 | -21.52 | 16.45 | -4.50 |
| All | -3.89 | 3.50 | -22.34 | 19.71 | -3.64 |
| IS GS | -3.24 | 3.45 | -18.48 | 17.49 | -2.45 |
| HS GS | -2.88 | 3.53 | -19.46 | 16.45 | -2.61 |
| | | | Dipole (debye) | | |
| All LS | 6.28 | 5.53 | 0.00 | 36.80 | 5.46 |
| All IS | 5.61 | 4.82 | 0.00 | 30.70 | 4.76 |
| All HS | 5.04 | 4.76 | 0.00 | 29.25 | 3.86 |
| All | 5.84 | 5.17 | 0.00 | 36.80 | 4.90 |
| IS GS | 5.64 | 5.89 | 0.00 | 30.70 | 4.01 |
| HS GS | 6.59 | 6.05 | 0.00 | 29.25 | 5.70 |
| | | | Lowdin metal partial charge (a.u.) | | |
| All LS | -0.15 | 0.39 | -1.97 | 1.18 | -0.09 |
| All IS | -0.12 | 0.40 | -2.03 | 1.12 | -0.04 |
| All HS | -0.11 | 0.36 | -1.73 | 1.03 | -0.01 |
| All | -0.14 | 0.39 | -2.03 | 1.18 | -0.06 |
| IS GS | -0.13 | 0.29 | -1.62 | 0.87 | -0.03 |
| HS GS | 0.02 | 0.30 | -1.23 | 0.65 | 0.12 |

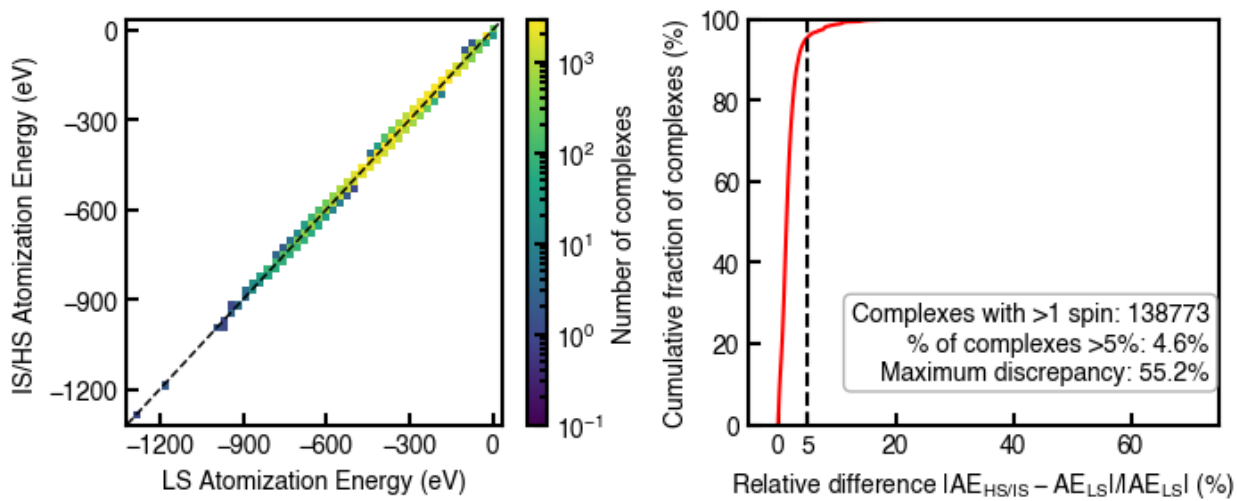

**Figure S10**. Spin state sensitivity of atomization energies. For each complex with at least two computed spin multiplicities, we report the difference in atomization energy between the LS and the highest available spin (IS or HS). The left panel shows a parity plot of LS versus IS/HS atomization energy and a density-colored scatter indicating how complexes cluster around and deviate from the perfect prediction y=x. The right panel shows the empirical cumulative distribution of the relative discrepancy ΔAE%, highlighting the threshold for large deviations for complexes with ΔAE% > 5%. An annotation reports the number of complexes with more than one computed spin state, the fraction exceeding 5%, and the maximum observed discrepancy.

**Table S38**. Largest HS/IS-LS atomization energy differences in eV. For each complex with at least two computed spin multiplicities, we compute the absolute difference and the relative change for the highest available spin state (IS or HS) in % w.r.t. the low-spin state. The table lists the four complexes with the largest relative change, along with the corresponding metal, charge, and the set of available spin multiplicities.

| Refcode | Metal | Charge | Spin mult. (2S+1) | AE(LS) | AE(IS/HS) | \|ΔAE\| | ΔAE (IS/HS–LS) % | Molecular formula | Symmetry | Coordinating atoms |
|---|---|---|---|---|---|---|---|---|---|---|
| NERWEM | Cu | -3 | 1,3 | -2.10 | -0.94 | 1.16 | 55.23 | $CuBr_4$ | tetrahedral | Br |
| HOXNAK | Cu | -3 | 1,3 | -2.36 | -1.16 | 1.20 | 50.74 | $CuI_4$ | tetrahedral | I |
| BINMUJ | Cu | -3 | 1,3 | -2.37 | -1.20 | 1.17 | 49.48 | $CuI_4$ | tetrahedral | I |
| JETRIN | Cu | -3 | 1,3 | -1.96 | -1.03 | 0.94 | 47.61 | $CuCl_4$ | tetrahedral | Cl |

**Table S39.** Metals with the most frequent large HS/IS-LS atomization energy differences in %. For each complex with at least two computed spin multiplicities, we compute the relative change of the atomization energy for the highest available spin state (IS or HS) in % w.r.t. the low-spin state. For each metal, we count the number of complexes for which ΔAE% > 5% and report the mean, median, and maximum value. The table lists the 10 metals with the largest counts within ΔAE% > 5%.

| Metal | Number of TMCs: ΔAE > 5% | Mean ΔAE% | Median ΔAE% | Max. ΔAE% |
|---|---|---|---|---|
| Cr | 257 | 12.76 | 0.84 | 15.73 |
| Mn | 336 | 6.86 | 1.09 | 20.90 |
| Fe | 760 | 6.59 | 1.70 | 17.05 |
| Co | 675 | 4.97 | 1.02 | 24.56 |
| Cu | 1226 | 5.08 | 1.77 | 55.23 |
| Zn | 1144 | 10.20 | 2.37 | 41.79 |
| Pd | 276 | 2.12 | 1.63 | 14.74 |
| Cd | 413 | 14.03 | 2.43 | 45.71 |
| Pt | 261 | 2.47 | 1.48 | 19.24 |
| Au | 608 | 11.44 | 0.86 | 29.34 |

**Table S40.** Relative atomization energy (relAE) w.r.t. molecular weight mean, median, and standard deviation for low-spin states (singlets and doublets) in eV mol g$^{-1}$ for BOS-TMC. The statistics are compared to the QM9 dataset.

| Statistic | TMC low-spin (eV mol g$^{-1}$) | QM9 (eV mol g$^{-1}$) |
|---|---|---|
| Count | 153781 | 130831 |
| Mean | -0.555 | -0.620 |
| Median | -0.569 | -0.622 |
| Standard deviation | 0.144 | 0.077 |
| Min | -1.286 | -1.070 |
| Max | -0.003 | -0.236 |

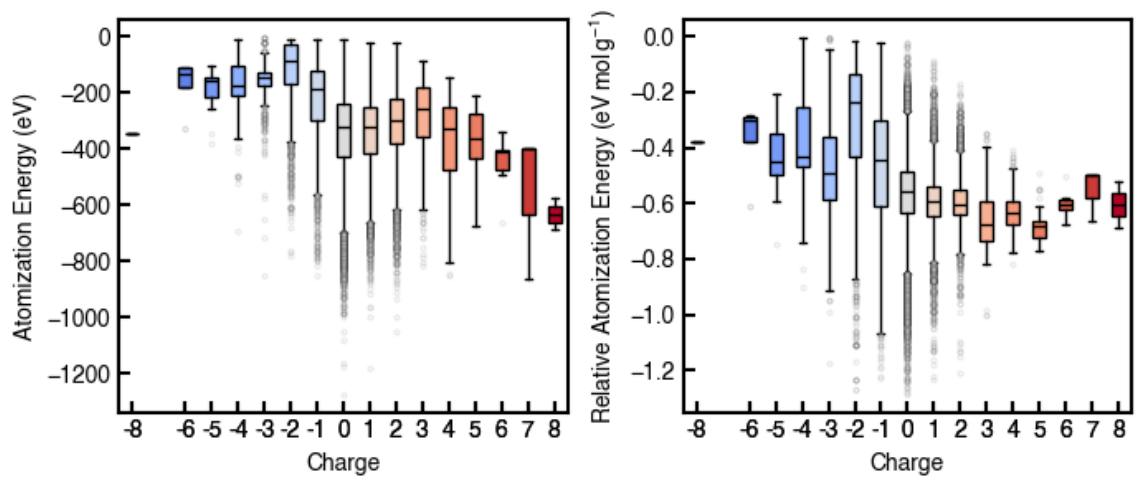

**Figure S11.** Atomization energy distributions by complex charge. The box plots summarize the distributions of (left) atomization energy in eV and (right) relative atomization energy relAE in eV mol g$^{-1}$, grouped by charge. Each box corresponds to all LS complexes at a given charge. Colors follow a cool-warm gradient, with more negative charges in blue and more positive charges in red.

**Table S41.** Comparison of $|q| > 1$ subset of 26,219 complexes that have singlet closed-shell relative atomization energies (relAEs) to a similarly sized subset and full set of the more neutral complexes. The data is given in eV.

|  | count | mean | std | min | max | median |
|---|---|---|---|---|---|---|
| \|q\|>1 | 26219 | -0.519 | 0.183 | -1.269 | -0.003 | -0.579 |
| \|q\|≤1 matched (size=\|q\|>1) | 26219 | -0.563 | 0.133 | -1.257 | -0.023 | -0.566 |
| \|q\|≤1 remainder | 101583 | -0.563 | 0.133 | -1.286 | -0.021 | -0.566 |

**Table S42.** Complexes with the least and most negative relative atomization energy (relAE) w.r.t. molecular weight in eV mol g$^{-1}$ for low-spin states. The table lists the five complexes for each extreme along with the corresponding refcode, metal, and charge. For the most negative relAE subset, LOBVAB and LOBTON correspond to identical structures.

| Group | Rank | Refcode | Charge | Metal | mw | relAE | Molecular formula | Symmetry | Coordinating atoms |
|---|---|---|---|---|---|---|---|---|---|
| Least negative relAE | 1 | BINMOD | -3 | Ag | 615.52 | -0.003 | $AgI_4$ | tetrahedral | I |
|  | 2 | HOXNAK | -3 | Cu | 571.16 | -0.004 | $CuI_4$ | tetrahedral | I |
|  | 3 | BINMUJ | -3 | Cu | 571.16 | -0.004 | $CuI_4$ | tetrahedral | I |
|  | 4 | ROKXUN | -4 | Ag | 1332.12 | -0.005 | $AgBiI_8$ | T shape | I |
|  | 5 | NERWEM | -3 | Cu | 383.16 | -0.005 | $CuBr_4$ | tetrahedral | Br |
| Most negative relAE | 1 | IGODIR | 0 | Re | 263.26 | -1.286 | $ReC_2H_5O_3$ | tetrahedral | C,O |
|  | 2 | SOWSON | 0 | Mo | 224.09 | -1284 | $MoC_6H_8O_3$ | tetrahedral | C,O |
|  | 3 | OCRPOL | 0 | Cr | 312.20 | -1.274 | $CrC_{12}H_8N_2O_5$ | trigonal bipyramidal | N,O |
|  | 4 / 5 | LOBTON (LOBVAB) | -2 | Mo | 203.97 | -1.269 | $MoCO_6$ | trigonal bipyramidal | O |
|  | 6 | BIWJEX | 0 | Tc | 338.06 | -1.266 | $TcC_8H_7N_4O_5$ | octahedral | N,O |

**Table S43.** Metals with the largest counts of the least and most negative relative atomization energies (relAEs) w.r.t. molecular weight in eV mol g$^{-1}$ for low-spin states. For each metal, we count the number of complexes for which the relative atomization energy falls within the bottom 5% (most negative tail) or the top 5% (least negative tail) of the overall relAE distribution. and the mean, median, and minimum/maximum value. The table lists 5 metals with the largest counts for each tail of the distribution in the periodic table order.

| Group | Metal | Number of TMCs | Mean relAE | Median relAE | Min. relAE | Max. relAE |
|---|---|---|---|---|---|---|
| Bottom 5% (most negative) | Ti | 1224 | -0.82 | -0.81 | -1.07 | -0.76 |
|  | V | 2005 | -0.89 | -0.88 | -1.24 | -0.76 |
|  | Fe | 337 | -0.80 | -0.78 | -1.05 | -0.76 |
|  | Co | 589 | -0.79 | -0.78 | -1.02 | -0.76 |
|  | Mo | 1571 | -0.93 | -0.93 | -1.28 | -0.76 |

| Top 5% (most positive) | Cu | 1243 | -0.14 | -0.14 | -0.30 | 0.00 |
| --- | --- | --- | --- | --- | --- | --- |
| | Zn | 966 | -0.15 | -0.14 | -0.30 | -0.04 |
| | Pt | 1098 | -0.24 | -0.26 | -0.30 | -0.05 |
| | Au | 1022 | -0.19 | -0.19 | -0.30 | -0.02 |
| | Hg | 517 | -0.19 | -0.22 | -0.30 | -0.04 |

**Table S44.** Summary of the number of structures converged in the manyDFA set.[a]

| | Low-spin | Intermediate-spin | High-spin |
| --- | --- | --- | --- |
| Initial set | 12,015 | 10,608 | 3,364 |
| PBE0/def2-SV(P) | 11,921 | 10,518 | 3,288 |
| PBE0/def2-TZVP | 11,901 | 10,498 | 3,280 |
| >= 6 DFAs converged | 11,883 | 10,490 | 3,277 |
| >= 6 DFAs not spin contaminated | 11,706 | 10,470 | 3,277 |
| >= 9 DFAs converged | 11,883 | 10,490 | 3,277 |
| >= 9 DFAs not spin contaminated | 11,660 | 10,458 | 3,276 |
| All 12 DFAs converged | 11,100 | 9,871 | 3,151 |
| 12 DFAs not spin contaminated | 10,780 | 9,797 | 3,143 |

[a]For spin-state pairs, we 10,290 IS-LS, 3,151 HS-LS, and 3,250 HS-IS where ≥6 DFAs converge and are not spin contaminated for both spin states; 10,236 IS-LS, 3,111 HS-LS, and 3,235 HS-IS for ≥9 DFAs; and 8,781 IS-LS, 2,318 HS-LS, and 3,039 HS-IS for all 12 DFAs.

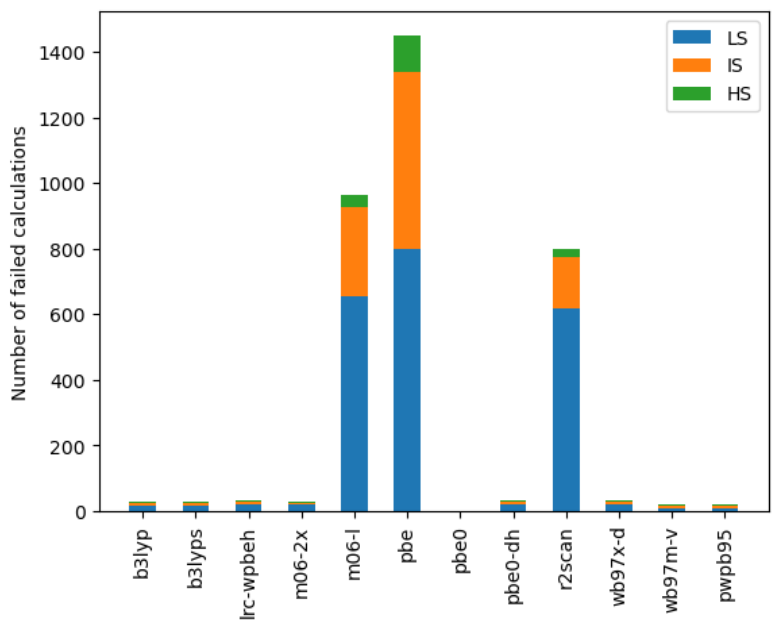

**Figure S12.** Summary of the number of failed calculations by DFA in the manyDFA set.

**Table S45.** Examples of structures with very large or small standard deviations among 12 functionals for the vertical spin-splitting energy. For each property, two examples are given for both a large and small shift for both low-spin to intermediate-spin and low-spin to high-spin transitions. We required each structure to be converged, not spin-contaminated, not an outlier for any property, and succeed for both relevant spin states across all functionals.

| Refcode | Metal | Oxidation State | Charge | Molecular Formula | Geometry | Coordinating Atoms | Spin States | Standard deviation in VSSE (kcal/mol) |
|---|---|---|---|---|---|---|---|---|
| HUQFAA | Fe | 2 | 2 | Fe C14 H25 N5 O | octahedral | C, N, N, N, N, N | Singlet-Triplet | 24.98 |
| KADVUK | Cu | 2 | 2 | Cu C8 H24 N4 O6 | octahedral | O, O, O, O, O, O | Doublet-Quartet | 22.76 |
| FEWRII | Ru | 2 | -1 | Ru B10 C12 H31 | linear | C, C, C, C, C, B, B, B, B (haptic) | Singlet-Triplet | 0.92 |
| GIRLAX | Pd | 2 | 1 | Pd C13 H20 N O | square planar | C, C, C, C, N, O (haptic) | Singlet-Triplet | 0.97 |
| XUBKAI | Ni | 2 | -2 | Ni C14 H14 N4 O4 | square planar | N, N, N, N | Singlet-Quintet | 39.62 |
| YIKCUR | Fe | 2 | 0 | Fe C6 H18 N2 O8 P2 | tetrahedral | N, N, P, P | Doublet-Sextet | 28.21 |
| OSIQIR | Co | 2 | 0 | Co C12 Cl2 H16 N4 | tetrahedral | Cl, Cl, N, N | Doublet-Sextet | 2.42 |
| SALNAY | Co | 2 | 0 | Co C14 H22 N4 | tetrahedral | N, N, N, N | Doublet-Sextet | 2.69 |

**Table S46.** Summary of the number of structures in the manyDFA set found to have different ground state spins depending on the functional chosen.

| | Number of structures |
|---|---|
| Run in at least two spin states | 10,498 |
| Converged and not spin contaminated for at least two functionals in all available spin states | 10,371 |
| Ground state spin changes based on functional | 141 |

**Table S47.** Summary statistics of the standard deviation among properties when grouped by DFA categories. The categories are: semilocal (PBE, r$^2$SCAN, M06-L), moderate HFX (B3LYP*, B3LYP, PBE0), high HFX (M06-2X, PBE0-DH, PWPB95), and range-separated hybrid (LRC-ωPBEh, ωB97X-D, ωB97M-V). We required each structure to be converged, not spin-contaminated, and not an outlier for any property across all functionals. For the dipole analysis, any structure with an average dipole moment across all 12 DFAs less than 0.1 D was omitted.

|  | mean | std | min | max | median |
|---|---|---|---|---|---|
| | Dipole (debye) | | | | |
| Semilocal | 0.148 | 0.129 | 0.006 | 6.014 | 0.126 |
| Moderate HFX | 0.077 | 0.084 | 1.1e-4 | 1.782 | 0.059 |
| High HFX | 0.062 | 0.082 | 1.7e-4 | 4.790 | 0.050 |
| RSH | 0.049 | 0.054 | 9.4e-5 | 2.446 | 0.041 |
| All | 0.291 | 0.271 | 0.001 | 3.918 | 0.219 |
| | Löwdin metal partial charge (a.u.) | | | | |
| Semilocal | 0.042 | 0.016 | 0.003 | 0.160 | 0.039 |
| Moderate HFX | 0.009 | 0.006 | 6.0e-5 | 0.096 | 0.008 |
| High HFX | 0.008 | 0.006 | 1.2e-4 | 0.166 | 0.007 |
| RSH | 0.012 | 0.006 | 1.3e-4 | 0.146 | 0.010 |
| All | 0.038 | 0.017 | 0.005 | 0.145 | 0.033 |
| | IS-LS VSSE (kcal/mol) | | | | |
| Semilocal | 1.82 | 1.55 | 0.01 | 10.51 | 1.26 |
| Moderate HFX | 1.68 | 1.81 | 2.4e-3 | 23.32 | 1.08 |
| High HFX | 3.67 | 3.03 | 0.04 | 29.28 | 2.55 |
| RSH | 1.83 | 1.70 | 0.06 | 42.37 | 1.57 |
| All | 4.97 | 3.20 | 0.92 | 24.98 | 4.20 |
| | HS-LS VSSE (kcal/mol) | | | | |
| Semilocal | 4.86 | 2.54 | 0.22 | 14.13 | 4.79 |
| Moderate HFX | 4.38 | 2.29 | 0.10 | 27.87 | 4.50 |
| High HFX | 11.02 | 6.33 | 0.56 | 50.76 | 10.54 |
| RSH | 2.87 | 1.88 | 0.19 | 20.80 | 2.61 |
| All | 9.28 | 3.13 | 2.42 | 39.62 | 9.34 |
| | Relative atomization energy (eV mol/g) | | | | |
| Semilocal | 7.2e-3 | 2.2e-3 | 5.2e-4 | 1.5e-2 | 7.2e-3 |
| Moderate HFX | 3.2e-3 | 9.9e-4 | 2.0e-4 | 9.9e-3 | 3.3e-3 |
| High HFX | 1.8e-3 | 1.1e-3 | 5.8e-6 | 1.2e-2 | 1.5e-3 |
| RSH | 2.0e-3 | 8.5e-4 | 6.4e-5 | 8.5e-3 | 1.9e-3 |
| All | 5.2e-3 | 1.5e-3 | 5.6e-4 | 1.2e-2 | 5.1e-3 |
| | Atomization energy (eV) | | | | |
| Semilocal | 2.43 | 0.69 | 0.12 | 4.93 | 2.46 |
| Moderate HFX | 1.10 | 0.37 | 0.06 | 3.19 | 1.08 |
| High HFX | 0.59 | 0.35 | 2.1e-3 | 2.91 | 0.52 |
| RSH | 0.69 | 0.31 | 0.02 | 2.50 | 0.65 |
| All | 1.74 | 0.46 | 0.19 | 3.45 | 1.76 |

**Table S48.** Structures with the highest standard deviation in IS-LS or HS-LS VSSEs when grouped by DFA categories. The categories are: semilocal (PBE, r$^2$SCAN, M06-L), moderate HFX (B3LYP*, B3LYP, PBE0), high HFX (M06-2X, PBE0-DH, PWPB95), and range-separated hybrid (LRC- ωPBEh, ωB97X-D, ωB97M-V).

| Refcode | Grouping | VSSE | St. Dev. | Metal | Ox. State | Charge | Formula | Geom. | Coord. Atoms | Spin States |
|---|---|---|---|---|---|---|---|---|---|---|
| MIGKET | Semilocal | IS-LS | 12.87 | Cu | 2 | 2 | Cu H16 N4 O2 | square planar | N, N, N, N | doublet-quartet |
| ZIYXIO01 | Moderate | IS-LS | 28.56 | Fe | 3 | 1 | Fe C20 H30 | linear | C, C, C, C, C, C, C, C, C, C (haptic) | doublet-quartet |
| URIXAX | High | IS-LS | 35.86 | Ni | 4 | 1 | Ni C9 H21 N6 O3 | octahedral | O, O, O, N, N, N | singlet-triplet |
| HUQFAA | RSH | IS-LS | 51.89 | Fe | 2 | 2 | Fe C14 H25 N5 O | octahedral | C, N, N, N, N, N | singlet-triplet |
| YAQGUU01 | Semilocal | HS-LS | 17.31 | Mn | 2 | 0 | Mn C14 Cl2 H18 N4 | octahedral | Cl, Cl, N, N, N, N | doublet-sextet |
| XUBKAI | Moderate | HS-LS | 34.14 | Ni | 2 | -2 | Ni C14 H14 N4 O4 | square planar | N, N, N, N | singlet-quintet |
| YIKCUR | High | HS-LS | 62.17 | Fe | 2 | 0 | Fe C6 H18 N2 O8 P2 | tetrahedral | N, N, P, P | singlet-quintet |
| EPOWAI | RSH | HS-LS | 25.48 | Fe | 3 | 1 | Fe C22 H34 | linear | C, C, C, C, C, C (haptic) | doublet-sextet |

**Table S49.** Examples of structures with very large or small standard deviations among 12 functionals for the magnitude of the dipole moment in the lowest possible spin state. Structures with trivially zero dipole moments (i.e., dipole magnitude less than 0.1 Debye) were excluded when identifying representative structures with small standard deviations. We required each structure to be converged, not spin-contaminated, and not an outlier for any property across all functionals.

| Refcode | Metal | Oxidation State | Charge | Molecular Formula | Geometry | Coordinating Atoms | Spin State | Standard deviation in dipole magnitude (Debye) |
|---|---|---|---|---|---|---|---|---|
| AKEHII | Cd | 2 | 0 | Cd Br C8 H16 N5 O5 | pentagonal bipyramidal | Br, N, O, O, O, O, O | singlet | 3.92 |
| KEJKEP | Cu | 2 | 0 | Cu C14 H10 N2 O5 | square planar | N, N, O, O | doublet | 3.46 |
| DUDGIV | Cu | 2 | 0 | Cu C9 Cl4 H12 N2 | square planar | Cl, Cl, Cl, N | Doublet | 3.07 |
| PANXEH | Pd | 2 | 4 | Pd C12 H32 N6 | square planar | N, N, N, N | singlet | 1.14e-3 |
| FEYTUZ | Ag | 1 | -3 | Ag Cl4 | tetrahedral | Cl, Cl, Cl, Cl | singlet | 1.43e-3 |
| PULTUN | Sc | 3 | 3 | Sc H14 O7 | pentagonal bipyramidal | O, O, O, O, O, O, O | singlet | 1.60e-3 |

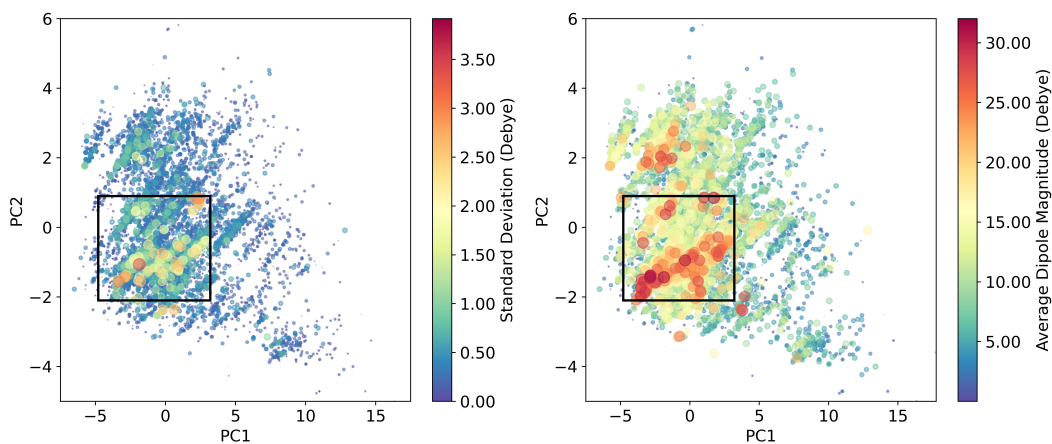

**Figure S13.** Visualization of the standard deviation (left) and average (right) in the dipole moment magnitude among a set of 12 functionals. Plotted over a PCA projection of depth-2 metal-centered RACs. Rectangular insets correspond to the same region of high standard deviation (left) and high average (right) points. We required each structure to be converged, not spin-contaminated, and not an outlier for any property across all functionals.

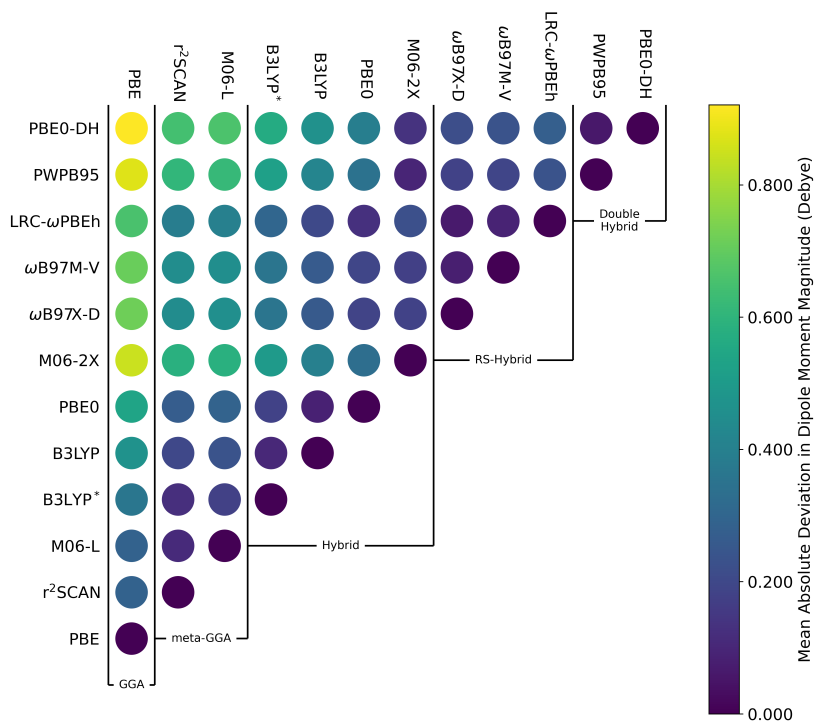

**Figure S14.** Plots of the mean absolute deviation in dipole moment magnitude among a set of 12 functionals. We required each structure to be converged, not spin-contaminated, and not an outlier for any property across all functionals.

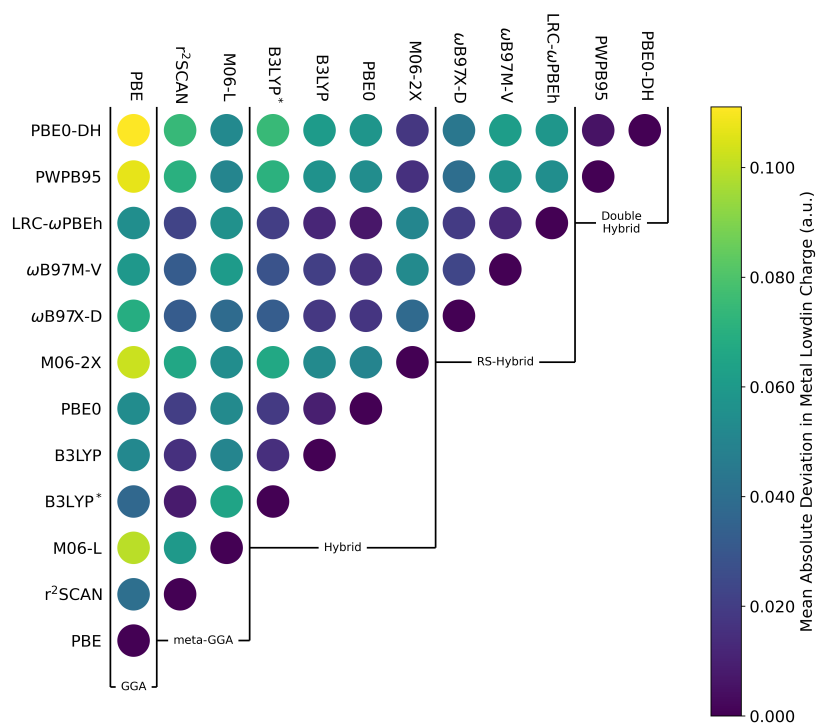

**Figure S15.** Plots of the mean absolute deviation in metal Löwdin charge among a set of 12 functionals. We required each structure to be converged, not spin-contaminated, and not an outlier for any property across all functionals.

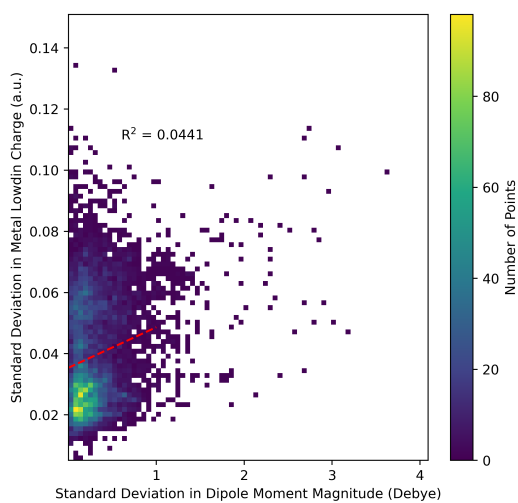

**Figure S16.** Standard deviation among 12 functionals in the metal Löwdin charge against standard deviation in the dipole moment magnitude. We required each structure to be converged, not spin-contaminated, and not an outlier for any property across all functionals.

**Table S50.** Average relative atomization energy (relAE, eV mol/g) for all twelve functionals studied, separated by spin state. We required each structure to be converged, not spin-contaminated, not an outlier for any property, and succeed for all lower-spin states across all functionals.

| Functional | Low Spin | Intermediate Spin | High Spin |
| --- | --- | --- | --- |
| B3LYP | -0.52 | -0.50 | -0.57 |
| B3LYP* | -0.53 | -0.50 | -0.58 |
| LRC-ωPBEh | -0.53 | -0.50 | -0.57 |
| M06-2X | -0.52 | -0.50 | -0.57 |
| M06-L | -0.52 | -0.50 | -0.57 |
| PBE | -0.54 | -0.51 | -0.59 |
| PBE0 | -0.53 | -0.50 | -0.57 |
| r$^2$SCAN | -0.53 | -0.50 | -0.58 |
| ωB97X-D | -0.52 | -0.50 | -0.57 |
| ωB97M-V | -0.53 | -0.50 | -0.57 |
| PBE0-DH | -0.52 | -0.50 | -0.57 |
| PWPB95 | -0.52 | -0.50 | -0.57 |

**Table S51.** Examples of structures with very large or small standard deviations among 12 functionals for atomization energy in the lowest possible spin state. We required each structure to be converged, not spin-contaminated, and not an outlier for any property across all functionals.

| Refcode | Metal | Oxidation State | Charge | Molecular Formula | Geometry | Coordinating Atoms | Spin State | Standard deviation in atomization energy (eV) |
| --- | --- | --- | --- | --- | --- | --- | --- | --- |
| WAGCOY | V | 4 | -2 | V N18 | octahedral | N, N, N, N, N, N | doublet | 3.45 |
| VUHPIZ | V | 5 | -2 | V N15 O | octahedral | N, N, N, N, N, O | singlet | 3.40 |
| BEYNUP | Ti | 4 | -2 | Ti N18 | octahedral | N, N, N, N, N, N | singlet | 3.29 |
| IGUKIH | Cu | 1 | 0 | Cu Cl | unknown | Cl | singlet | 0.19 |
| PEDVEZ | Ag | 1 | -2 | Ag Br3 | trigonal planar | Br, Br, Br | singlet | 0.20 |
| SELWOZ01 | Ag | 1 | -1 | Ag Br2 | linear | Br, Br | singlet | 0.20 |

**Table S52.** Examples of structures with very large or small standard deviations among 12 functionals for relative atomization energy in the lowest possible spin state. We required each structure to be converged, not spin-contaminated, and not an outlier for any property across all functionals.

| Refcode | Metal | Oxidation State | Charge | Molecular Formula | Geometry | Coordinating Atoms | Spin State | Std. dev. relAE (eV mol/g) |
|---|---|---|---|---|---|---|---|---|
| HIWGAV | Co | 3 | -1 | Co C4 O4 | tetrahedral | C, C, C, C | singlet | 1.2e-2 |
| VUHPIZ | V | 5 | -2 | V N15 O | octahedral | N, N, N, N, N, O | singlet | 1.2e-2 |
| DOBVOI | V | 5 | 0 | V Cl3 O | seesaw | Cl, Cl, Cl, O | singlet | 1.1e-2 |
| PEDVEZ | Ag | 1 | -2 | Ag Br3 | trigonal planar | Br, Br, Br | singlet | 5.6e-4 |
| IFIWAV | Cd | 2 | -2 | Cd Br4 | tetrahedral | Br, Br, Br, Br | singlet | 5.9e-4 |
| WOJYUS | Zn | 2 | -2 | Zn Br4 | tetrahedral | Br, Br, Br, Br | singlet | 6.6e-4 |

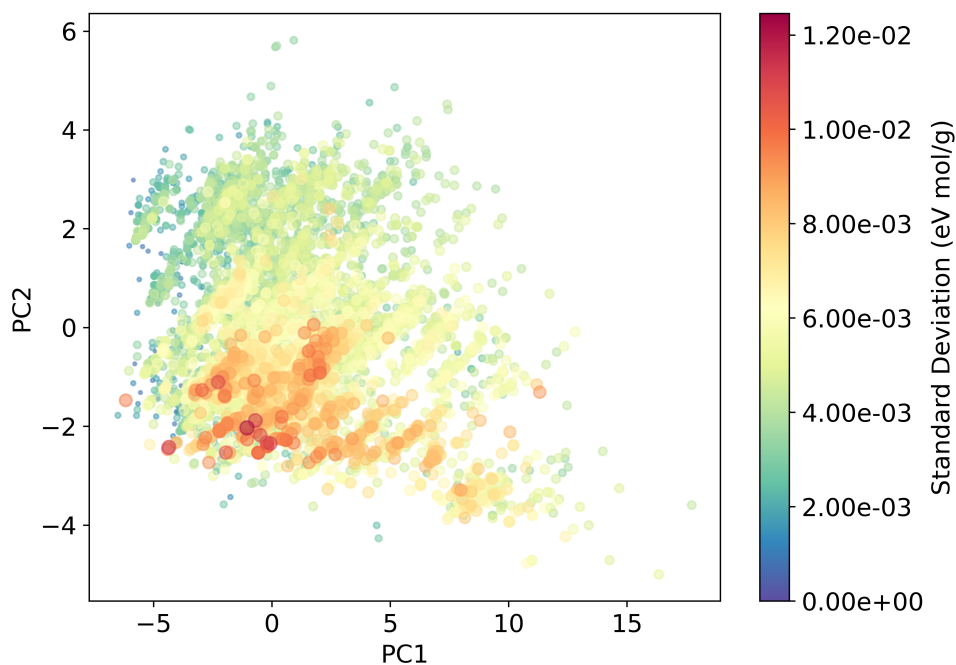

**Figure S17.** Visualization of the standard deviation in the relative atomization energy (relAE) among a set of 12 functionals. Plotted over a PCA projection of depth 2 metal-centered RACs. We required each structure to be converged, not spin-contaminated, not an outlier for any property, and succeed for all lower spin states across all functionals.

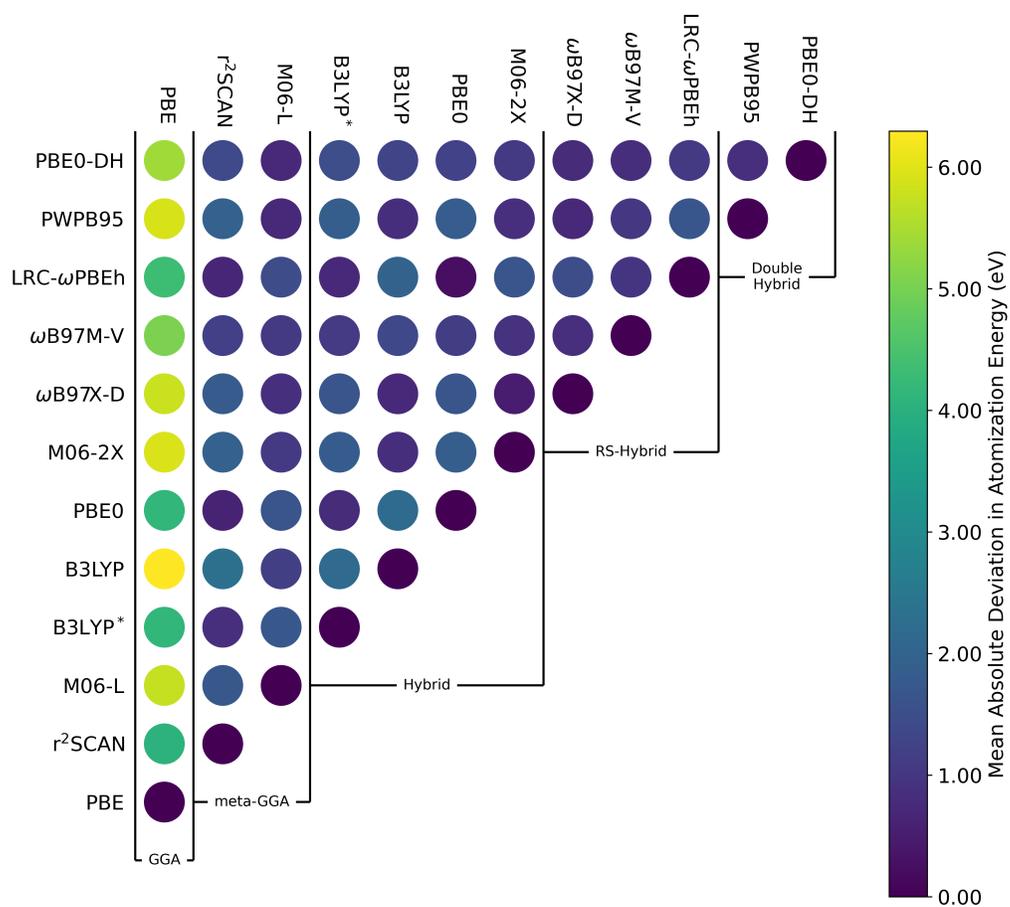

**Figure S18.** Plots of the mean absolute deviation in atomization energy among a set of 12 functionals. We required each structure to be converged, not spin-contaminated, and not an outlier for any property across all functionals.

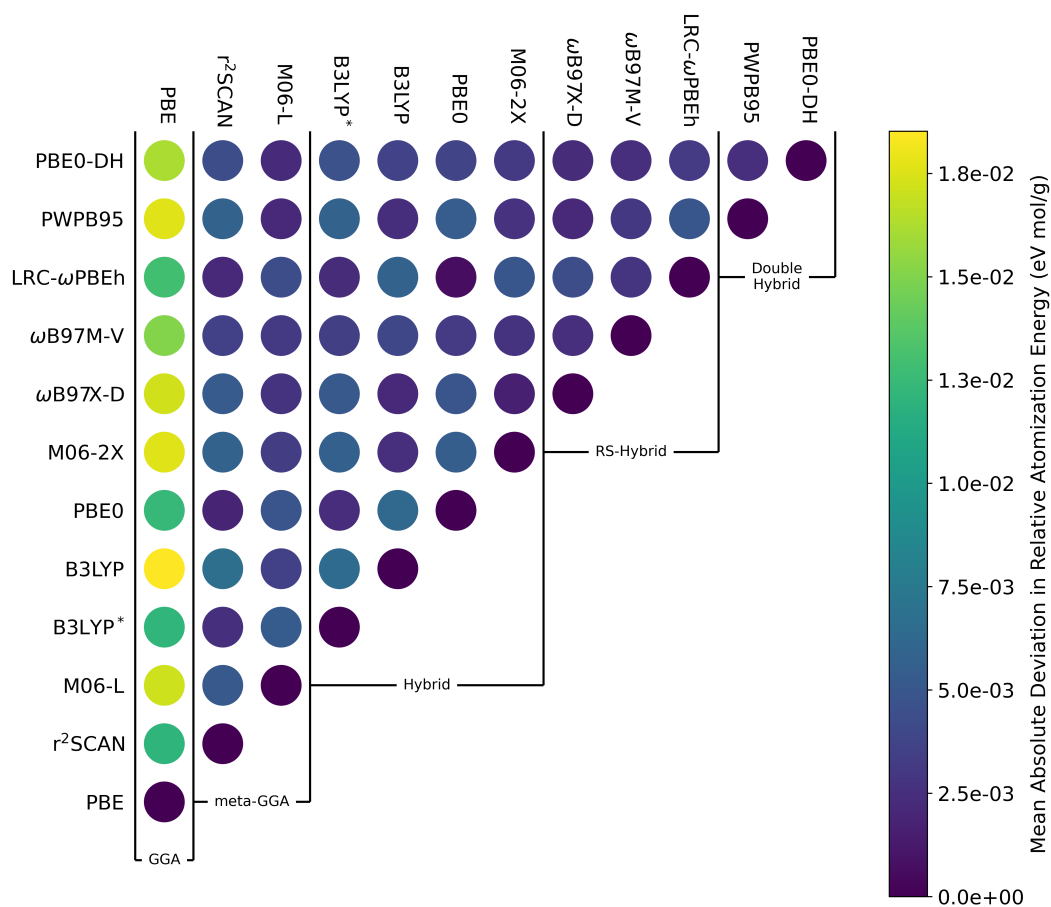

**Figure S19.** Plots of the mean absolute deviation in relative atomization energy (relAE) among a set of 12 functionals. We required each structure to be converged, not spin-contaminated, and not an outlier for any property across all functionals.

Page S47

**Table S53.** Structures with the largest deviations in IS-LS VSSE, dipole moment, metal Löwdin charge, and relative atomization energy, determined by the points with the lowest summed rank of standard deviations in these properties. Smaller ranks indicate that structures had higher standard deviations in the associated property.

| Refcode | Dipole rank | Charge rank | VSSE rank | rel. AE rank | Metal | Ox. State | Charge | Formula | Geom. | Coord. Atoms | Spin state |
|---|---|---|---|---|---|---|---|---|---|---|---|
| KOCXEG | 44 | 1 | 342 | 39 | Co | 3 | 2 | Co C8 H11 N8 S2 | Square Planar | N, N, N, N | Singlet |
| KOGXOT | 354 | 123 | 351 | 679 | Cu | 2 | 0 | Cu C4 H12 N4 O6 | Square Planar | O, O, N, N | Doublet |
| JEYPOS | 199 | 23 | 957 | 405 | Ni | 2 | 0 | Ni C8 H18 N6 S2 | Square Planar | S, S, N, N | Singlet |
| NBPYCU | 191 | 371 | 862 | 525 | Cu | 2 | 0 | Cu C10 H8 N4 O4 | Square Planar | N, N, O, O | Doublet |
| SESCID | 156 | 542 | 284 | 1087 | Cu | 2 | 0 | Cu C10 H9 N5 O5 | Square Pyramidal | N, N, O, O, O | Doublet |

**Table S54.** Pairs of functionals with the lowest and highest Spearman's rank correlation coefficient among several properties.

| Property | Functional 1 | Functional 2 | Correlation coefficient |
|---|---|---|---|
| Dipole moment | LRC- ωPBEh | ωB97X-D | 0.99996 |
| Metal Löwdin charge | PBE0-DH | PWPB95 | 0.99975 |
| IS-LS VSSE | LRC- ωPBEh | ωB97X-D | 0.99950 |
| HS-LS VSSE | LRC- ωPBEh | ωB97X-D | 0.99940 |
| Relative atomization energy | PBE0 | LRC- ωPBEh | 0.99999 |
| Atomization energy | PBE0 | LRC- ωPBEh | 0.99999 |
| Dipole moment | PBE | M06-2X | 0.99319 |
| Metal Löwdin charge | PBE | PBE0-DH | 0.96716 |
| IS-LS VSSE | PBE | M06-2X | 0.91908 |
| HS-LS VSSE | PBE | M06-2X | 0.87594 |
| Relative atomization energy | PBE | M06-2X | 0.99913 |
| Atomization energy | PBE | M06-2X | 0.99930 |

**References**

(1) Garrison, A. G.; Kulik, H. J. System-Specific Reparameterization of Density Functionals with Machine Learning: Application to Spin-Splitting Energies of Transition Metal Complexes. *J. Chem. Theory Comput.* **2026,** *22*, 2243-2260.